\documentclass[aps,prd,preprintnumbers,twocolumn,secnumarabic,balancelastpage,amsmath,amssymb,amsfonts,nofootinbib,floatfix,superscriptaddressletterpaper]{revtex4-2}
\usepackage{graphicx}
\usepackage{braket}
\usepackage{adjustbox}
\usepackage{array}
\usepackage{xcolor}

\begin{document}

\title{Constraints on dark matter with future MeV gamma-ray telescopes}

\preprint{MIT-CTP/5792}

\author{Kayla E.~O'Donnell}
\email{kaylaeod@mit.edu}
\affiliation{Center for Theoretical Physics, Massachusetts Institute of Technology, Cambridge, Massachusetts 02139, USA}

\author{Tracy R.~Slatyer}
\email{tslatyer@mit.edu}
\affiliation{Center for Theoretical Physics, Massachusetts Institute of Technology, Cambridge, Massachusetts 02139, USA}
\affiliation{Department of Physics, Harvard University, Cambridge, Massachusetts 02138, USA}
\affiliation{Radcliffe Institute for Advanced Study at Harvard University, Cambridge, Massachusetts 02138, USA}

\begin{abstract}
    A number of new balloon or space-based $\gamma$-ray observatories have been proposed to close a ``MeV gap'' in sensitivity to $\gamma$ rays in the MeV--GeV energy band. One aspect of the science case for these instruments is their ability to constrain or discover decaying or annihilating dark matter. In this work, we forecast the sensitivity of these instruments for dark matter annihilation or decay for a range of possible Standard Model final states and compare to existing bounds.
\end{abstract}

\maketitle

\section{Introduction}

There are many different experimental approaches to detecting interactions between dark matter and the known particles of the Standard Model. One common distinction is between ``direct detection," where dark matter interacts directly with the apparatus, and ``indirect detection," where the apparatus detects signs of dark matter, such as its annihilation or decay products. Historically, direct detection efforts have been primarily focused on GeV--TeV range weakly interacting massive particles. However, the past decade has seen an increase in well-motivated sub-GeV dark matter candidates, which has spurred an increase in interest in the direct detection of sub-GeV dark matter, resulting in many promising new detector technologies (e.g., Ref.~\cite{2022arXiv220907426C} and references therein).

On the indirect detection front, however, there is still a notable ``sensitivity gap" for $\gamma$ rays in the MeV range, which could be produced by the annihilation or decay of modestly heavier dark matter. To combat this, many instruments capable of detecting MeV-range $\gamma$ rays have been proposed. These include the Compton Spectrometer and Imager (COSI), which has already been selected to run (for indirect-detection-sensitivity forecasts for COSI, see \cite{Caputo:2022dkz}), as well as the Galactic Explorer with a Coded Aperture Mask Compton Telescope (GECCO), the enhanced ASTROGAM (e-ASTROGAM), the All-sky Medium Energy Gamma-ray Observatory (AMEGO), the Advanced Energetic Pair Telescope (AdEPT), the PAir-productioN Gamma-ray Unit (PANGU), the Gamma-Ray and AntiMatter Survey (GRAMS), and the Massive Argon Space Telescope (MAST). Reference \cite{Coogan:2021sjs} explores the constraints that these instruments will be able to place on the annihilation cross section of MeV-range dark matter under the Higgs and vector-portal models, while Ref.~\cite{PhysRevLett.126.171101} explores the constraints that these instruments will be able to place on the fraction of dark matter composed of primordial black holes. Other works in the literature have explored the individual sensitivity of specific instruments to indirect detection signals, e.g.~\cite{PhysRevD.107.023022, Bartels_2017, GUO2023137853}.

In this work, we forecast model-independent constraints that these proposed instruments will be able to place on the dark matter decay lifetime and annihilation cross section for a range of Standard Model final states including photons, electrons, muons, and pions and compare the forecasted sensitivity of these proposed instruments to existing constraints in the literature. Section~\ref{sec:methods} describes our calculations. The signal and background models that we assume are discussed in Secs.~\ref{subsec:signal} and ~\ref{subsec:background}, respectively. In Sec.~\ref{subsec:fisher}, we describe the statistical method we used to derive our projected constraints from the signal and background models and the specifications of the instrument in question. In Sec.~\ref{subsec:instruments}, we describe each of the instruments considered in this work. We then present and discuss our results in Sec.~\ref{sec:results}; we also discuss systematic uncertainties and some internal consistency checks. We conclude in Sec.~\ref{sec:conclusion}. In Appendix \ref{sec:draco}, we include some supplementary results for Draco and M31 targets. In Appendix \ref{sec:atmospheric}, we explore the effect of including atmospheric backgrounds.

\section{Methods}\label{sec:methods}

In order to forecast constraints that these instruments may place on dark matter decay and annihilation, we must determine what decay rates or annihilation cross sections would generate a photon signal that is statistically distinguishable from background using these instruments as a function of the dark matter mass. Since the parameters of our background models have some uncertainty, we use the Fisher-matrix method described in \cite{Edwards_2018}. The Fisher information matrix characterizes the curvature of the likelihood near its maximum, where the likelihood is a function of the parameters governing the signal and background models (a review can be found in the context of cosmology in e.g., Ref.~\cite{albrecht2009findingsjointdarkenergy}). Inverting the Fisher matrix yields the covariance matrix for the relevant parameters, and so can be used to forecast the approximate statistical significance of a signal of a given normalization, if the likelihood is approximately Gaussian. This method quantifies the degeneracy between the dependence of our observed spectra on the signal and background parameters, allowing us to extract limits on the signal parameters that would be distinguishable from background uncertainty. To use this method, we begin by describing models for the signal and background.

\subsection{Signal modeling} \label{subsec:signal}

The observation targets we consider are the Milky Way Galactic Center, the Draco dwarf galaxy, and the M31 galaxy. For each of these targets, we assume the dark matter follows a Navarro-Frenk-White (NFW) density profile \cite{1996ApJ...462..563N}:

\begin{equation} \label{eq:nfw}
    \rho(r) = \frac{\rho_0}{\frac{r}{r_s}\left(1+\frac{r}{r_s}\right)^2},
\end{equation}

where $r$ is the distance to the center of the galaxy, $\rho_0$ is a constant with units of density, and $r_s$ is a constant with units of length.

The observed photon flux, as a function of photon energy, from the decay/annihilation of dark matter particles of mass $m_\chi$ is given by

\begin{equation}
    \frac{d\Phi}{dE_\gamma}(E_\gamma) = \frac{1}{4\pi m_\chi^a} \cdot \left[\int \left[\rho(r)\right]^a d\Omega d\ell \right] \cdot \Gamma \cdot \frac{dN}{dE_\gamma}(E_\gamma),
\end{equation}

where $a=1$ for decay and $a=2$ for annihilation. The factor in brackets is called the $D$ factor in the decay case or the $J$ factor in the annihilation case. The integral covers the entire observation region, which consists of all points in space that are within a particular angular distance to the center of the object from the point of view of the instrument. In this work, we consider an observation region of $10^\circ$ for both decay and annihilation. This is on the same order as the maximum angular resolution for the instruments under consideration (see Fig.~\ref{fig:specs}). Together, the first two factors represent the number of particles in the observation region available to decay or the number of pairs of particles in the observation region available to annihilate. The third factor, $\Gamma$, represents the interaction rate per particle or pair of particles. This means that $\Gamma=1/\tau$ in the decay case and $\Gamma=\left<\sigma v\right>/(2f_\chi)$ in the annihilation case, where $f_\chi=1$ if the particles are self-conjugate and $2$ otherwise. In this work, we assume the particles are self-conjugate. Finally, $dN/dE_\gamma(E_\gamma)$ represents the distribution of photons per decay/annihilation event as a function of photon energy.

Table \ref{table:profileparameters} lists the profile parameters used to derive the $J$ and $D$ factors used in this work and the references from which these results are taken. Table \ref{table:jdfactors} lists the $J$ and $D$ factors used in this work.

\begin{table}
    \begin{center}
        \begin{tabular}{c c c c c} \hline \hline
            Target & $d\ (\mathrm{kpc})$ & $r_s\ (\mathrm{kpc})$ & $\rho_0\ (\mathrm{MeV/cm^3})$ & Ref. \\ \hline
            Galactic Center & $8.122$ & $11$ & $839$ & \cite{PhysRevD.107.023022}, \cite{de_Salas_2019} \\
            Draco & $76$ & $10^{0.32}$ & $976$ & \cite{PhysRevD.107.023022} \\
            M31 & $770$ & $34.6$ & $84.7$ & \cite{10.1093/pasj/psv042} \\ \hline \hline
        \end{tabular}
    \end{center}
    \caption{The NFW profile parameters used to compute our $J$ (annihilation) and $D$ (decay) factors, where $d$ is the distance from Earth to the object and the other profile parameters are as in Eq.~(\ref{eq:nfw}).}
    \label{table:profileparameters}
\end{table}

\begin{table}
    \begin{center}
        \begin{tabular}{c c c} \hline \hline
            Target & $J(10^\circ)$ & $D(10^\circ)$ \\ \hline
            Galactic Center & $7.119\times{10}^{28}$ & $1.158\times{10}^{25}$ \\
            Draco & $1.940\times{10}^{25}$ & $6.949\times{10}^{22}$ \\
            M31 & $6.437\times{10}^{24}$ & $2.309\times{10}^{23}$ \\ \hline \hline
        \end{tabular}
    \end{center}
    \caption{$J$ (annihilation) and $D$ (decay) factors, computed using the NFW profile parameters in Table~\ref{table:profileparameters}. The $J$ factors are in units of $\mathrm{MeV^2\ cm^{-5}}$, while the $D$ factors are in units of $\mathrm{MeV\ cm^{-2}}$.}
    \label{table:jdfactors}
\end{table}

Since this work focuses on MeV-range dark matter, we consider only the following six decay/annihilation final states: a photon pair, a photon and neutral pion, a neutral pion pair, an electron-positron pair, a muon/antimuon pair, and a pair of charged pions. In more realistic dark matter models, interactions between dark matter and standard matter are more complicated (and can be analyzed using tools such as those developed in Refs.~\cite{10.21468/SciPostPhys.8.6.092,Coogan_2022}), but the final state will often be expressible as a linear combination of the states we consider in this work. Because neutral pions decay with a high branching ratio into $\gamma \gamma$, final states involving neutral pions can also stand in for situations where the annihilation/decay produces a new mediator that decays to two photons (although the results in this case will depend somewhat on the mediator mass if it is produced nonrelativistically). In all six cases, we consider the final-state-radiation (FSR) photon spectrum as well as the photon spectra from any radiative decay channels with a branching ratio (BR) greater than 0.01. In the case of dark matter decaying or annihilating directly to two photons, we simply have
\begin{equation}
    \frac{dN}{dE_\gamma}(E_\gamma)=2\delta\left(E_\gamma-\frac{am_\chi}{2}\right),
\end{equation}
where once again $a=1$ for decay and $a=2$ for annihilation.

In the case of dark matter decaying or annihilating to neutral pions, which decay directly to a photon pair, we simply have $dN/dE_\pi=2\delta(E_\gamma-m_\pi/2)$ in the rest frame of the neutral pions, where $E_\pi$ is the energy of the photon in the neutral pion's frame. We now boost this signal to the observation frame, which we assume is stationary with respect to the original dark matter particles. Let $E_\chi$ be the energy of the photon in the rest frame of the dark matter particles. From Appendix B of Ref.~\cite{PhysRevD.91.103531}, we have (assuming isotropic scalar interactions)

\begin{equation}
    \frac{dN_\gamma}{dE_\chi} = \int_{t_\mathrm{min}}^{t_\mathrm{max}} \frac{\epsilon}{\sqrt{1-\epsilon^2}}\frac{dE_\pi}{E_\pi}\frac{dN_\gamma}{dE_\pi},
    \label{eq:transform}
\end{equation}
where $\epsilon \equiv 2m_\pi/(am_\chi)$,
\begin{equation}
    t_\mathrm{min} \equiv \frac{E_\chi}{\epsilon}\left(1-\sqrt{1-\epsilon^2}\right),
\end{equation}
and
\begin{equation}
    t_\mathrm{max} \equiv \mathrm{min}\left[\frac{m_\pi}{2},\frac{E_\chi}{\epsilon}\left(1+\sqrt{1-\epsilon^2}\right)\right].
\end{equation} Plugging our $\delta$-function signal into Eq.~(\ref{eq:transform}), we obtain
\begin{equation}
    \frac{dN_\gamma}{dE_\chi} = \frac{8}{am_\chi\sqrt{1-\epsilon^2}}
\end{equation}
for
\begin{equation}
    \left(1-\sqrt{1-\epsilon^2}\right) < \frac{4E_\chi}{am_\chi} < \left(1+\sqrt{1-\epsilon^2}\right),
\end{equation}
and $dN_\gamma/dE_\chi=0$ otherwise.

For dark matter decaying or annihilating to a photon and neutral pion, we have \cite{PhysRevD.92.023533}
\begin{equation}
    \frac{dN_\gamma}{dE_\chi} = \delta(E-E_0) + \frac{2}{\Delta E}\left(\Theta(E-E_-)-\Theta(E-E_+)\right),
\end{equation}
where
\begin{align}
    E_0 = \Delta E &= \frac{(am_\chi)^2+m_{\pi^0}^2}{2am_\chi}, \\
    E_\pm &= \frac{am_\chi}{2} \left(\left(1+\frac{m_{\pi^0}^2}{(am_\chi)^2}\right) \pm \left(1-\frac{m_{\pi^0}^2}{(am_\chi)^2}\right)\right).
\end{align}

The FSR spectrum for dark matter decay/annihilation into a lepton pair ($l \in \{e,\mu\}$) is given in \cite{PhysRevD.103.063022} as follows:

\begin{equation}
    \frac{dN}{dE} = \frac{2 \alpha}{\pi \beta (3-\beta^2) a m_\chi} \left[\mathcal{A} \ln \left[\frac{1+R(\nu)}{1-R(\nu)}\right] - 2 \mathcal{B} R(\nu)\right]
\end{equation}

where

\begin{equation}
    \mathcal{A} \equiv \frac{(1+\beta^2)(3-\beta^2)}{\nu} - 2(3-\beta^2) + 2\nu,
\end{equation}

\begin{equation}
    \mathcal{B} \equiv \frac{(1-\nu)(3-\beta^2)}{\nu} + \nu,
\end{equation}

Here, $\alpha$ is the fine-structure constant, and we have defined the following quantities: $\mu \equiv m_l/(am_\chi)$, $\beta^2 \equiv 1-4\mu^2$, $\nu \equiv 2E_\gamma/(am_\chi)$, and $R(\nu) \equiv \sqrt{1-4\mu^2/(1-\nu)}$.

In the case of dark matter decaying/annihilating to muons, however, we must also consider the photons that originate from muon radiative decay: $\mu^- \to e^- \bar{\nu}_e \nu_\mu \gamma$. In the rest frame of the muon, this decay process yields the following photon spectrum (per muon decay event) \cite{PhysRevD.103.063022}:

\begin{equation}
    \begin{split}
        \frac{dN}{dE_\gamma} = \frac{\alpha(1-x)}{36\pi E_\gamma} \left[12\left(3-2x(1-x)^2\right)\log\left[\frac{1-x}{r}\right]\right.\\
        +x(1-x)(46-55x)-102\bigg],
    \end{split}
    \label{eq:muonrad}
\end{equation}

where $x \equiv 2E_\gamma/m_\mu$ and $r \equiv (m_e/m_\mu)^2$. We use Eq.~(\ref{eq:transform}), replacing $\pi$ with $\mu$, to boost this signal into the observation frame.

The FSR spectrum for charged pions is given in \cite{PhysRevD.103.063022} as follows:

\begin{equation}
    \begin{split}
        \frac{dN}{dE_\gamma} &= \frac{4\alpha}{\pi \beta a m_\chi} \left[\left(\frac{\nu}{\beta^2} - \frac{1-\nu}{\nu}\right) R(\nu)\right. \\
        &\ \ \ \ + \left(\frac{1+\beta^2}{2\nu} - 1\right) \ln \left(\frac{1+R(\nu)}{1-R(\nu)}\right)\Bigg].
    \end{split}
\end{equation}

Charged pions decay radiatively to muons (BR$>0.999$) or electrons (BR$<0.001$), with each of these processes producing photon spectra given in \cite{PhysRevD.103.063022}. However, the dominant radiative decay signal is the signal from the subsequent muon decay into electrons given in Eq.~(\ref{eq:muonrad}). This signal is boosted into the charged pion's frame as follows:

\begin{equation}
    \frac{dN_\gamma}{dE_\pi} = \int_{t_\mathrm{min}}^{t_\mathrm{max}} \frac{\epsilon'}{1-{\epsilon'}^2} \frac{dE_\pi}{E_\pi} \frac{dN_\gamma}{dE_\mu},
\end{equation}
where $E_\mu$ is the photon's energy in the muon's frame, $E_\pi$ is the photon's energy in the charged pion's frame, $\epsilon' \equiv m_\mu / m_\pi$, $t_\mathrm{min} = E_\pi \epsilon'$, and $t_\mathrm{max} = \mathrm{min}(m_\mu / 2, E_\pi / \epsilon')$. After boosting into the charged pion's frame, we then use Eq.~(\ref{eq:transform}) to boost this signal into the dark matter's rest frame.

We now have our photon flux as a function of dark matter interaction rate for each of the decay/annihilation modes considered in this work. To model the data we can obtain from a real instrument with a nonzero energy resolution, we implement Gaussian smearing on a logarithmic scale and make the conservative assumption that the instruments have no chance of detecting photons immediately outside of their nominal energy range. This forms the function, with normalization parametrized by $\Gamma$, that we use as input to our Fisher matrix.

\subsection{Background modeling} \label{subsec:background}

For all three of the targets we consider, we take the background models from \cite{PhysRevD.107.023022}. For Draco and M31, we use the two-parameter background model

\begin{equation} \label{eq:powerlaw}
    \frac{\partial^2 \Phi}{\partial E_\gamma \partial \Omega} = A \left(\frac{E_\gamma}{1 \mathrm{\ MeV}}\right)^{-\alpha},
\end{equation}

with fiducial parameters $A = 2.4 \times 10^{-3}\ \mathrm{\ MeV^{-1}\ cm^{-2}\ s^{-1}\ sr^{-1}}$, $\alpha = 2$. This model captures the galactic and extragalactic backgrounds and has an estimated systematic error of 15\% \cite{PhysRevD.92.023533}. Since our analysis uses the square root of the background flux, this contributes an estimated systematic error in our analysis of about 7\%.

\begin{figure}
    \includegraphics[width=\linewidth]{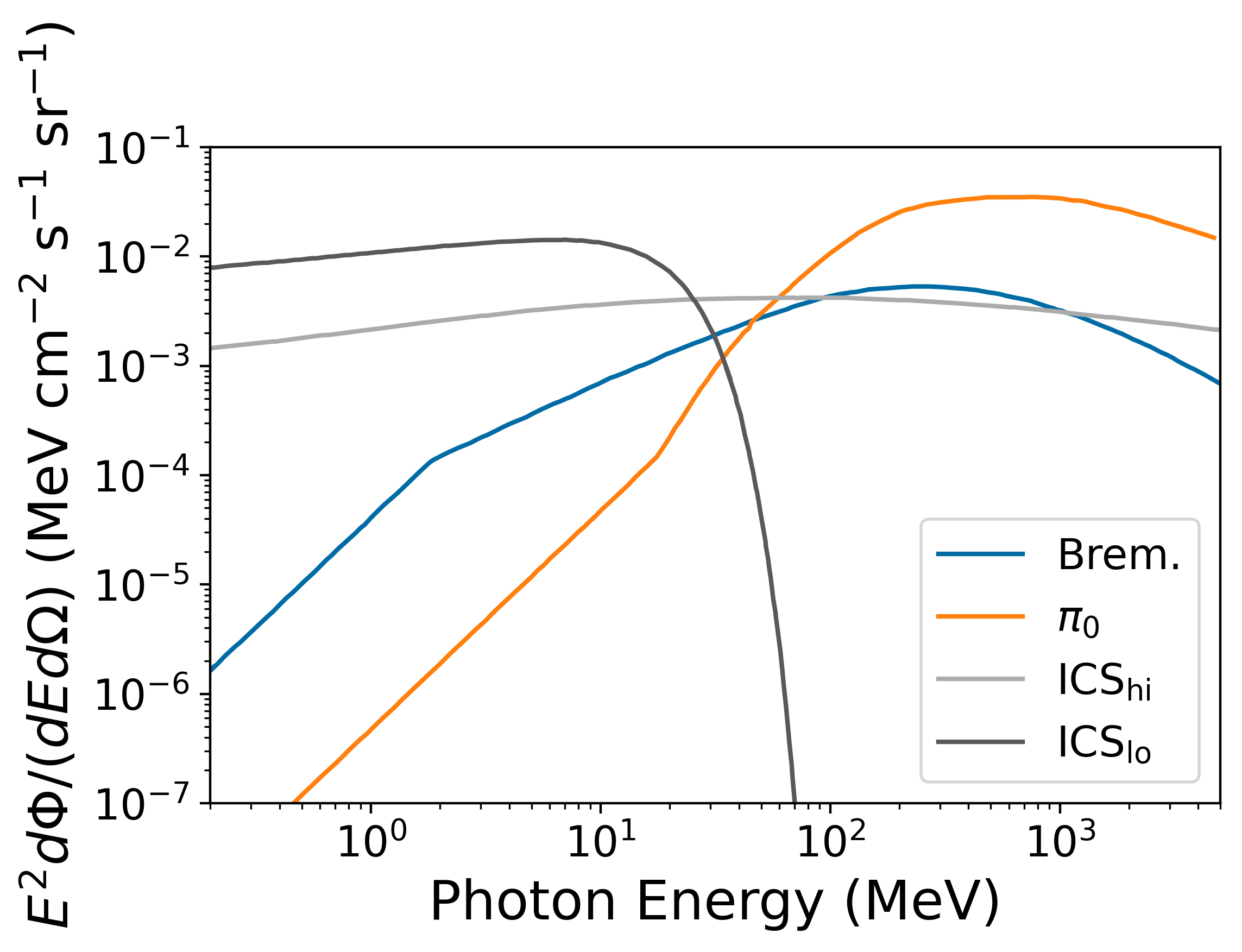}
    \caption{The background photon signal in the direction of the Galactic Center from Fig.~2 in Ref.~\cite{Bartels_2017}, rescaled to the $10^\circ$ cone ROI using data from model A from Ref.~\cite{Calore_2015} at 1 GeV.}
    \label{fig:background}
\end{figure}

For the Galactic Center, we model the galactic and extragalactic backgrounds separately. We split the galactic background into four components: a bremsstrahlung component, a $\pi^0$ component, an inverse-Compton-scattering (ICS) component fit to Fermi data at higher energies, and an ICS component fit to Compton Telescope (COMPTEL) and the Energetic Gamma Ray Experiment Telescope (EGRET) data at lower energies. The spectrum of each of these components is given in Fig.~2 in Ref.~\cite{Bartels_2017} for a $b \leq 5^\circ, l \leq 5^\circ$ region of interest (ROI). In order to rescale each of these components to the $10^\circ$ cone ROI that we are considering, we use all-sky models of the same physical emission components to estimate the ratio of the integrated flux between the two ROI. Specifically, for this purpose we use model A from Ref.~\cite{Calore_2015}. This model is broken up into two components: the ICS component and the $\pi^0$/bremsstrahlung component. For each of the two model A components, we find the ratio of the flux from the $10^\circ$ cone ROI to the flux from the $b \leq 5^\circ, l \leq 5^\circ$ ROI, and this becomes our fiducial rescaling factor for that model A component. These rescaling factors are close to independent of photon energy, so we simply compute them for 1 GeV photons. We then rescale the bremsstrahlung component and $\pi^0$ components from Ref.~\cite{Bartels_2017} by the $\pi^0$/bremsstrahlung rescaling factor from model A, and we rescale the ICS components from Ref.~\cite{Bartels_2017} by the ICS rescaling factor from model A. The rescaled plot is shown in Fig.~\ref{fig:background}.

We use an extragalactic background model of the same form as Eq.~(\ref{eq:powerlaw}), with fiducial parameters $A_\text{e.g.}=0.004135\ \mathrm{\ MeV^{-1}\ cm^{-2}\ s^{-1}\ sr^{-1}}$, $\alpha=2.8956$ \cite{PhysRevD.104.023516}. This model underestimates the extragalactic background at energies above a few MeV (see Fig.~10 of \cite{Ackermann_2015}), but is consistent with observation in the region where the extragalactic background is significant. In the fit, we vary the normalization $A_\text{e.g.}$ of this model but hold the slope $\alpha$ fixed.

We only consider the irreducible backgrounds from cosmic $\gamma$ rays in our analysis. In general, we also expect there to be instrumental backgrounds and backgrounds arising from Earth's atmosphere (especially for balloon experiments). However, most of the experiments we consider have not studied these backgrounds (and their ability to adjust sensitivity cuts to exclude them). For this reason, we defer such an analysis to a future point where more of the instrument concepts have been developed in depth (for example e-ASTROGAM has performed detailed background/instrument modeling in Ref.~\cite{DEANGELIS20181}), but caution that our results may be somewhat optimistic as a result. The GRAMS study in Ref.~\cite{ARAMAKI2020107} finds atmospheric photon backgrounds somewhat larger than our Galactic Center background flux, but not by an enormous factor (within an order of magnitude or so).

\subsection{Statistical analysis} \label{subsec:fisher}

We derive forecasted bounds on the amplitude of the signal from dark matter decay/annihilation using the Fisher-matrix method previously employed in \cite{Edwards_2018}. We first define the vector $\vec{\theta}$ encoding the signal and background models:
\begin{equation}
    \vec{\theta} = (\Gamma, A, \alpha)
\end{equation}
for the two-parameter background models, and
\begin{equation}
    \vec{\theta} = (\Gamma, A_\mathrm{brem}, A_{\pi^0}, A_{\mathrm{ICS_{hi}}}, A_{\mathrm{ICS_{lo}}}, A_{\mathrm{e.g.}})
\end{equation}
for the Galactic Center background model. Here each $A_X$ parameter describes a rescaling of the normalization for the component $X$ (as shown in Fig.~\ref{fig:background} for all the components except the extragalactic one), and the full background model is the sum of these components. The total differential flux is
\begin{equation}
    \phi(\vec{\theta}) = \frac{\partial^2 (\Phi_\chi + \Phi_\mathrm{bg})(\vec{\theta})}{\partial E_\gamma \partial \Omega}.
\end{equation}
The Fisher matrix, which we note is symmetric, is then defined as follows:
\begin{equation}
    \mathcal{F}_{ij} = \int dE_\gamma d\Omega T_\mathrm{obs} A_\mathrm{eff}(E_\gamma) \left(\frac{1}{\phi} \frac{\partial \phi}{\partial \theta_i} \frac{\partial \phi}{\partial \theta_j}\right)_{\vec{\theta}=\vec{\theta}_\mathrm{fid}},
\end{equation}
where $\vec{\theta}_\mathrm{fid}$ are the fiducial background parameters as given in the previous subsection with $\Gamma_\mathrm{fid}$ set to zero. $A_\mathrm{eff}$ represents the effective area of the instrument as a function of incident photon energy, and $T_\mathrm{obs}$ represents the projected observation time. In this work, we take $T_\mathrm{obs} = 10^6 \mathrm{\ s}$ for illustration (around 11.6 days or 280 h, which could be achieved by a balloon flight with a large-field-of-view instrument or a pointed observation by a satellite). We note that the strength of our constraints scales with $\sqrt{T_\mathrm{obs}}$.
We now have
\begin{equation}
    \Gamma_{\chi,\mathrm{max}} = N_\sigma \sqrt{(\mathcal{F}^{-1})_{11}},
\end{equation}
where $N_\sigma$ indicates the desired statistical significance of the bounds, i.e. the degree to which the signal hypothesis would be excluded in the fiducial case where the null hypothesis (no dark matter signal) is true. In this work, we present the $2\sigma$ forecast bounds for comparison with current $2\sigma$ constraints; the limit can simply be rescaled by e.g.~$5/2$ if one instead wishes to forecast the $5\sigma$ upper limit.

Our approach does not use the full spatial information to distinguish signal and background, only spectral information; an analysis using spatial information could potentially improve the forecast constraints, especially for scenarios where the spectrum is less peaked and so more degenerate with the background. In particular, our constraints may not capture the full power of instruments with excellent angular resolution.

\section{Instrument Details} \label{subsec:instruments}
\begin{figure*}
    \includegraphics[width=.48\textwidth]{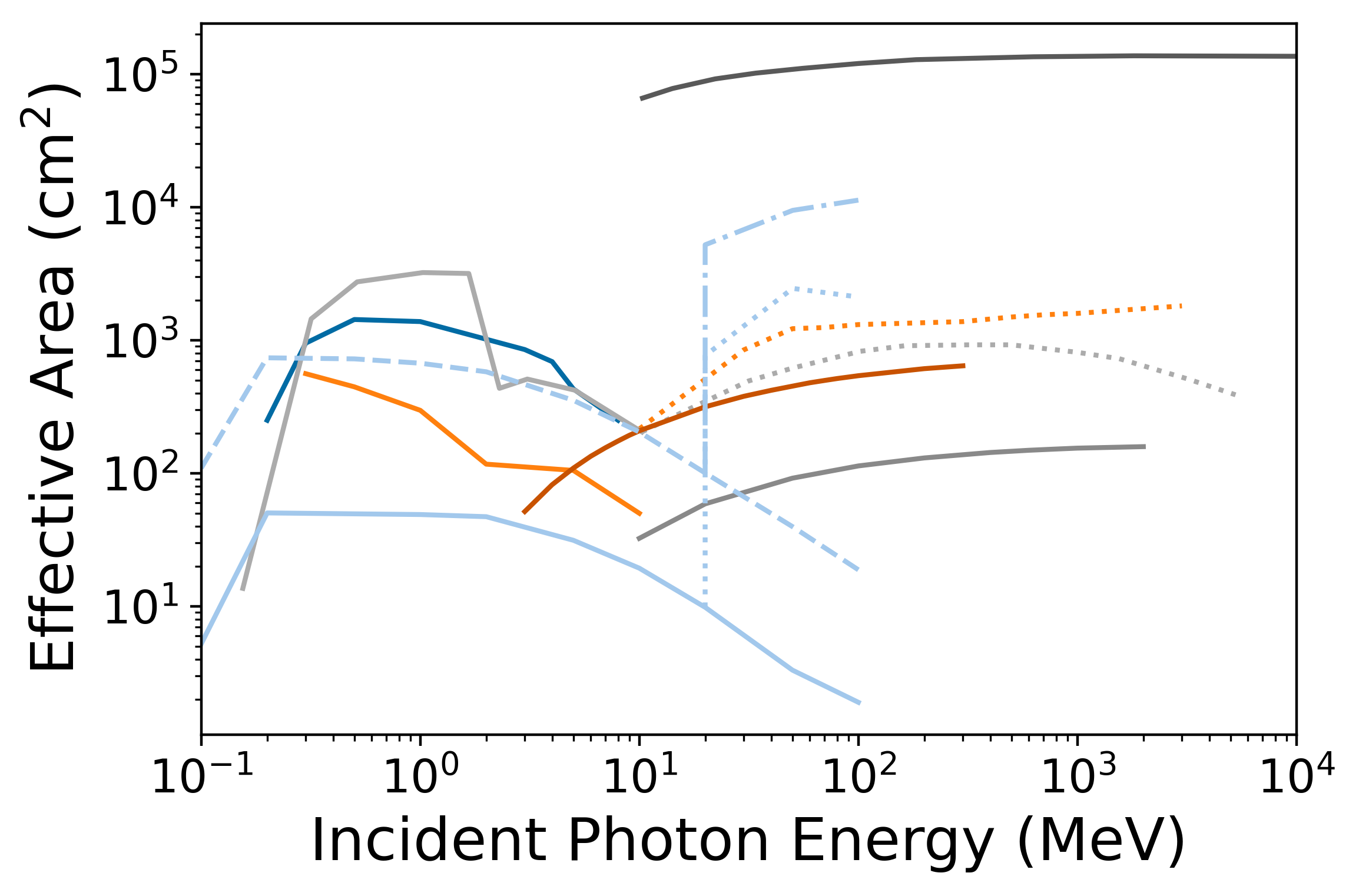} \hfill
    \includegraphics[width=.48\textwidth]{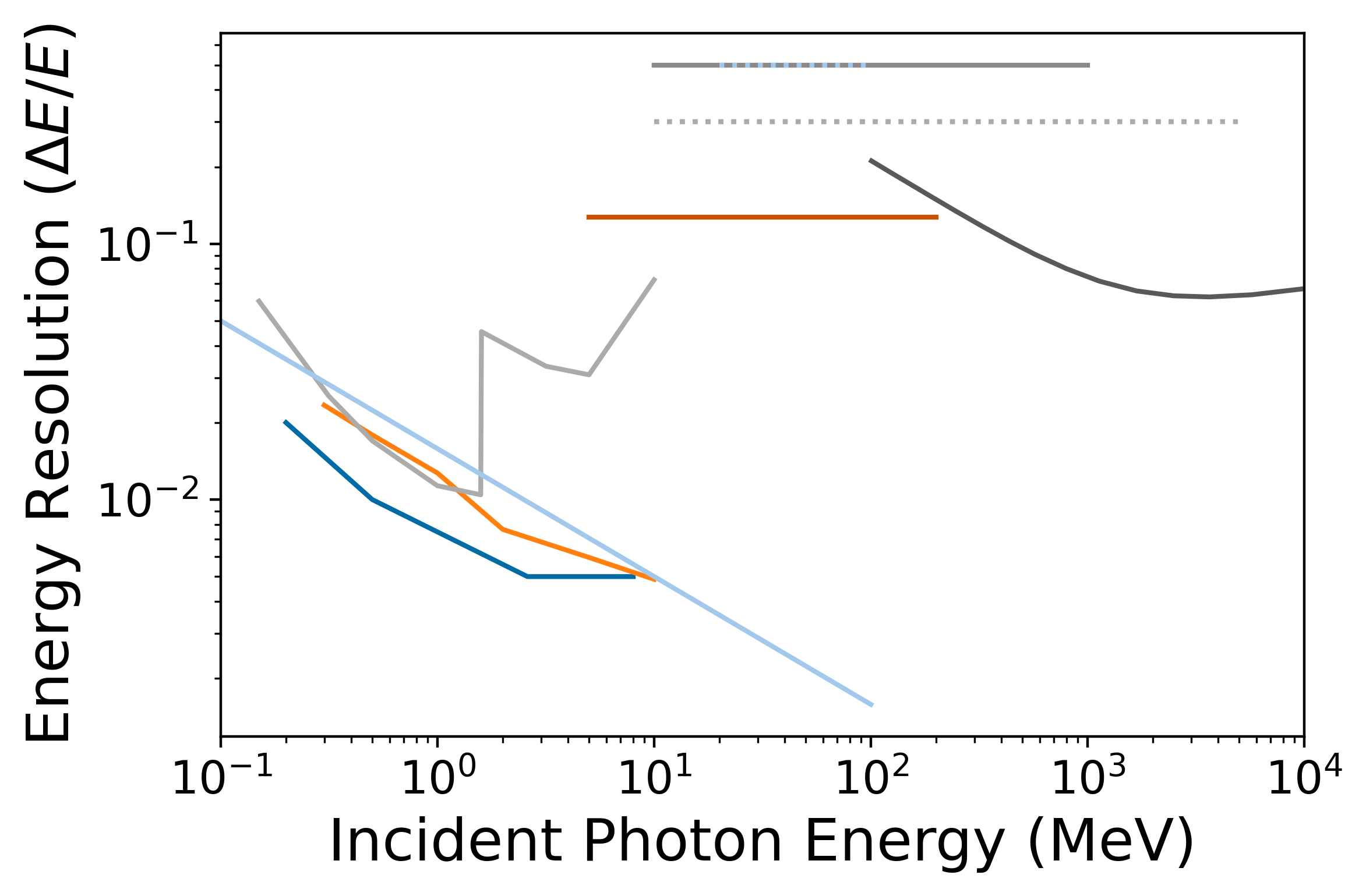} \hfill
    \includegraphics[width=.6\textwidth]{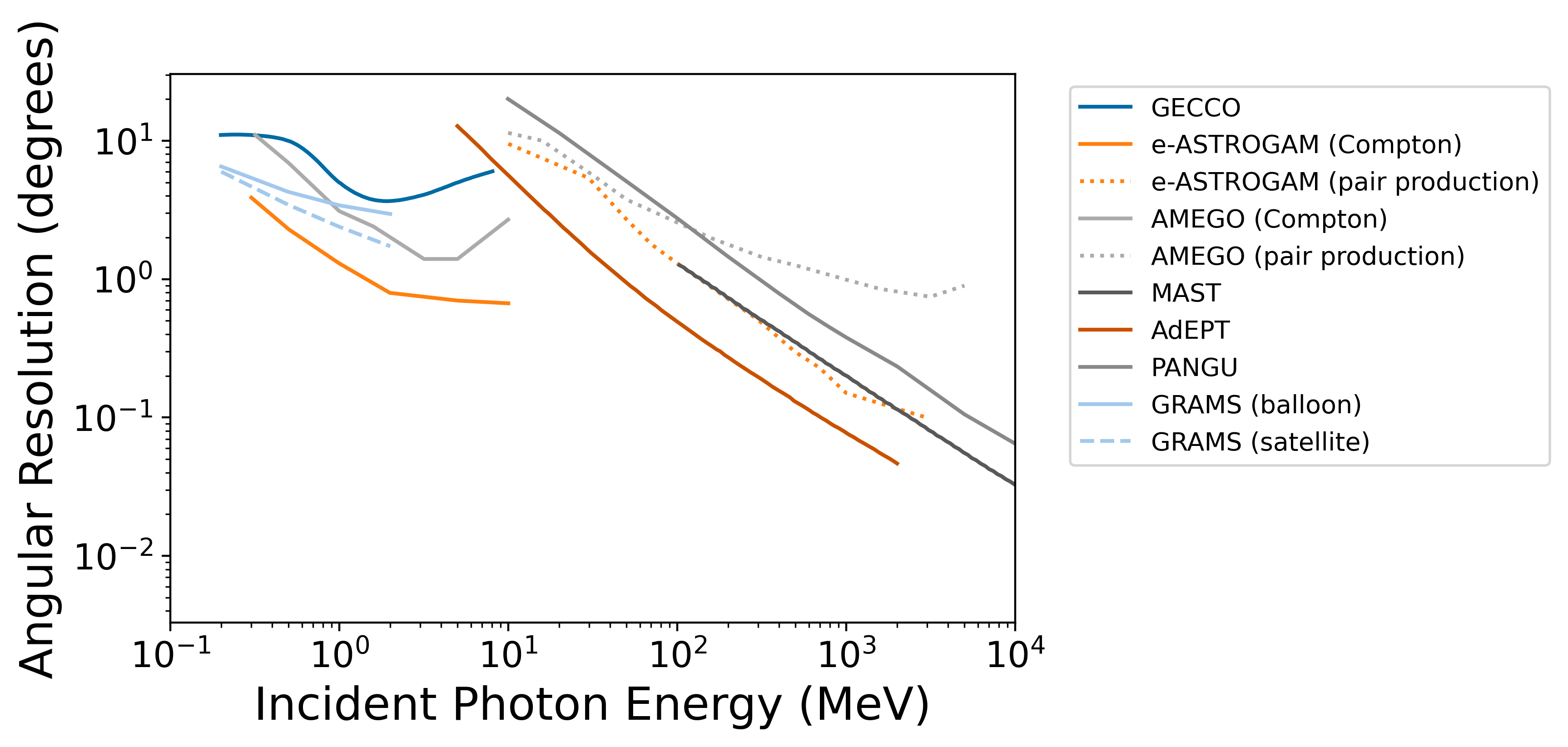}
    \caption{The effective area, energy resolution, and angular resolution as a function of incident photon energy for each of the instruments under consideration. Dotted lines represent the pair-production mode for instruments that have both a Compton mode and a pair-production mode. For GRAMS, we consider the balloon version in our analysis, but there is a proposed upgrade to a satellite version that we also show above using dashed lines for the Compton mode and dot-dashed lines for the pair-production mode. Many of the instruments did not have available a model of the dependence of energy resolution on incident photon energy. For these instruments, we simply used the largest energy resolution given across the entire energy range. The precise definition of angular resolution for each instrument is given in the reference listed for that instrument (see Sec.~\ref{subsec:instruments} for references).}
    \label{fig:specs}
\end{figure*}

Here, we list the instrument concepts whose constraints we are forecasting, along with the assumptions we are making about each instrument. In general, unless otherwise noted, we used log interpolation from the data provided in each paper to form our effective area and energy resolution functions. The effective area, energy resolution, and angular resolution as a function of incident photon energy for each instrument is shown in Fig.~\ref{fig:specs}. For instruments that did not give an explicit dependence of the energy resolution on incident photon energy, we took the largest value within the range that was given. Each of these instruments has a field of view between 1 and $2\pi$ sr.

\subsection{GECCO}
GECCO combines a Compton telescope with a coded mask telescope for high energy and angular resolution in the range of 0.2--8 MeV. The expected energy resolution is $<1^\circ$ between 0.5 and 5 MeV, and the expected angular resolution is $\sim 1$ arc min in the $4 \times 4\ \mathrm{deg^2}$ coded aperture mask data analysis and $4^\circ$--$8^\circ$ in the $60 \times 60\ \mathrm{deg^2}$ Compton telescope analysis. An overview of GECCO is given by \cite{Orlando_2022}.

\subsection{e-ASTROGAM}
e-ASTROGAM has two modes: a Compton scattering and a pair-production mode. The Compton scattering mode has a fine (order 1\%) energy resolution, while the pair-production mode has a coarser (20\%--30\%) energy resolution. For our bounds, we assumed an energy resolution of 30\% ($\Delta E / E$) throughout the pair-production regime. An overview of e-ASTROGAM is given by \cite{DEANGELIS20181}.

\subsection{AMEGO}
AMEGO also has a Compton scattering and a pair-production mode. The Compton scattering mode is divided into two parts: untracked Compton and tracked Compton. This division is responsible for the abrupt jump in the projected decay lifetime constraints. The energy resolution of the Compton mode is given in \cite{Kierans:2020otl} for each photon energy; for the pair-production mode, we used a conservative assumption of 30\% energy resolution ($\Delta E / E$) throughout. An overview of AMEGO is given by \cite{Kierans:2020otl}. There is a more recent proposal for a similar instrument, AMEGO-X \cite{Caputo_2022}; AMEGO-X has broadly similar effective area and energy resolution to AMEGO, and so we would expect comparable sensitivity, but we show AMEGO results due to the availability of more detailed information on effective area and resolution.

\subsection{MAST}
MAST is a liquid argon time-projection chamber and covers an energy range of 100 MeV--1 TeV. MAST has $\sim 10\%$ energy resolution and $<1^\circ$ angular resolution (with much better angular resolution at higher energies) in the energy range that we consider in this work. An overview of MAST is given by \cite{DZHATDOEV20191}.

\subsection{AdEPT}
AdEPT is a pair-production telescope covering the energy range of 5--200 MeV. The effective area at each energy is detailed in \cite{HUNTER201418}; this paper also gives an energy resolution of 30\% FWHM at 70 MeV. For our calculations, we assumed that the instrument will have this energy resolution throughout its energy range. An overview of AdEPT is given by \cite{HUNTER201418}.

\subsection{PANGU}
PANGU is a pair-production telescope covering the energy range of 10 MeV--1 GeV. Reference \cite{Wu:2014tya} gives an upper bound on the energy resolution of 50\% ($\Delta E / E$); we used that value throughout its energy range in our calculations. An overview of PANGU is given by \cite{Wu:2014tya}.

\subsection{GRAMS}
GRAMS uses a liquid argon time-projection chamber and covers the 100 keV--100 MeV energy range. Its energy resolution has been estimated to be of order 1\%, with a slightly coarser energy resolution at lower energies \cite{ARAMAKI2020107}. At higher energies, GRAMS has a pair-production mode that would provide increased effective area, but worsened energy resolution. For this mode we take the energy resolution to be $50\%$, consistent \cite{privcomm_aramaki} with the continuum sensitivity analysis of Ref.~\cite{ARAMAKI2020107}.

An overview of GRAMS is given by \cite{Aramaki:2021o5}. An engineering balloon flight with a small-scale liquid argon time-projection chamber has recently been conducted \cite{Nakajima:2024fgx}, paving the way for a planned satellite version of GRAMS with detector upgrades. In Fig.~\ref{fig:specs}, we plot the effective area and angular resolution for both the balloon (solid for Compton scattering mode, dotted for pair-production mode) and the satellite (dashed for Compton mode, dot-dashed for pair-production mode) versions of GRAMS. The balloon and satellite versions have the same projected energy resolution. In our analysis, we compute the projected sensitivity of the balloon version of GRAMS. The satellite version would improve all sensitivity estimates by a factor of a few due to the increased effective area.

\section{Results and Discussion} \label{sec:results}

\subsection{Current limits and forecast sensitivity}

All of our projected constraints assume an observation time of $T_{\mathrm{obs}} = 10^6 \mathrm{\ s}$. The strength of the constraints simply scales with the square root of the observation time.

\begin{figure*}
    \includegraphics[width=.48\textwidth]{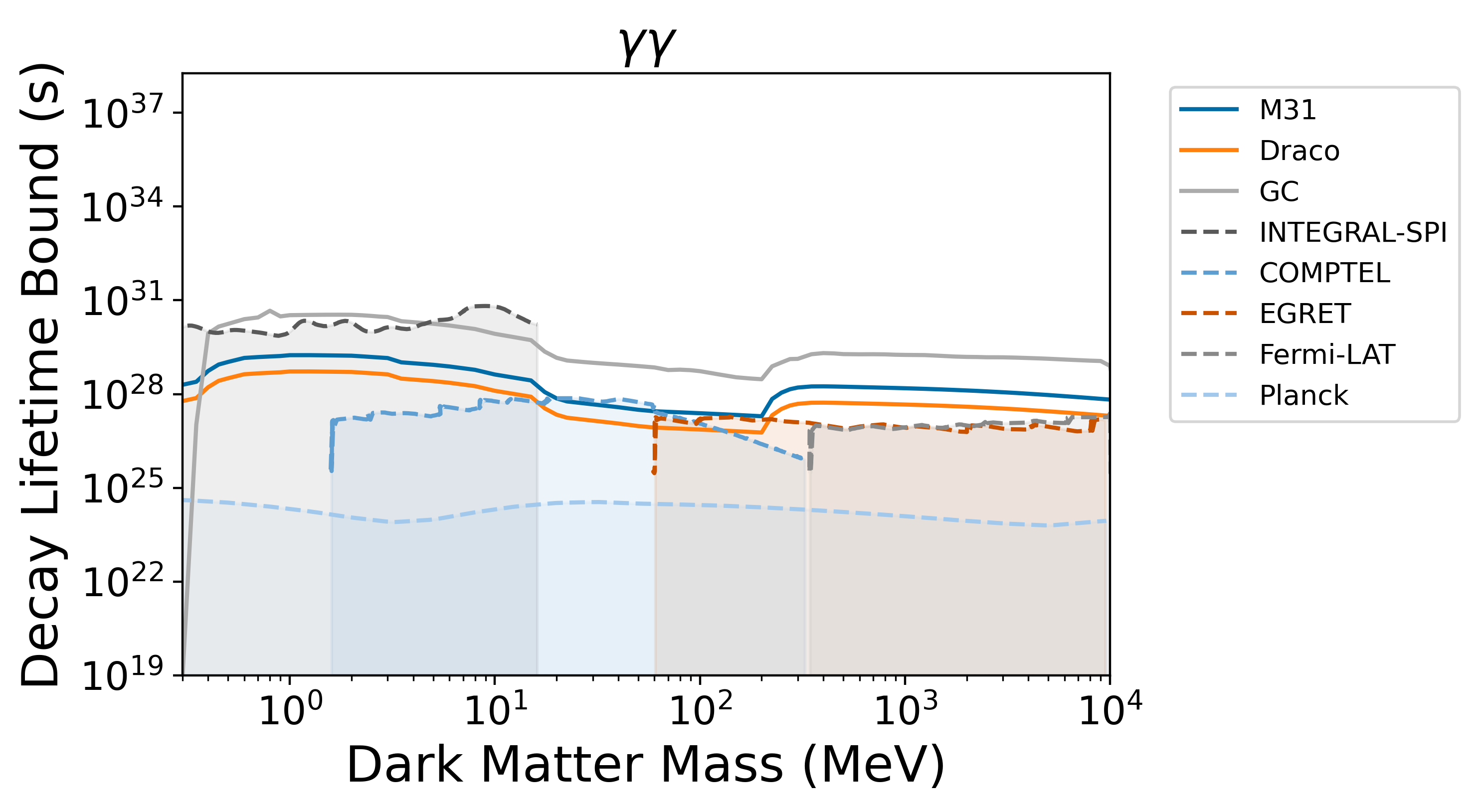} \hfill
    \includegraphics[width=.48\textwidth]{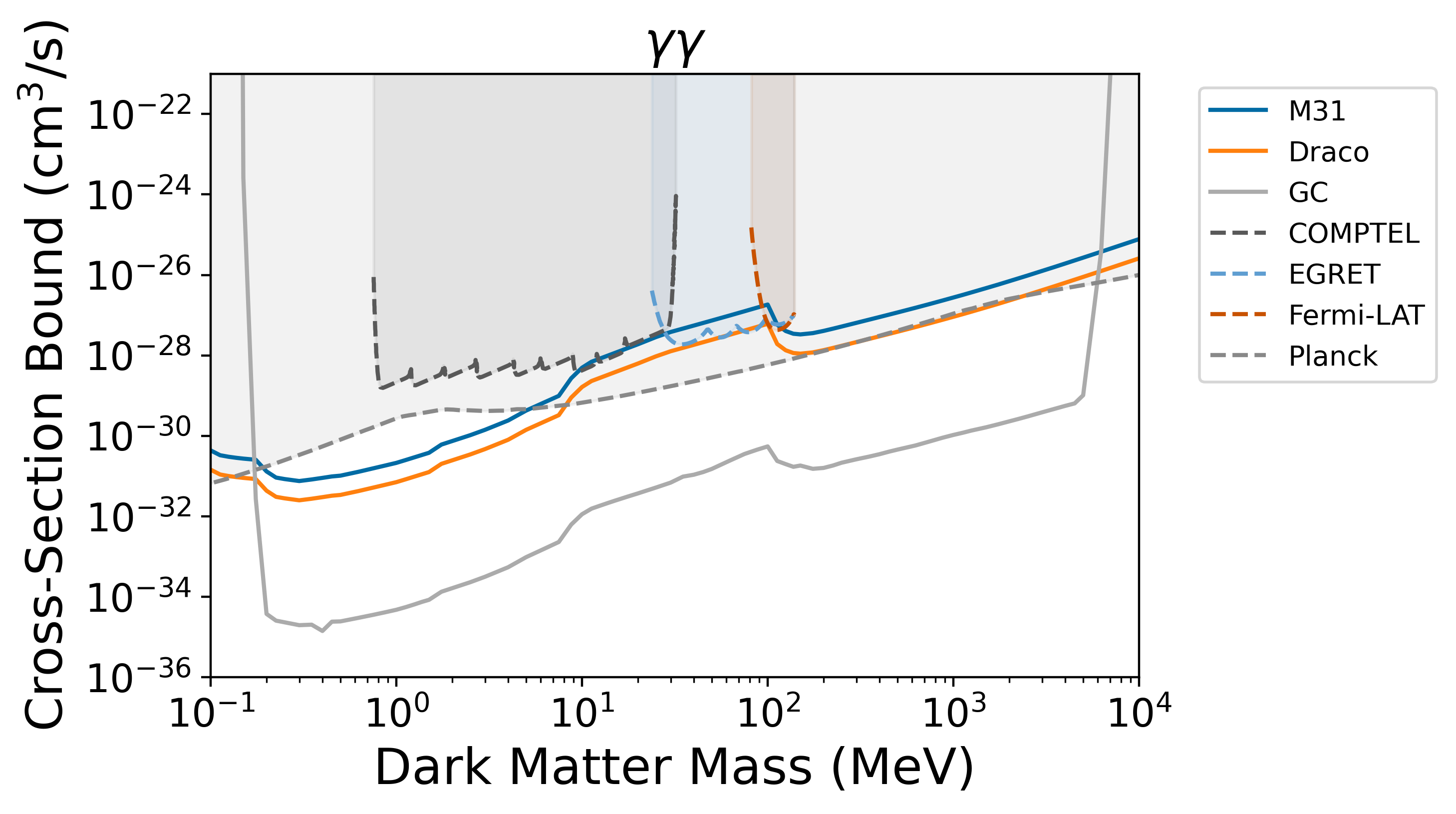}
    \includegraphics[width=.48\textwidth]{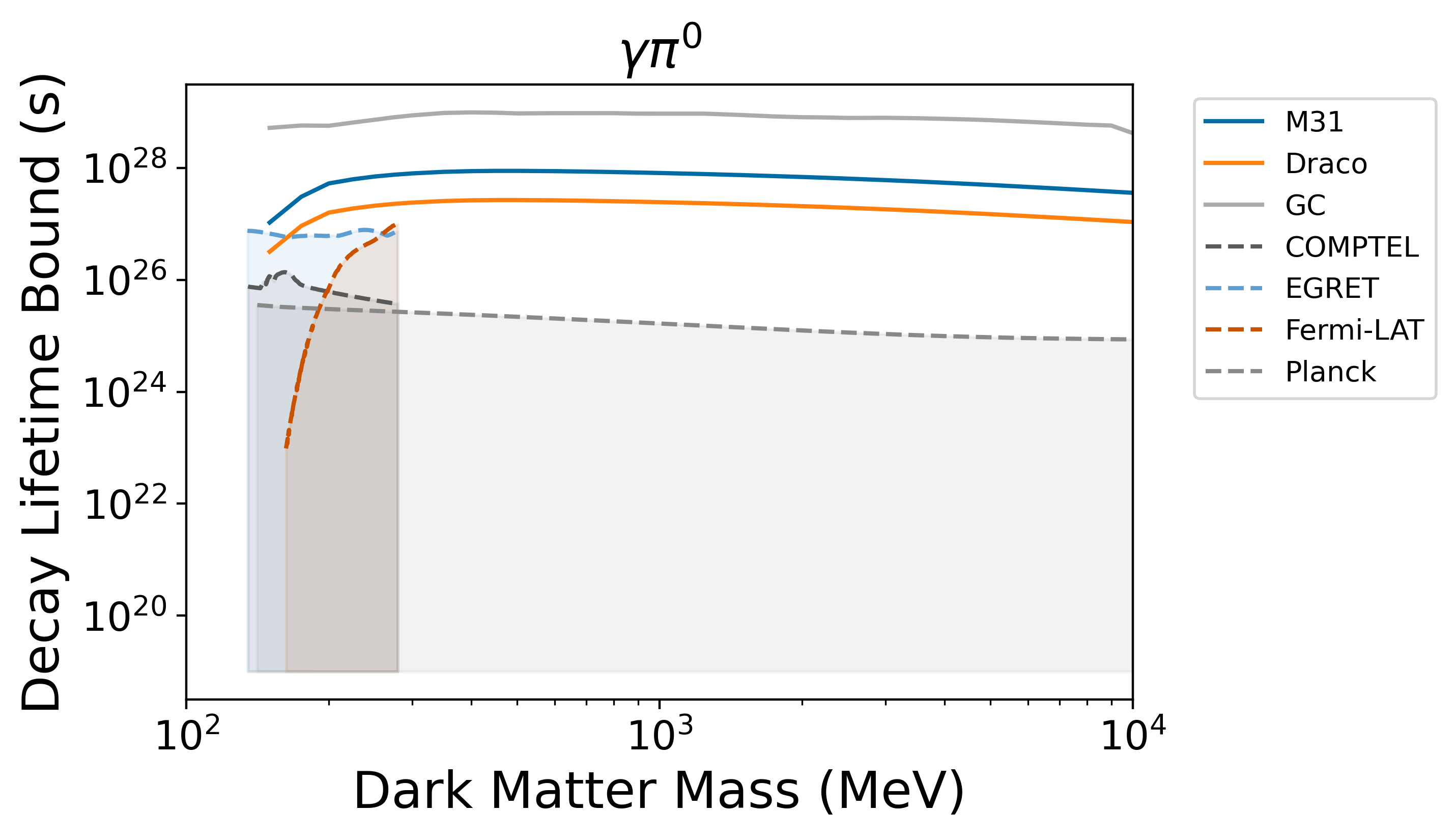} \hfill
    \includegraphics[width=.48\textwidth]{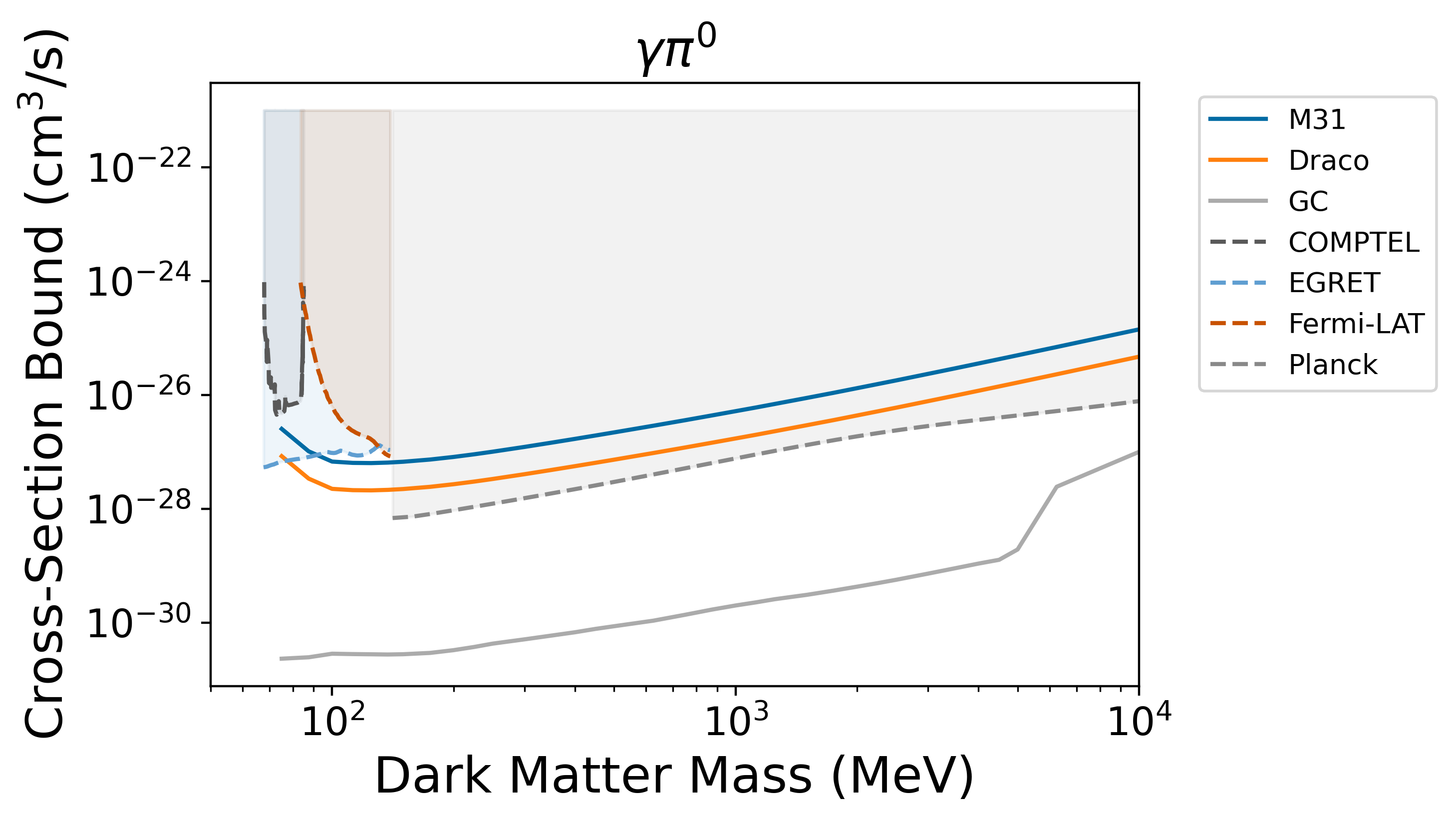}
    \includegraphics[width=.48\textwidth]{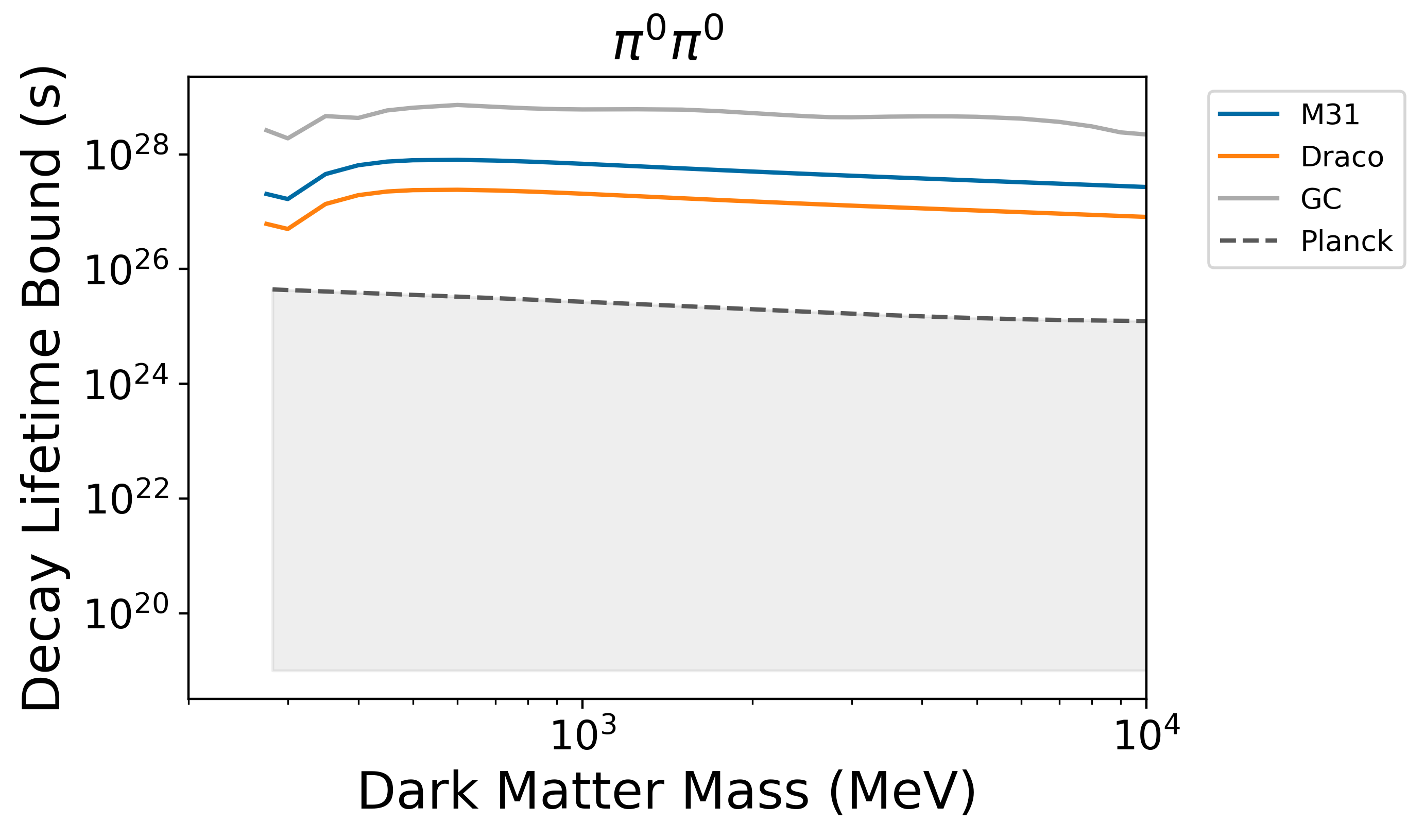} \hfill
    \includegraphics[width=.48\textwidth]{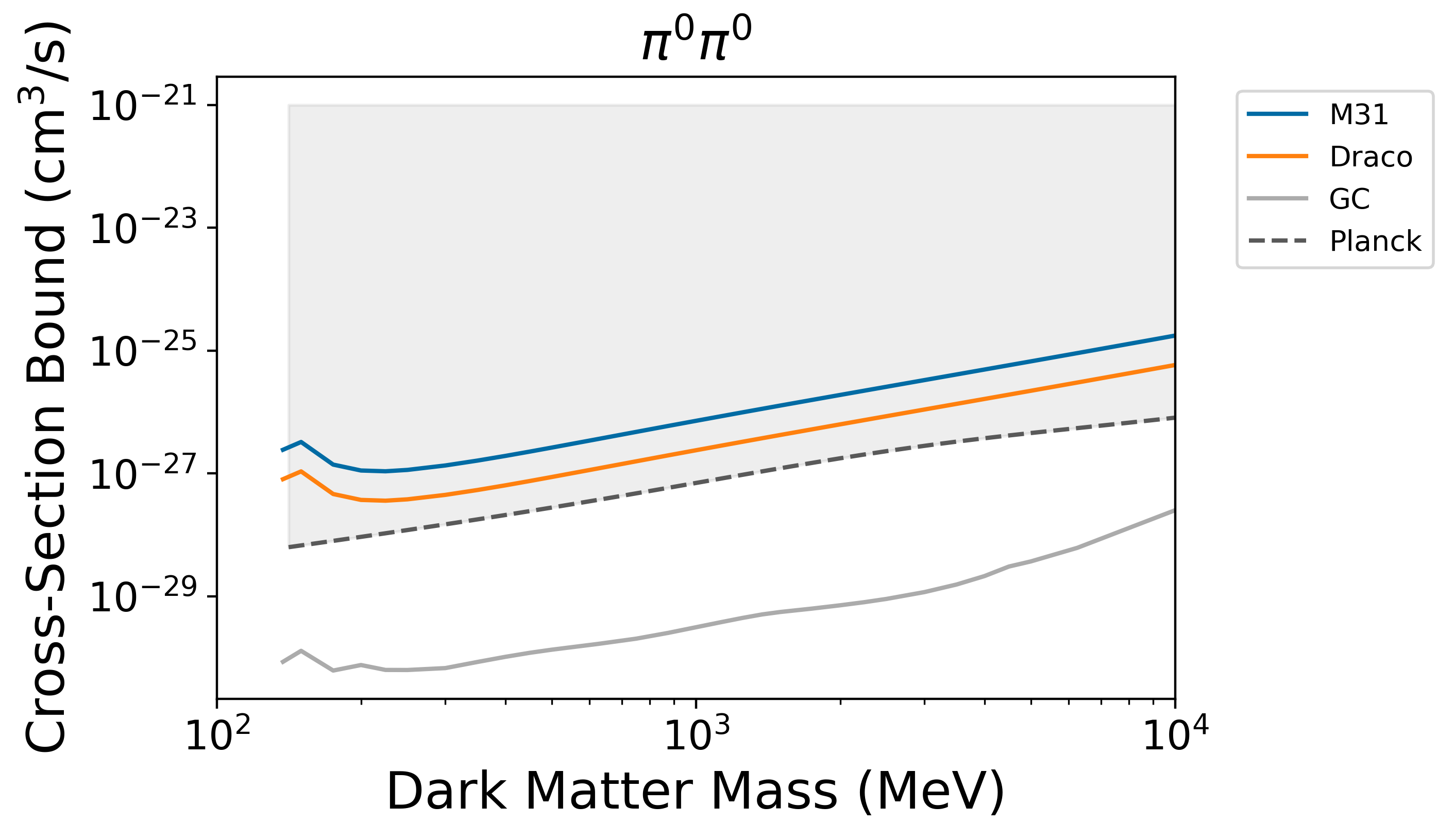}
    \caption{The best constraints across all the instruments considered for dark matter decay and annihilation into (from top to bottom) two photons, a photon and a neutral pion, and two neutral pions, assuming an observation time of $T_{\mathrm{obs}}=10^6 \mathrm{\ s}$. These are compared against the current constraints described in the text of Sec.~\ref{sec:results}.}
    \label{fig:main_neutral}
\end{figure*}

\begin{figure*}
    \includegraphics[width=.48\textwidth]{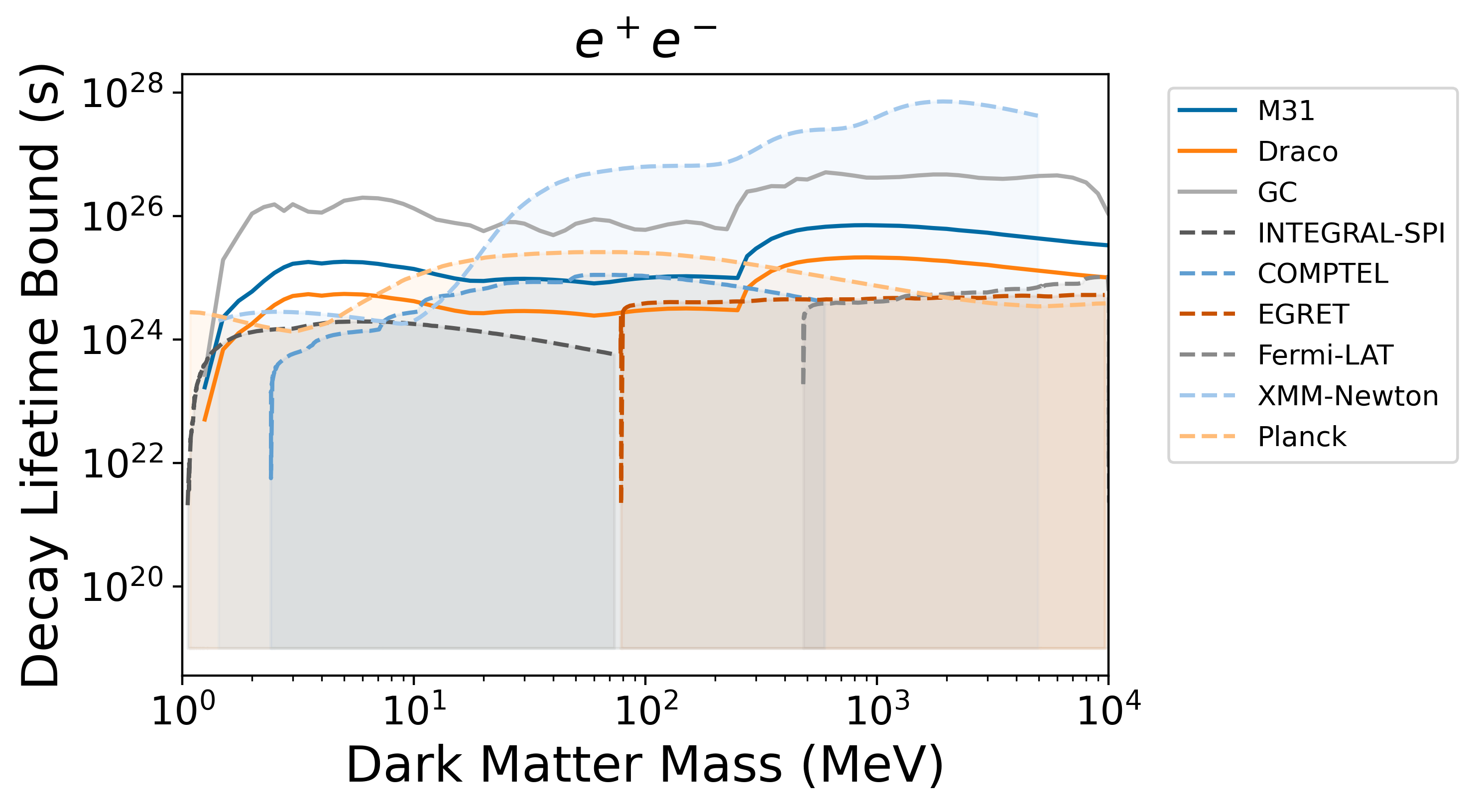} \hfill
    \includegraphics[width=.48\textwidth]{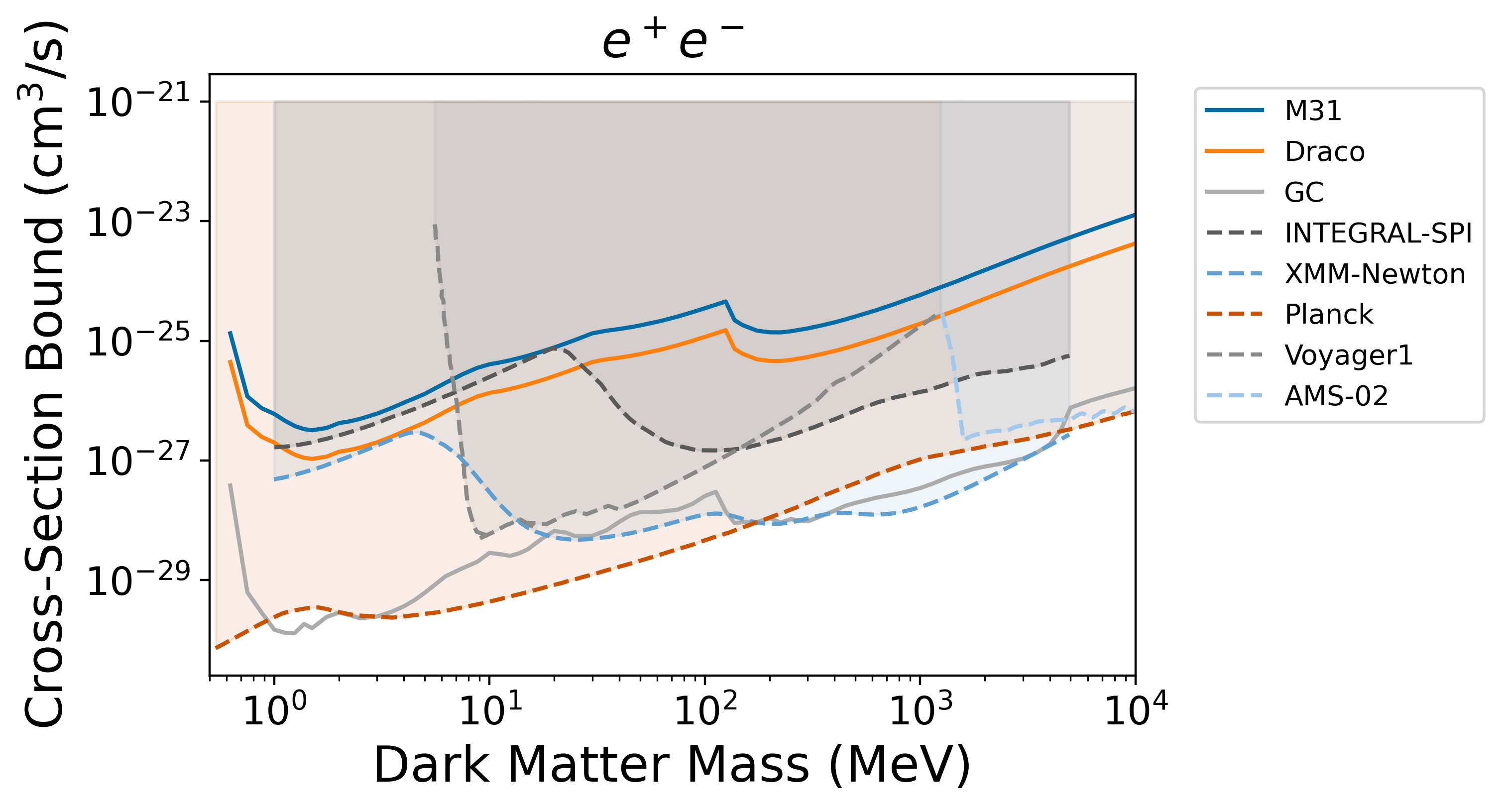}
    \includegraphics[width=.48\textwidth]{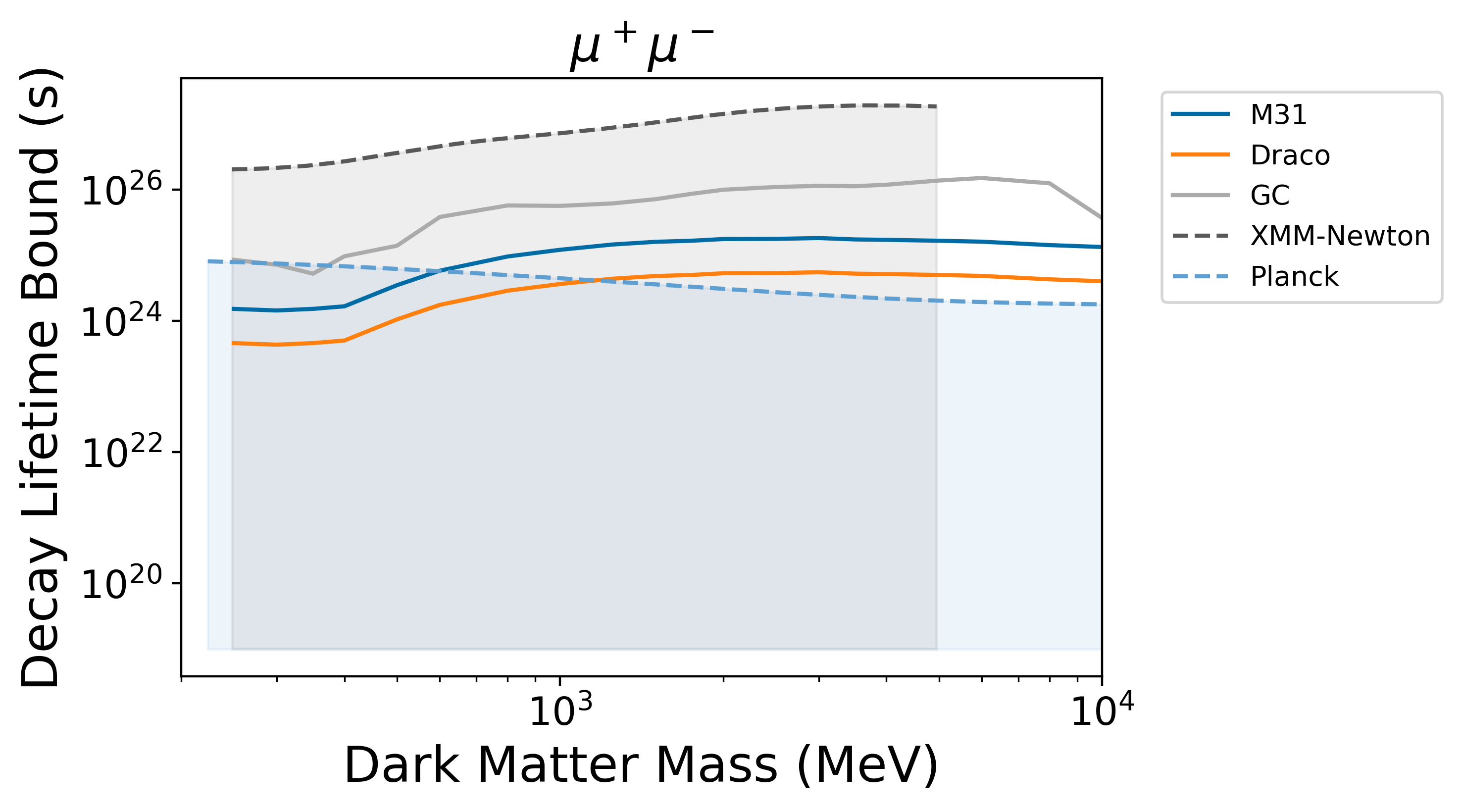} \hfill
    \includegraphics[width=.48\textwidth]{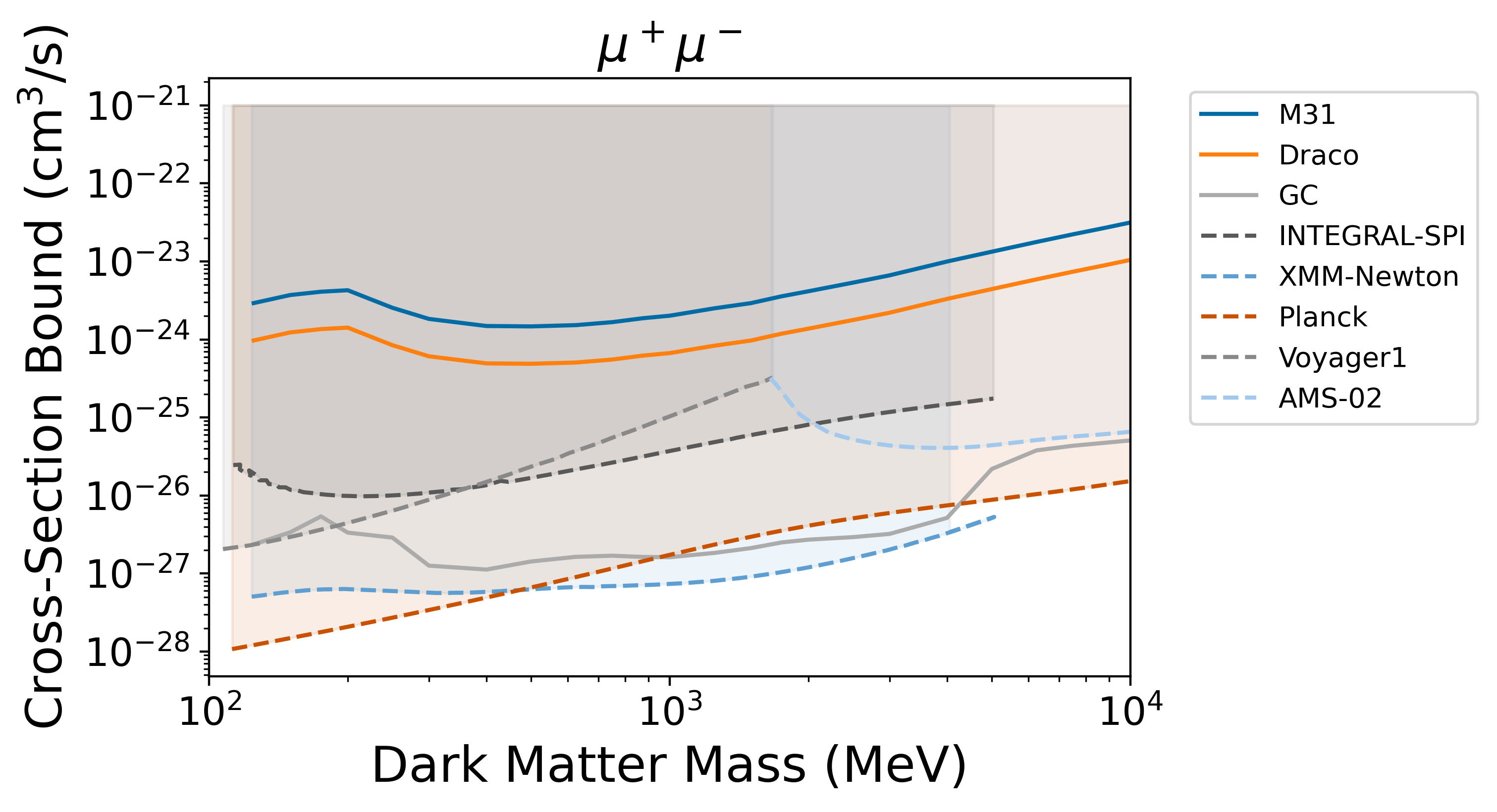}
    \includegraphics[width=.48\textwidth]{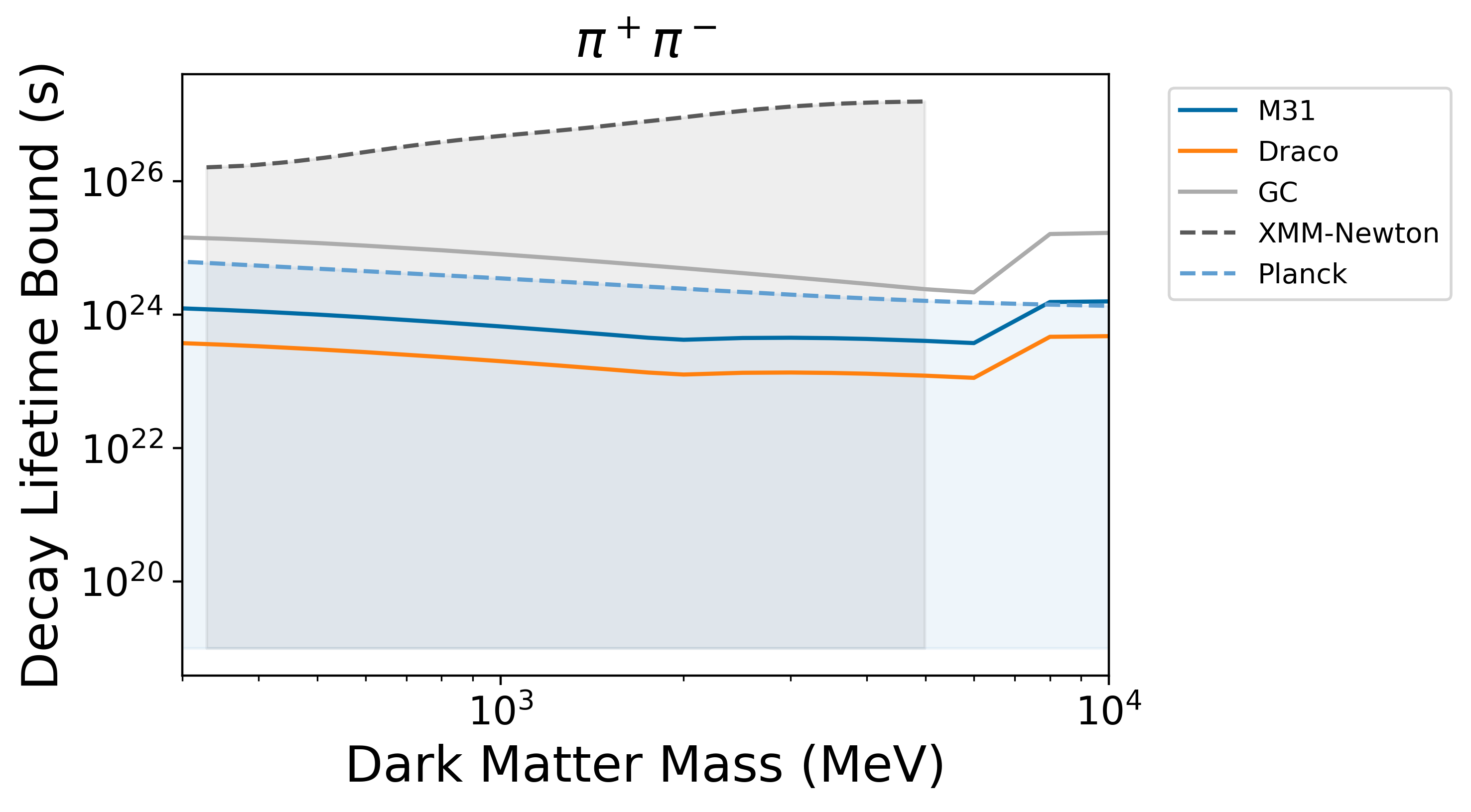} \hfill
    \includegraphics[width=.48\textwidth]{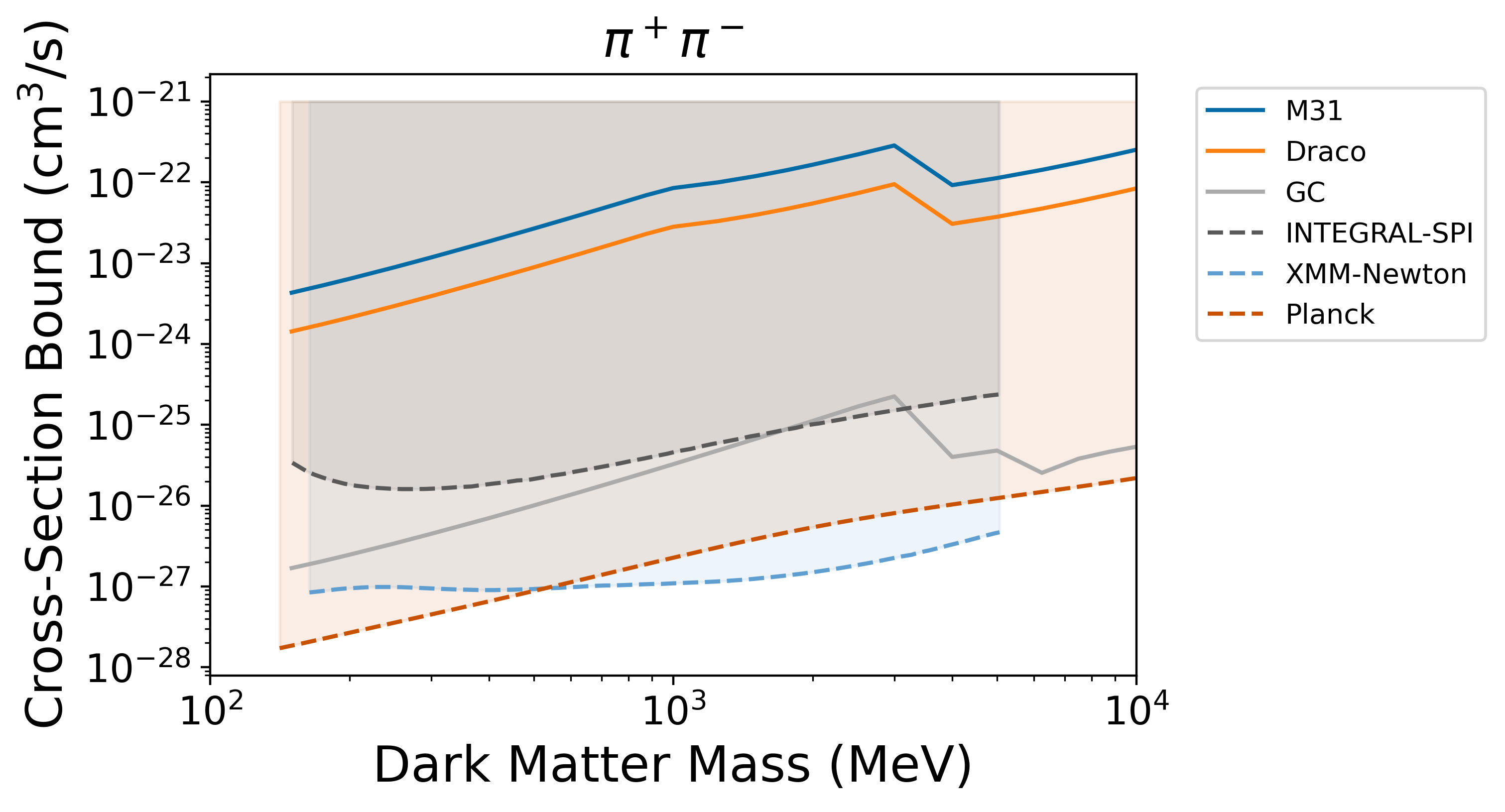}
    \caption{The best constraints across all the instruments considered for dark matter decay and annihilation into the final states (from top to bottom) $e^+e^-$, $\mu^+\mu^-$, and $\pi^+\pi^-$, assuming an observation time of $T_{\mathrm{obs}}=10^6 \mathrm{\ s}$. These are compared against the current constraints described in the text of Sec.~\ref{sec:results}.}
    \label{fig:main_charged}
\end{figure*}

In Figs.~\ref{fig:main_neutral} and \ref{fig:main_charged}, we present the best constraints, across all the instruments considered, for each decay/annihilation mode and target considered.

\begin{figure*}
    \includegraphics[width=.48\textwidth]{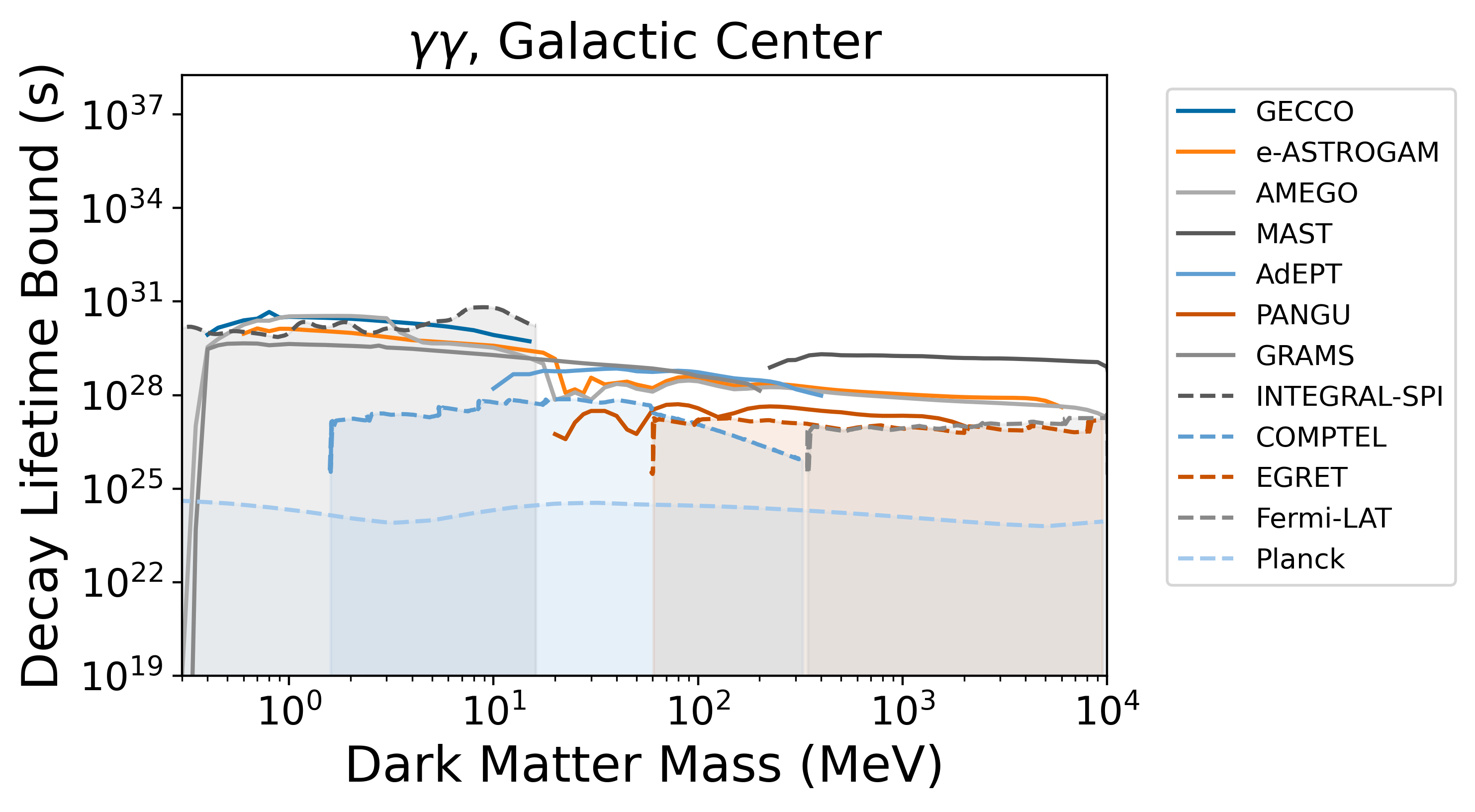} \hfill
    \includegraphics[width=.48\textwidth]{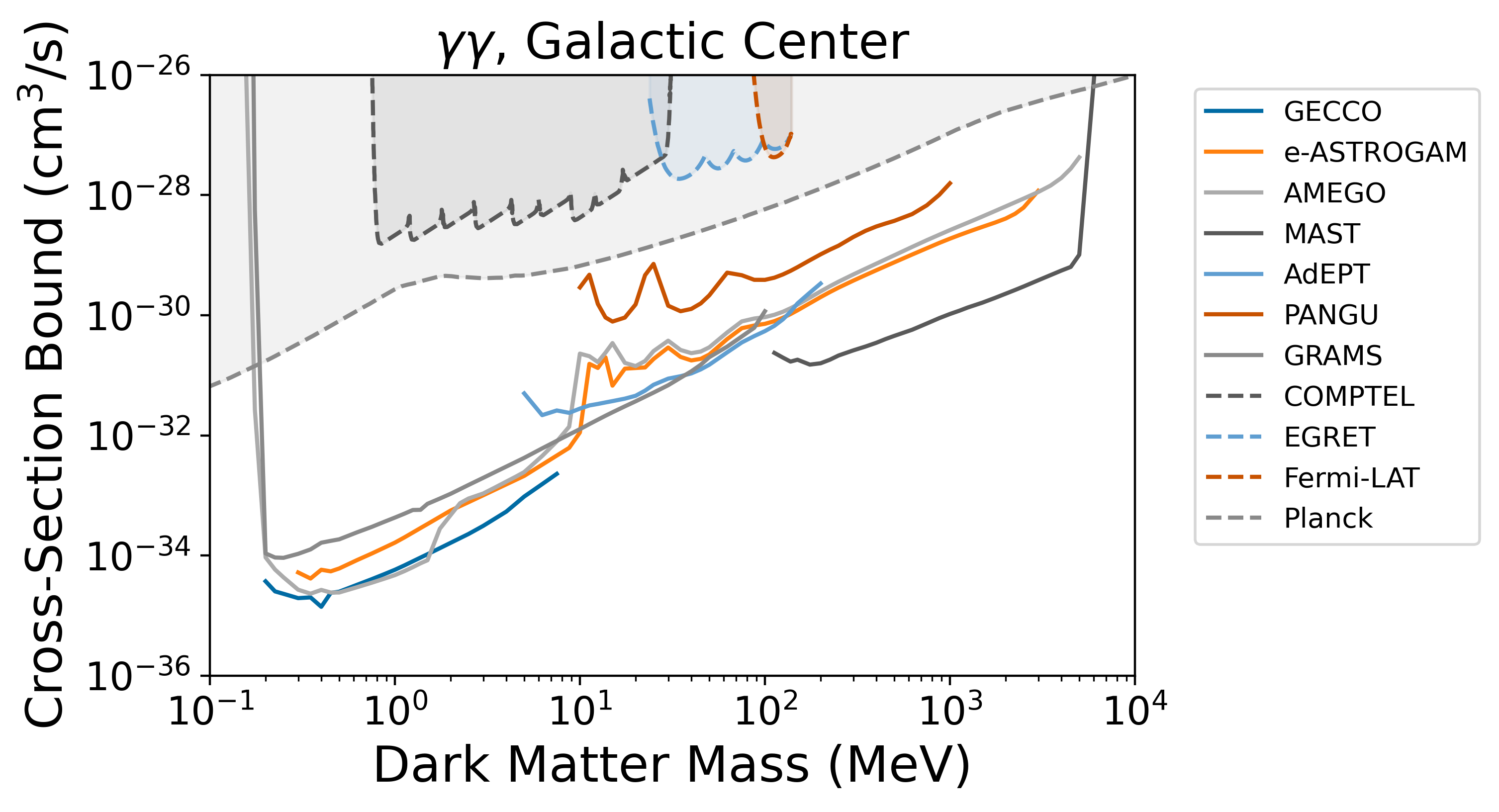}
    \includegraphics[width=.48\textwidth]{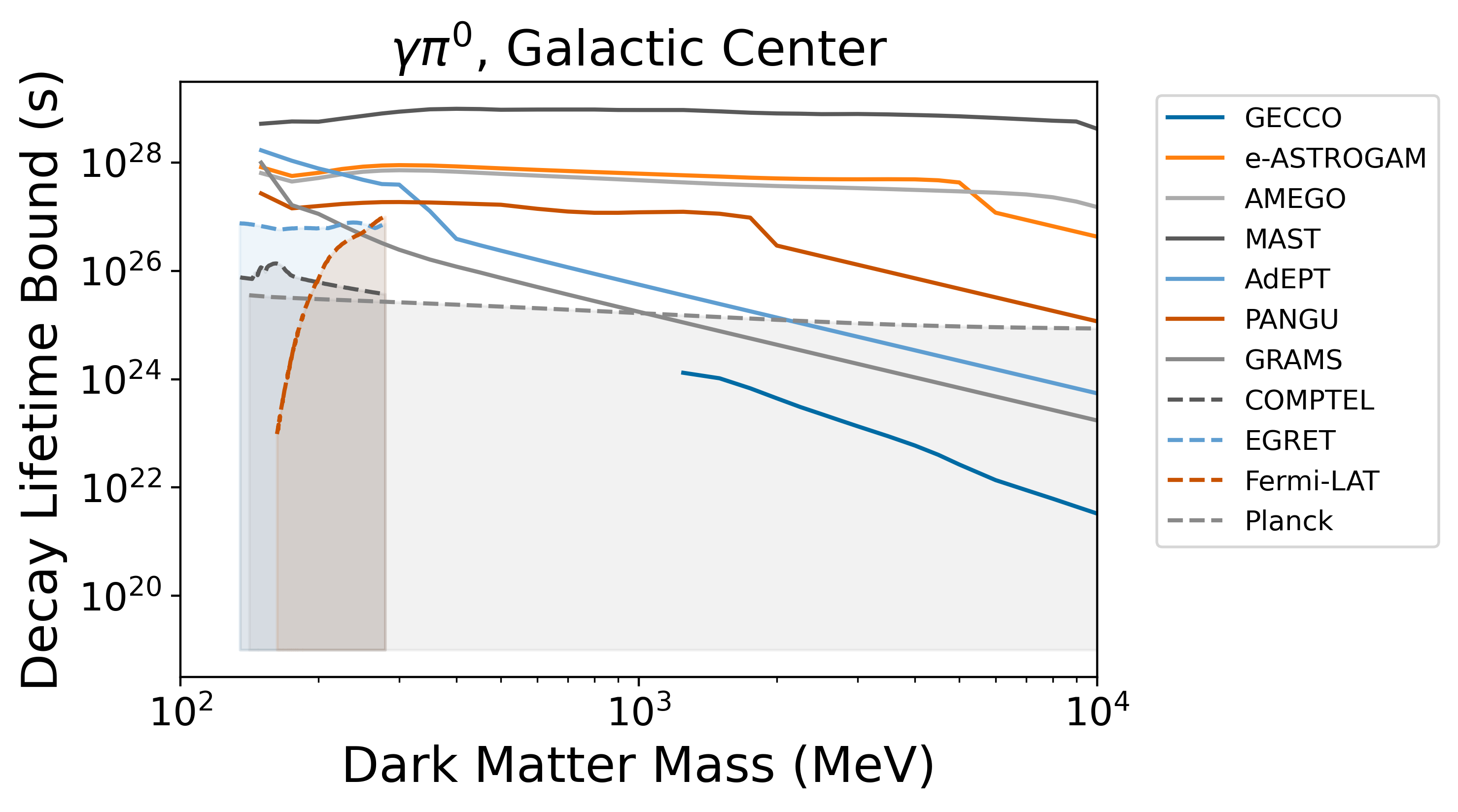} \hfill
    \includegraphics[width=.48\textwidth]{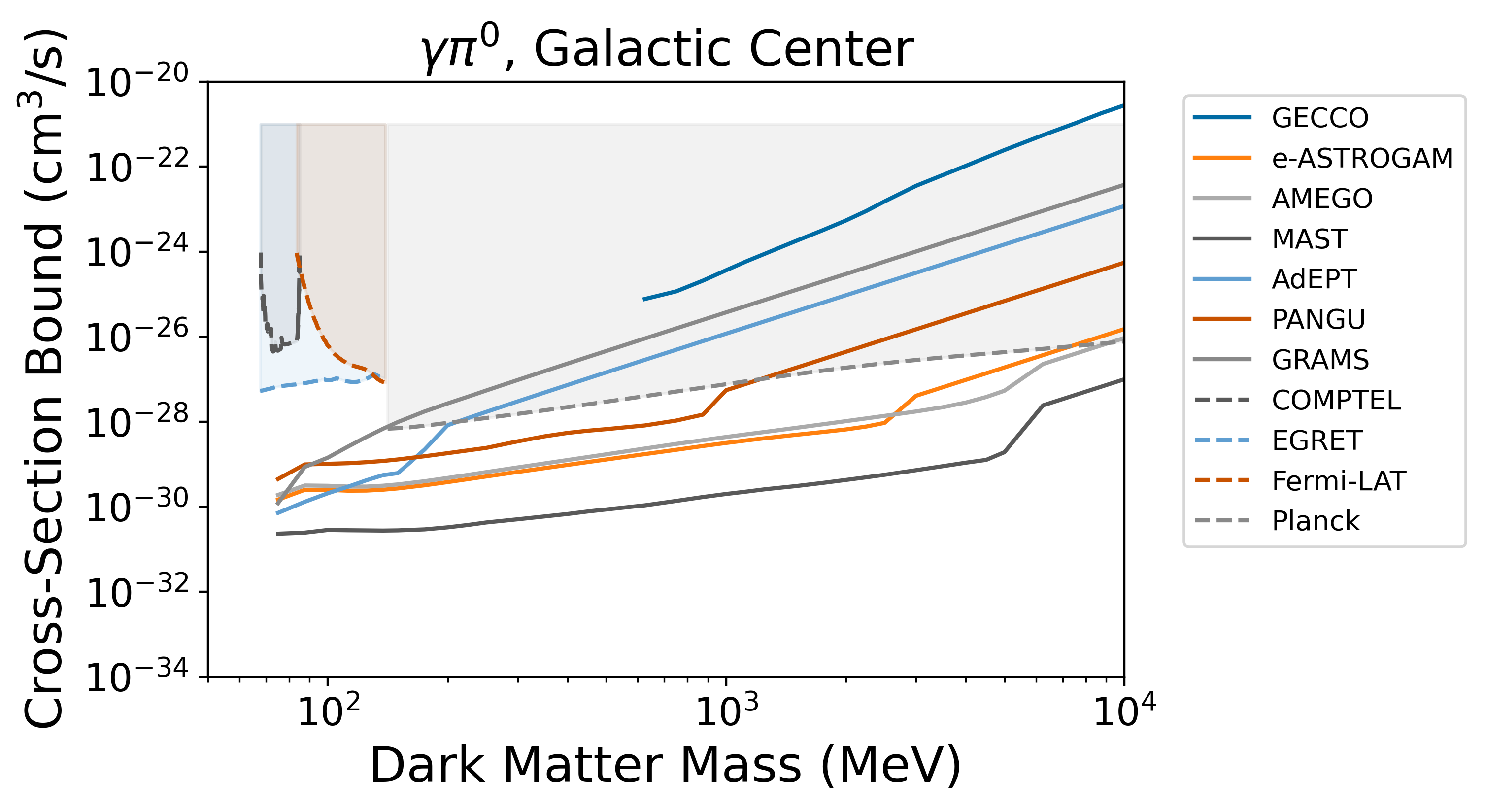}
    \includegraphics[width=.48\textwidth]{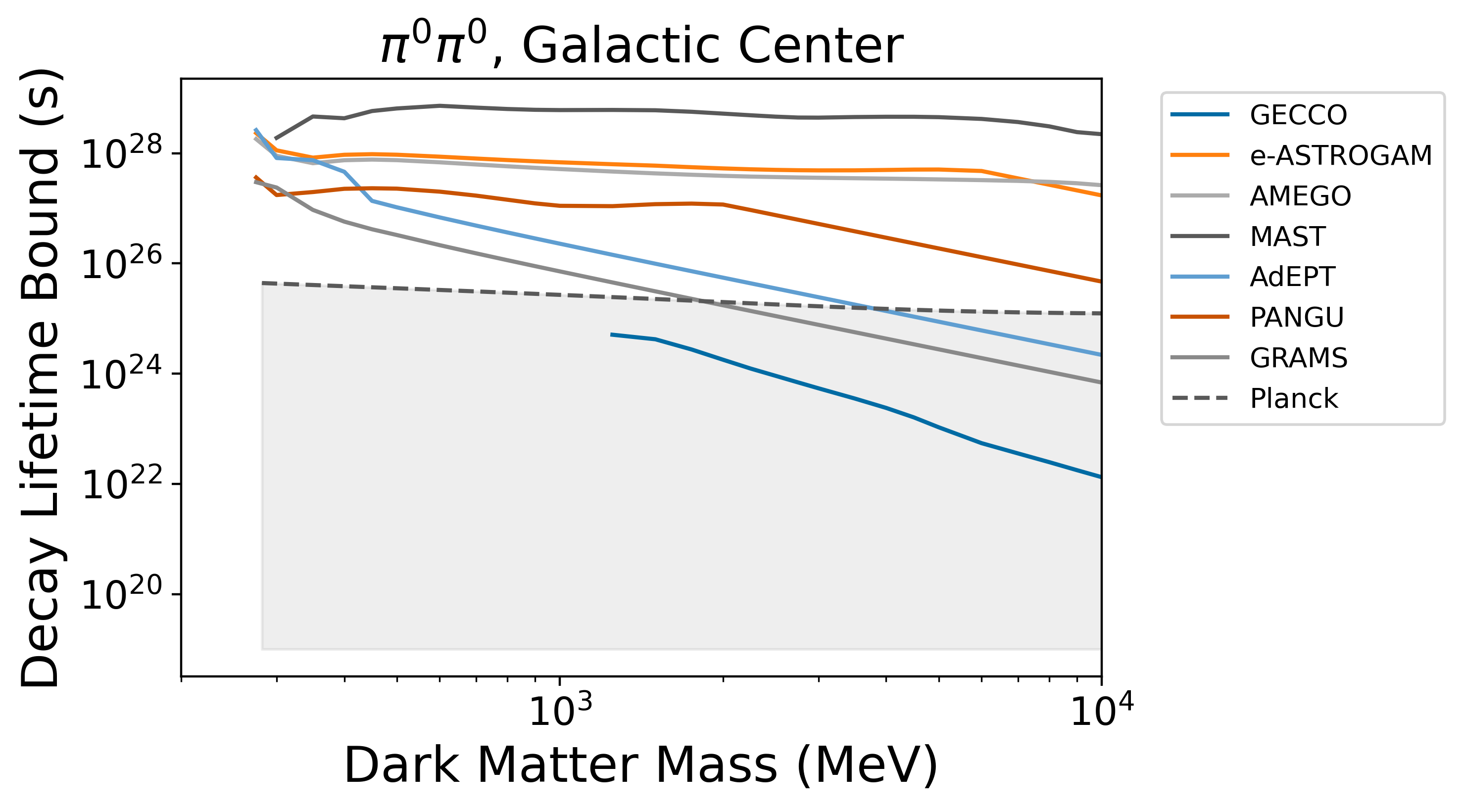} \hfill
    \includegraphics[width=.48\textwidth]{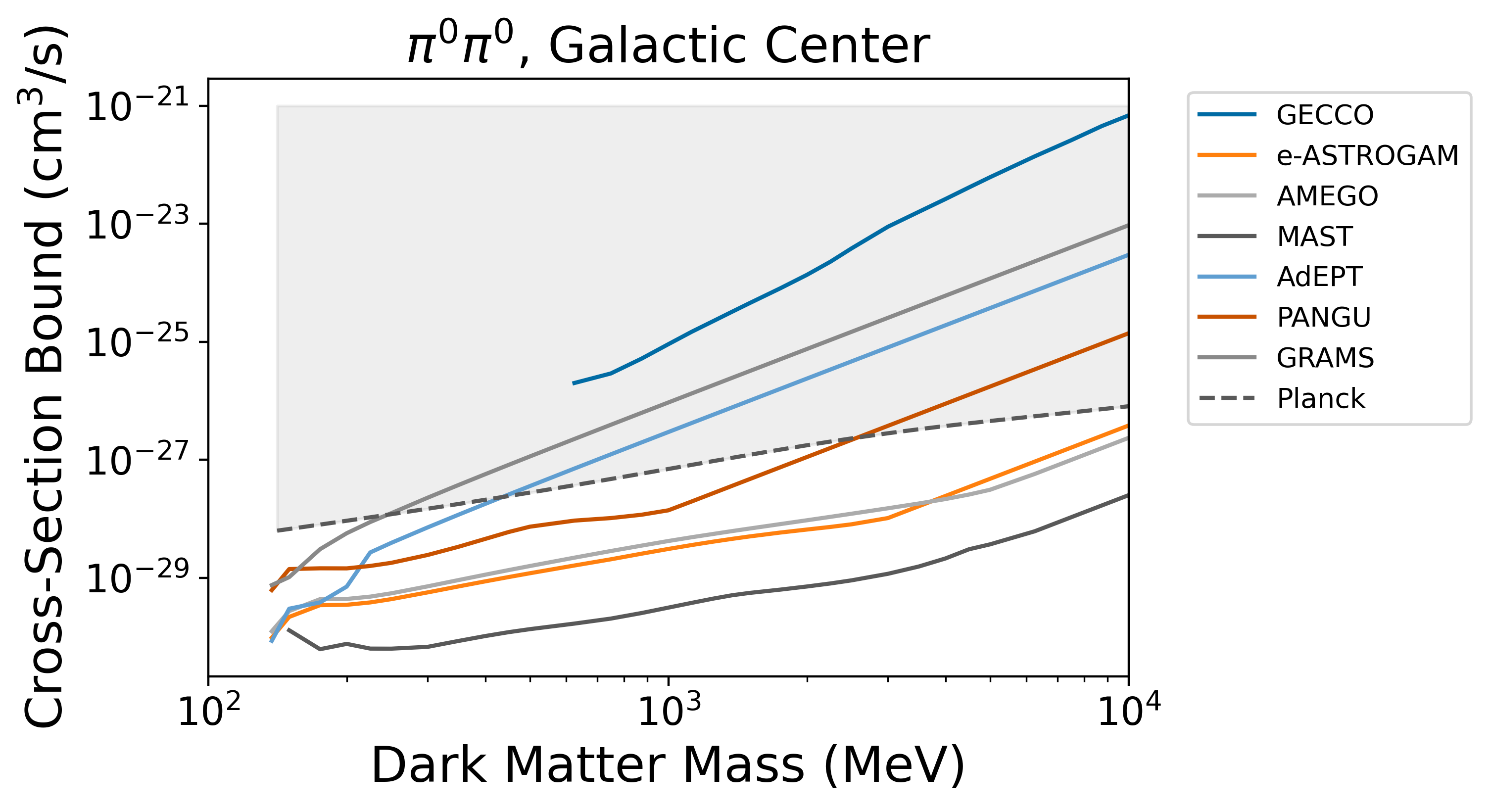}
    \caption{The constraints each instrument, pointed at the Galactic Center, can place on dark matter decay and annihilation into (from top to bottom) two photons, a photon and a neutral pion, and two neutral pions, assuming an observation time of $T_{\mathrm{obs}} = 10^6 \mathrm{\ s}$. These are compared against the current constraints described in Sec.~\ref{sec:results}.}
    \label{fig:gc_neutral}
\end{figure*}

\begin{figure*}
    \includegraphics[width=.48\textwidth]{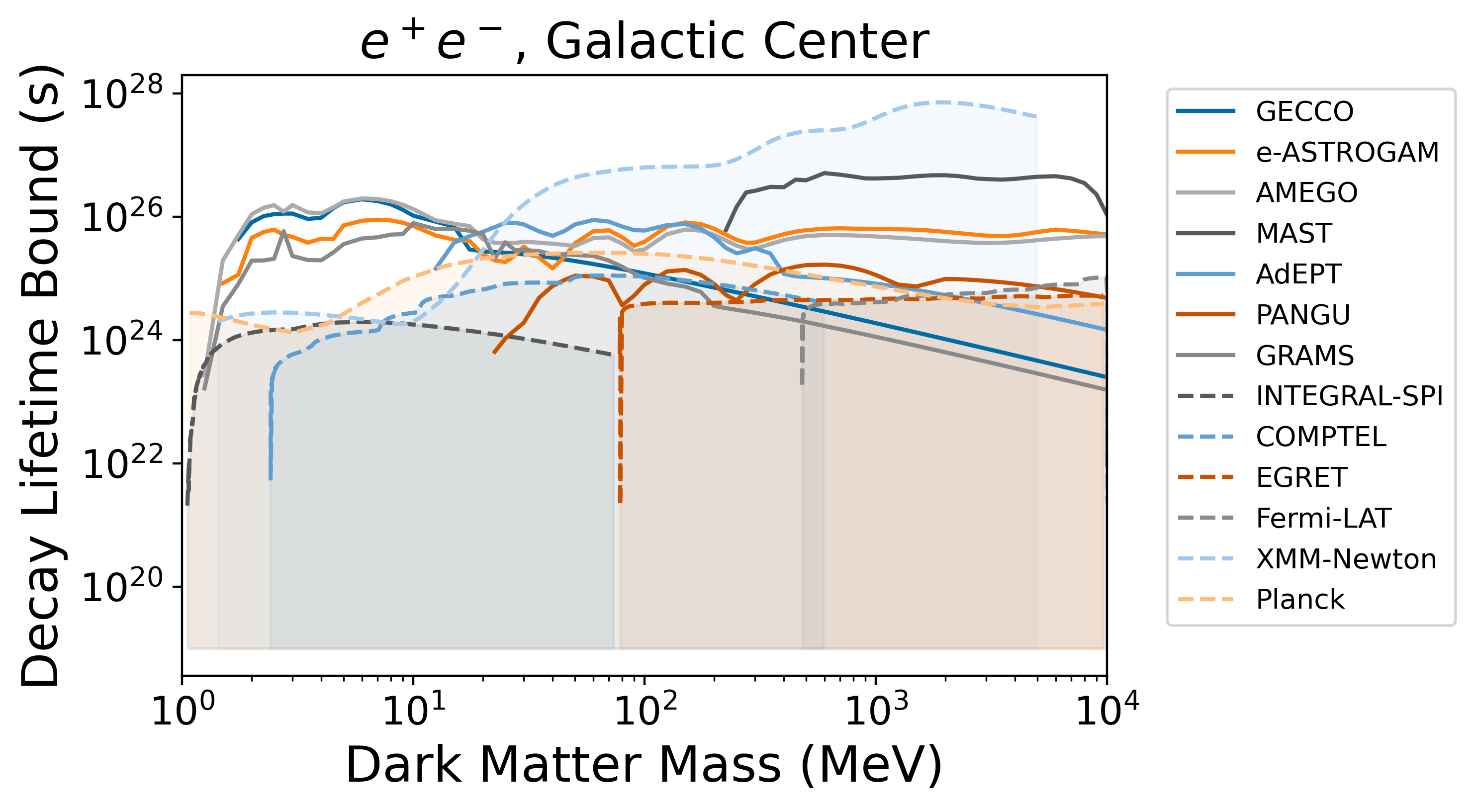} \hfill
    \includegraphics[width=.48\textwidth]{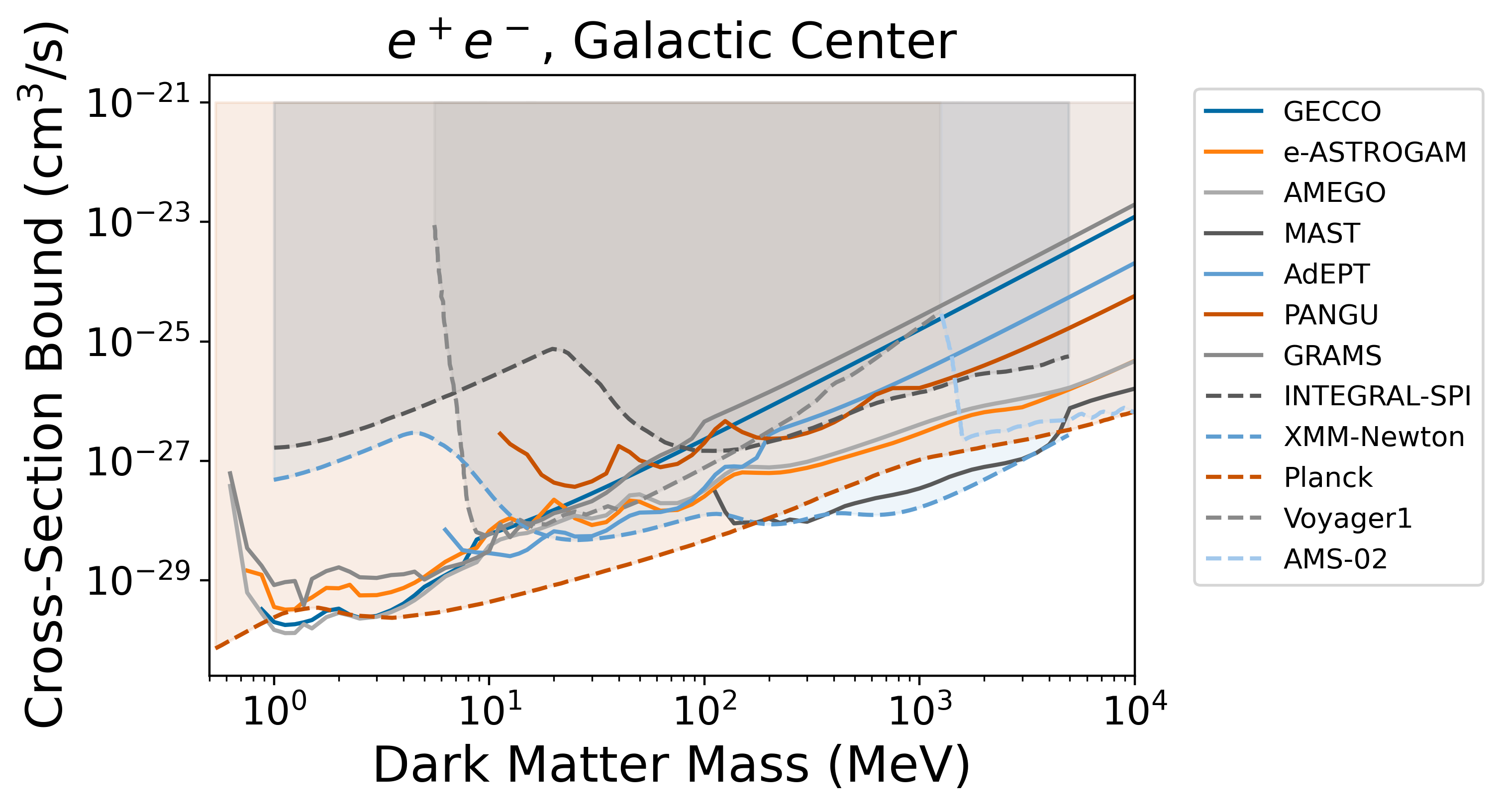}
    \includegraphics[width=.48\textwidth]{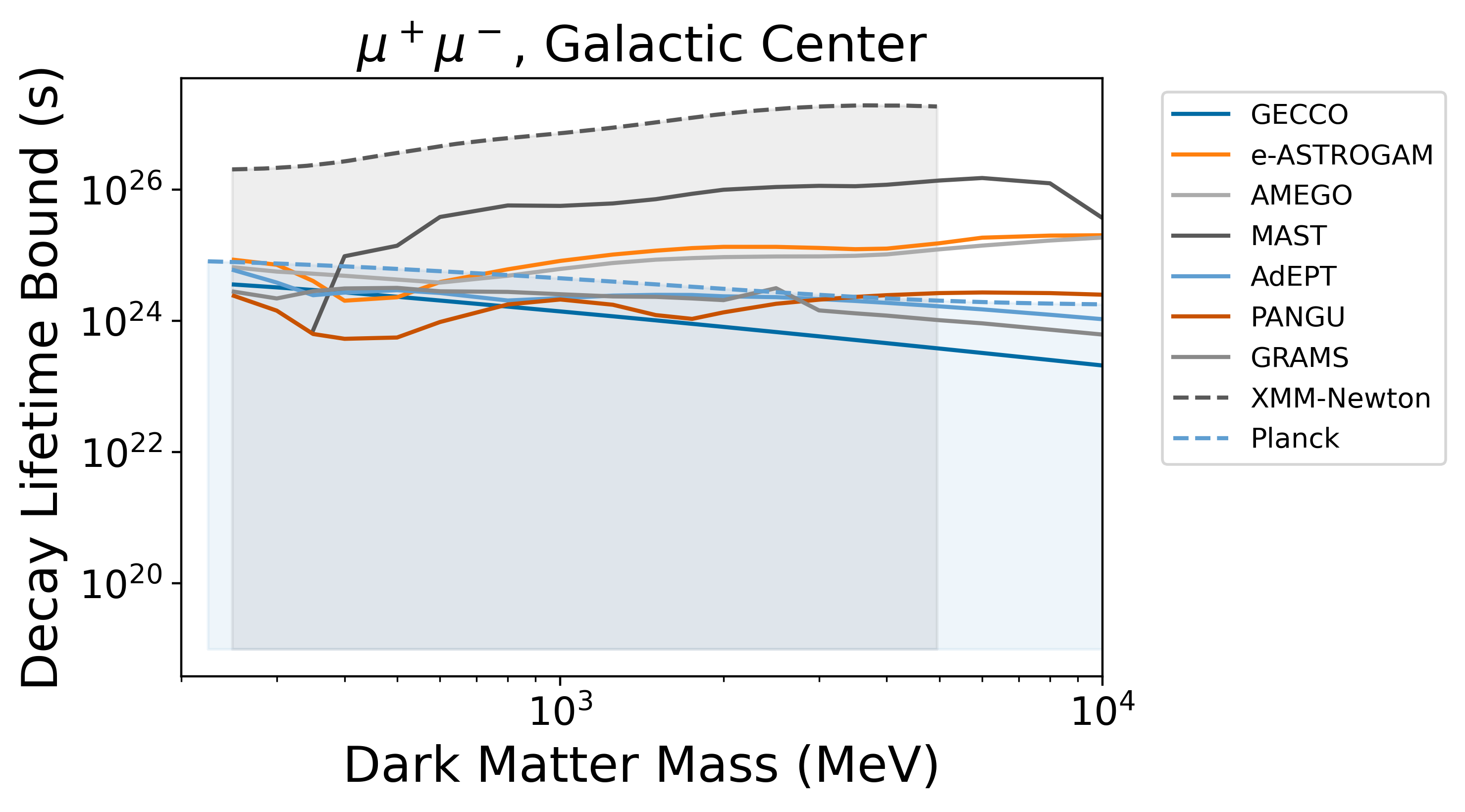} \hfill
    \includegraphics[width=.48\textwidth]{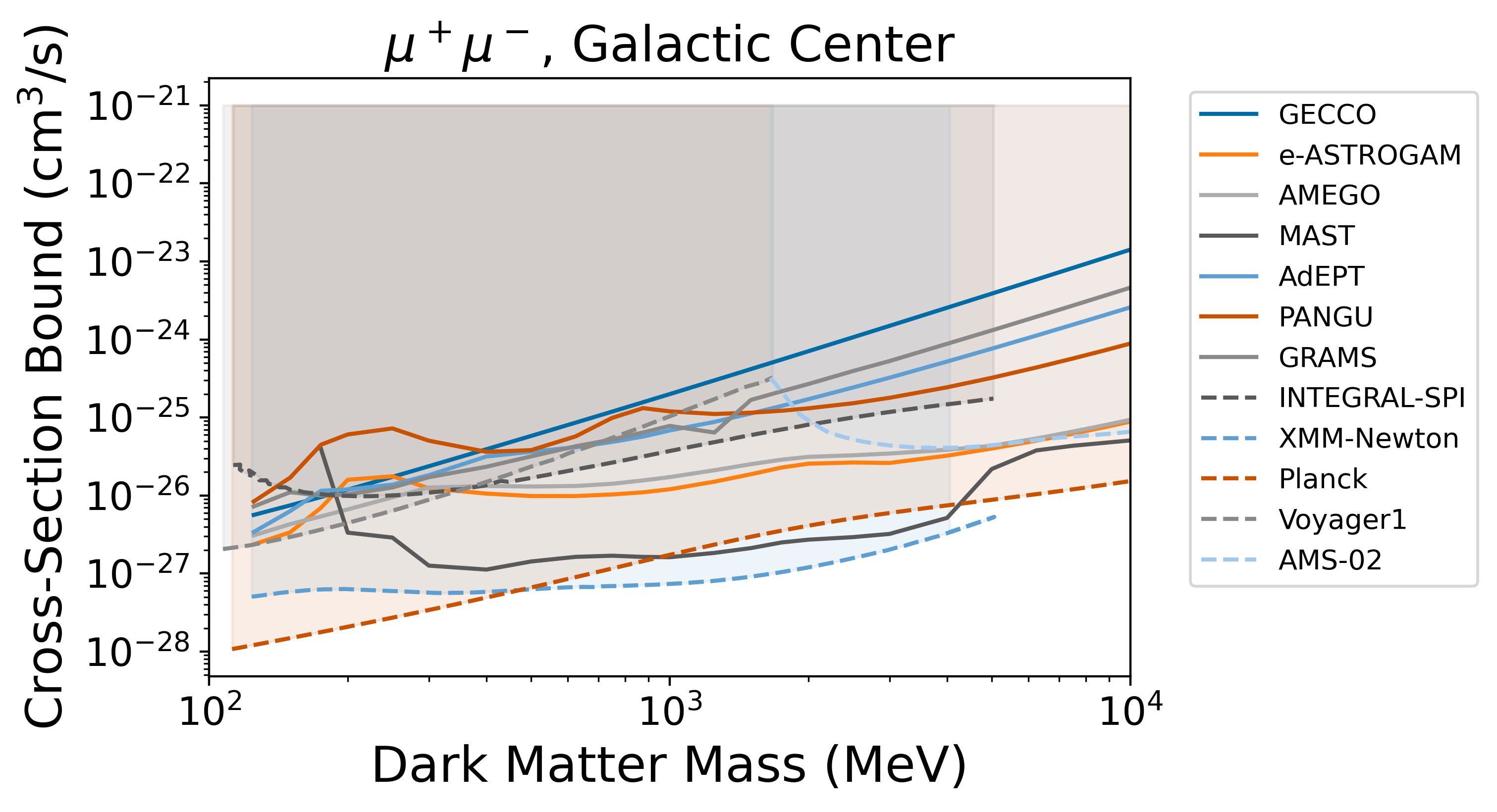}
    \includegraphics[width=.48\textwidth]{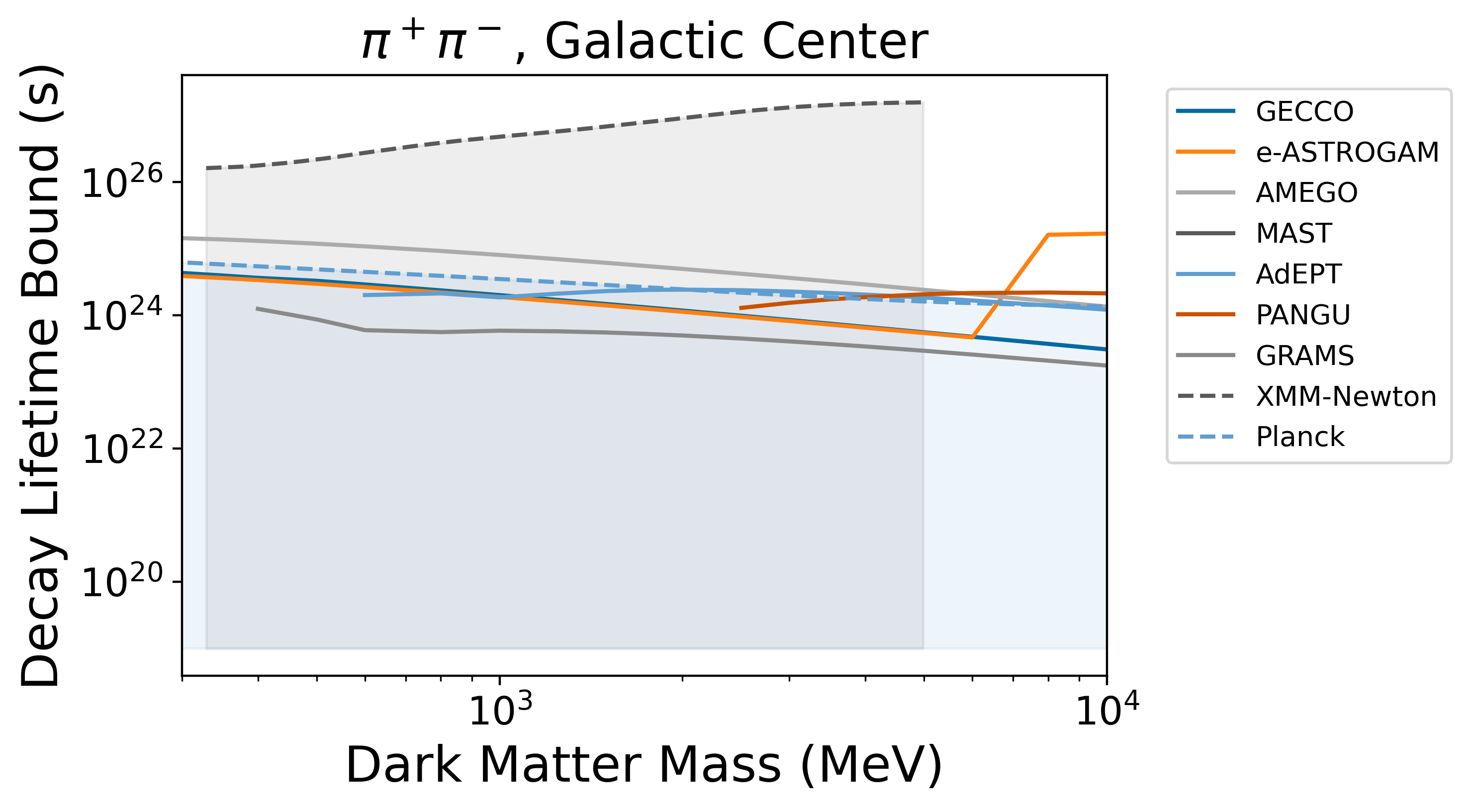} \hfill
    \includegraphics[width=.48\textwidth]{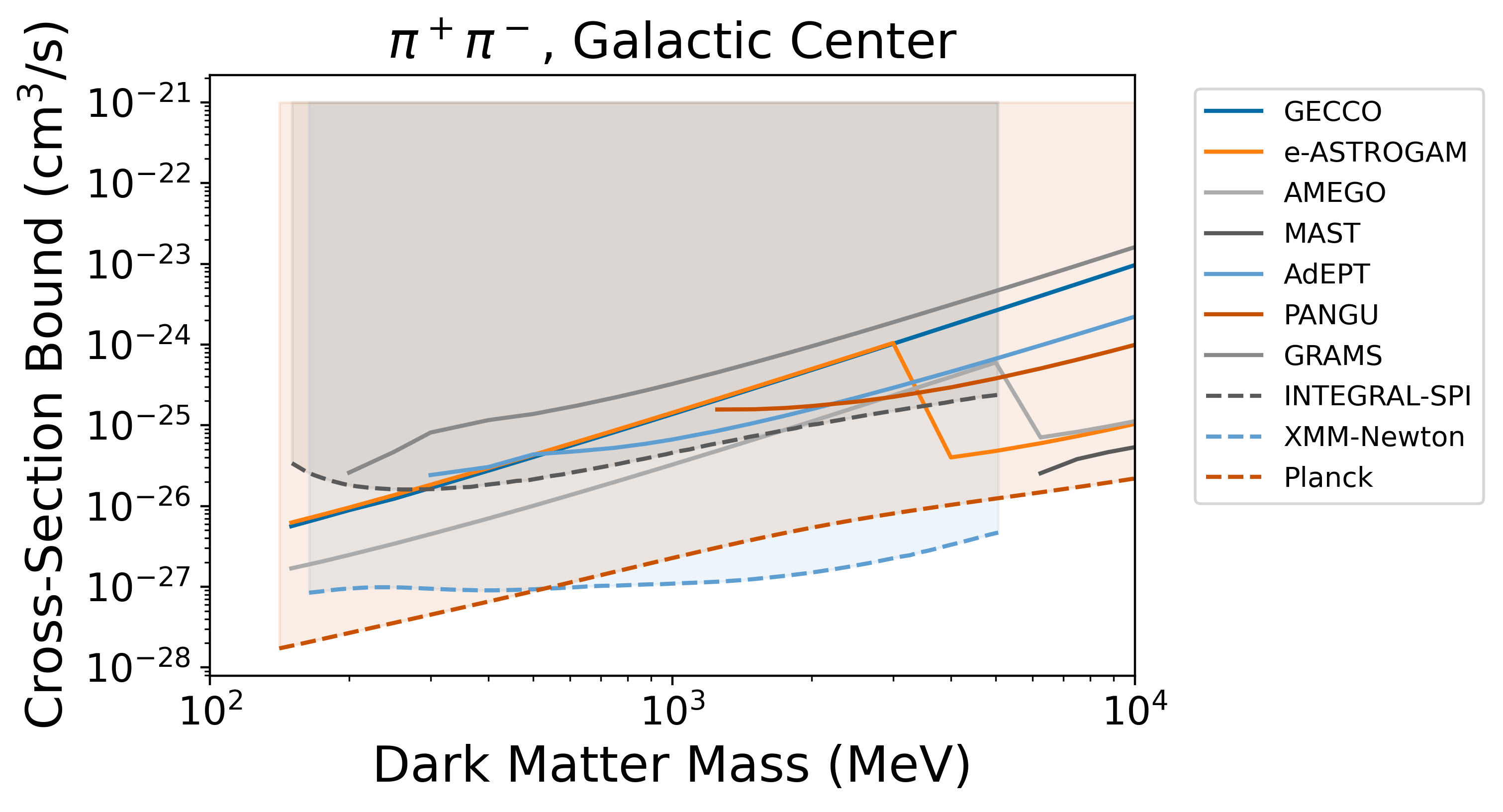}
    \caption{The constraints each instrument, pointed at the Galactic Center, can place on dark matter decay and annihilation into the final states (from top to bottom) $e^+e^-$, $\mu^+\mu^-$, and $\pi^+\pi^-$, assuming an observation time of $T_{\mathrm{obs}} = 10^6\mathrm{\ s}$. These are compared against the current constraints described in the text of Sec.~\ref{sec:results}.}
    \label{fig:gc_charged}
\end{figure*}

In Figs.~\ref{fig:gc_neutral} and \ref{fig:gc_charged}, we present our results for each instrument and decay/annihilation mode for observation of the Galactic Center, since this target generally provides the strongest constraints out of all the targets we considered. Results for Draco can be found in Appendix \ref{sec:draco}, and results for M31 can be found simply by rescaling the results for Draco based on the $J$ or $D$ factors of the two targets, since we used the same background model for both targets.

For all decay and annihilation channels we considered, current constraints from cosmic microwave background data from the \textit{Planck} satellite are given in Ref.~\cite{PhysRevD.93.023527} ($s$-wave annihilation) and Ref.~\cite{PhysRevD.95.023010} (decay). For lepton-producing decay and annihilation channels, we use the INTEGRAL constraints from Ref.~\cite{PhysRevD.103.063022}, the XMM-Newton constraints from Ref.~\cite{Cirelli_2023}, and the Voyager-1 and AMS-02 constraints given by \cite{Boudaud:2016mos}. Additionally, for decay to electrons and photons, we use the INTEGRAL-SPI (electrons only), COMPTEL, EGRET, and Fermi-LAT constraints given in Ref.\cite{essig}. For constraints from INTEGRAL-SPI for decay to photons, we use the data corresponding to the total spectral analysis in 1D parameter space given in Ref.~\cite{Calore:2022pks}. For the COMPTEL, EGRET, and Fermi-LAT constraints for annihilation to photons as well as for decay or annihilation to a photon and a neutral pion, we use the constraints from Ref.~\cite{PhysRevD.92.023533}. All of the current constraints that we considered are at the 95\% confidence limit.

\subsection{Comparison of different experiments}

We now discuss our results. In this section, we will provide comparisons of the reach of the proposed experiments (for a fixed observation time of $10^6$ s), but we caution that these proposals are at different stages of development and may have quite different timescales and degrees of ambition; thus this discussion should not be taken as a ``ranking'' of experiments.

Our forecast constraints from the Galactic Center are stronger than our constraints from M31 / Draco across the board. As one might expect, the Galactic Center seems to be a more promising target for indirect detection of dark matter if the profile is NFW; Draco and M31 provide backup in the event of central densities that are markedly suppressed relative to the NFW profile. Forecast constraints for dark matter decay and annihilation to uncharged particles are generally much stronger than current constraints (except for the current INTEGRAL-SPI constraints given in Ref.~\cite{Calore:2022pks} in the case of decay to low-energy photons and the current \textit{Planck} constraints given in Ref.~\cite{PhysRevD.93.023527} in the case of annihilation to neutral pions, both when considering Draco as a target). Forecast constraints from decay and annihilation to charged particles are weaker than the XMM-Newton constraints given in Ref.~\cite{Cirelli_2023} (which include inverse Compton emission peaked at lower energies) for decay of dark matter particles with mass higher than about 20 MeV. Additionally, for decay and annihilation to charged particles, forecast constraints from all instruments except MAST are weaker than the \textit{Planck} constraints given in Ref.~\cite{PhysRevD.93.023527} for annihilation of dark matter particles with mass higher than about 10 MeV. However, a suppression of annihilation at low velocities, e.g.~due to the dominant channel being $p$ wave, would relax the existing \textit{Planck} constraints relative to those studied in this work. 

For the channels including monochromatic photons, MAST offers the most striking sensitivity improvement within the $>100$ MeV energy range, primarily due to its very large projected effective area. With the exception of MAST, the constraints provided by the other instruments for a given energy range and integration time are all within about an order of magnitude of each other, except for PANGU, which has a particularly large energy resolution. For the instruments that detect lower-energy photons ($<10$ MeV), GECCO has the highest projected sensitivity due to its excellent energy resolution in this band. For the instruments that detect photons with energies between 10 and 100 MeV, AdEPT and GRAMS have the highest projected sensitivity. e-ASTROGAM and AMEGO are also forecast to perform well throughout their energy range. All instruments are forecast to be competitive with or provide improvement over current constraints if pointed at the Galactic Center.

For decay and annihilation channels involving neutral pions, MAST once again offers the strongest forecast constraints due to its large effective area, and e-ASTROGAM and AMEGO are also forecast to perform relatively well. The instruments that are best suited to detect lower-energy photons (GECCO and GRAMS) are not forecast to provide very strong constraints for channels involving neutral pions, primarily because the photon spectrum from these interactions does not overlap much with their energy range.

For the channels involving annihilation to leptons, MAST is forecast to be competitive with current constraints within its energy range if pointed at the Galactic Center. However, it is not competitive with the XMM-Newton constraints from Ref.~\cite{Cirelli_2023} for decay. For the channels involving decay or annihilation to electrons, the instruments we consider provide the most promising increase in sensitivity for dark matter masses below about 10 MeV. In this range, particularly in the case of decay or $p$-wave annihilation, GECCO, AMEGO, e-ASTROGAM, and GRAMS are all forecast to offer a sizable increase in sensitivity to decay or annihilation to electrons when pointed at the Galactic Center (relative to the existing bounds we have considered). None of the instruments considered are particularly well suited to provide a sensitivity increase for decay or annihilation to charged pions.

Overall, MAST provides by far the strongest projected constraints within its energy range, AdEPT provides the strongest projected constraints outside of MAST's energy range, and GECCO provides the strongest projected constraints outside of AdEPT's energy range. e-ASTROGAM and AMEGO also provide reasonably competitive forecast constraints for all decay and annihilation channels across all energy ranges considered. However, we caution again that these instruments are not all at the same scale or stage of development, and are not direct competitors.

We have focused on general final states rather than specific dark matter models in this work. One might ask whether there is a benchmark decay rate or annihilation cross section that these experiments should aim to detect or exclude. We are not aware of such a model-independent target for decay lifetime, and so have preferred to simply show how the forecast sensitivity compares to existing bounds. For $p$-wave annihilation, a possible benchmark would be the cross section that would yield the correct relic density through thermal freeze-out. This cross section is somewhat model dependent, but if freeze-out was completely dominated by $p$-wave annihilation and this contribution also dominated the signal today, we would expect the present-day cross section to be of order $\sim 10^{-26} (v_\text{galactic}/v_\text{freeze-out})^2$ cm$^3$/s, where $v_\text{galactic}$ and $v_\text{freeze-out}$ denote the typical velocity of dark matter in our Galaxy and at the time of freeze-out respectively. Taking $v_\text{freeze-out}^2 \sim 0.1$ and  $v_\text{freeze-out}^2 \sim 10^{-6}$, the target cross section should be in the neighborhood of $\sim 10^{-31}$ cm$^3$/s. This cross section seems very difficult to reach for leptonic final states but may be attainable for photon-rich final states (e.g.~involving neutral pions as well as direct annihilation to photons).

\subsection{Systematic uncertainties}

We begin our discussion of systematic uncertainty with our signal modeling. Errors in the parameters used to determine the Galactic Center $J$ factors are of order 20\% \cite{de_Salas_2019}, errors in the parameters used to determine the M31 $J$ factors are of order 10\% \cite{10.1093/pasj/psv042}, and errors in the parameters used to determine the Draco $J$ factors are about 5\%--15\% \cite{10.1111/j.1365-2966.2010.16753.x}. We assume throughout that dark matter follows an NFW profile, and there may be larger $J$ factor uncertainties associated with changing the model for the profile. Our signal modeling also assumes that the signal flux comes from one dark matter component which makes up the bulk of dark matter. If this is not the case, our decay / annihilation bounds for any individual dark matter component are too strong by a factor of the fraction of dark matter corresponding to that component (squared in the case of annihilation). For our background modeling, our Fisher analysis takes into account statistical fluctuations and correlations between the background and signal, but does not account for the systematic uncertainties in the background model itself, which are estimated at 15\% \cite{PhysRevD.92.023533}.

Finally, systematic errors in the estimation of the effective areas and energy resolutions that we referenced would contribute to systematic errors in our data. Since most of these concepts have not yet been prototyped, the accuracy of these figures to the specifications of the final product is not guaranteed. Furthermore, many of the proposals we referenced gave a range of energy resolutions rather than specifying the energy resolution as a function of incident photon energy. For these instruments, we used the worst energy resolution in the range for our calculations, contributing a systematic error that weakens our bounds.

We also remind the reader that as discussed previously, our approach does not use spatial information within the ($10^\circ$ radius) ROI, and so very likely underestimates the constraining power of instruments with superior angular resolution. On the other hand, we have neglected instrumental and atmospheric backgrounds, which may degrade the sensitivity estimates in this work if they cannot be reduced below the astrophysical $\gamma$-ray signal.

\subsection{Comparison with the literature and cross-checks}

We note briefly that there are discrepancies between our projected constraints for GECCO and those from \cite{PhysRevD.107.023022}; our results differ by various constant  factors [typically $\mathcal{O}(1)$ and in all cases less than one order of magnitude]. 

To derive a cross-check on our results, let us consider a simplified version of our analysis. Since the energy resolution of GECCO is quite narrow, the energy spectrum of photons can be approximated by a $\delta$ function, for purposes of estimating the background to a line signal. In this limit, the ratio of the Galactic Center constraints to the M31 constraints should be approximately equal to the ratio of the decay factors divided by the square root of the ratio of the background models, evaluated at the desired mass point. We also find in this limit that, for the method we used,
\begin{equation}
    \frac{\langle \sigma v \rangle}{f_\chi} = \frac{1}{\tau} \sqrt{\frac{\Omega_a}{\Omega_d}} \frac{J_d}{J_a} \frac{2m_a^2}{m_d},
\end{equation}
where $\langle \sigma v \rangle$ is the annihilation constraint calculated at $m_\chi = m_a$ with an observation angle of $\Omega_a$ and a $J$ factor of $J_a$, $\tau$ is the decay constraint calculated for the same target and integration time at $m_\chi = m_d = 2 m_a$ with an observation angle of $\Omega_d$ and a $J$ factor of $J_d$, and $f_\chi=1$ if the particles are self-conjugate and $f_\chi=2$ otherwise. We find that our calculations for all of the constraints for the photon annihilation and decay modes, in cases where the telescope energy resolution is good and these approximations should be valid, are consistent with these relations.

We also compared our results for AdEPT with Draco observations to those of Ref.~\cite{PhysRevD.92.023533}. They assume a much longer run time (5 yr vs $10^6$ s) and also take an energy-independent value for the angular resolution that is smaller than our prescription across much of the energy range, allowing them to set a small ROI for their dwarf analysis which consequently reduces background. The $J$ and $D$ factors they assume for Draco are also modestly different from ours. Once these effects are taken into account, our forecast limits appear roughly consistent for the decay case and a factor of a few weaker for the annihilation case; we do employ a different statistical method so this may be responsible for at least some of the difference.

\section{Conclusions} \label{sec:conclusion}
All of the instruments we considered would markedly improve our current constraints for dark matter decay or annihilation directly to photons in their respective energy ranges when pointed at the Galactic Center. MAST, e-ASTROGAM, AMEGO, PANGU, and AdEPT would all also improve our current constraints for decay or annihilation channels involving neutral pions when pointed at the Galactic Center. However, since all of the instruments we considered are photon instruments, they are not able to place constraints that are as strong onto lepton decay or annihilation modes, as we have focused only on photons produced directly in the annihilation/decay (via radiative corrections), rather than lower-energy secondary photons that would require modeling of the charged particles' propagation. That being said, for lower-energy dark matter, GECCO, GRAMS, e-ASTROGAM, and AMEGO are all forecast to be able to place competitive constraints on decay or annihilation to electrons. Additionally, MAST is forecast to be able to improve constraints on annihilation to leptons if pointed at the Galactic Center given a long enough integration time. All the constraints shown were for an integration time of $10^6$ s, and it is reasonable to extend this integration time by a couple of orders of magnitude in the event that one of these instruments is mounted on a satellite, given that none of the instruments are pointed.

\section*{Acknowledgments}

The authors thank Stefano Profumo, Tsuguo Aramaki, and Regina Caputo for helpful discussions, and Dave Tucker-Smith for pointing out a plotting error in an earlier version of this work. K.E.O was supported by the MIT Undergraduate Research Opportunities Program, by the Mariana Polonsky Slocum (1955) Memorial Fund, and by the DeFlorez Endowment Fund. T.R.S. was supported in part by a Guggenheim Fellowship; the Edward, Frances, and Shirley B. Daniels Fellowship of the Harvard Radcliffe Institute; the Bershadsky Distinguished Fellowship of the Harvard Physics Department; and the Simons Foundation (Grant No.~929255). T.R.S. thanks the Kavli Institute for Theoretical Physics (KITP), the Aspen Center for Physics, and the Mainz Institute for Theoretical Physics for their hospitality during the completion of this work; this research was supported in part by Grant No.~NSF PHY-2309135 to KITP and performed in part at the Aspen Center for Physics, which is supported by National Science Foundation Grant No.~PHY-2210452. This work was supported by the U.S. Department of Energy, Office of Science, Office of High Energy Physics of U.S. Department of Energy under Contract No.~DE-SC0012567. 

\bibliographystyle{unsrt}
\bibliography{bibliography.bib}

\appendix

\section{Results for Draco and M31}
\label{sec:draco}
\begin{figure*}
    \includegraphics[width=.48\textwidth]{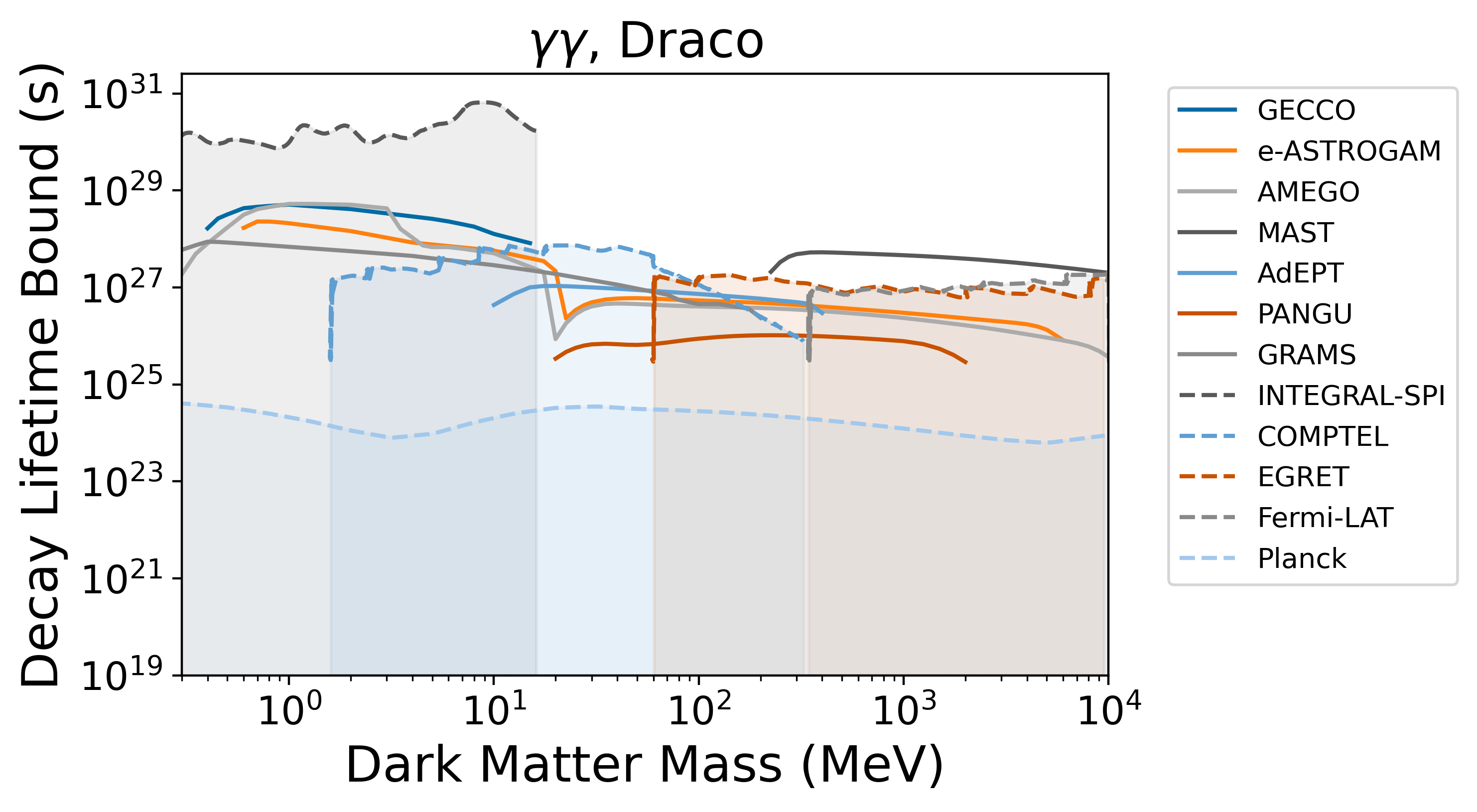} \hfill
    \includegraphics[width=.48\textwidth]{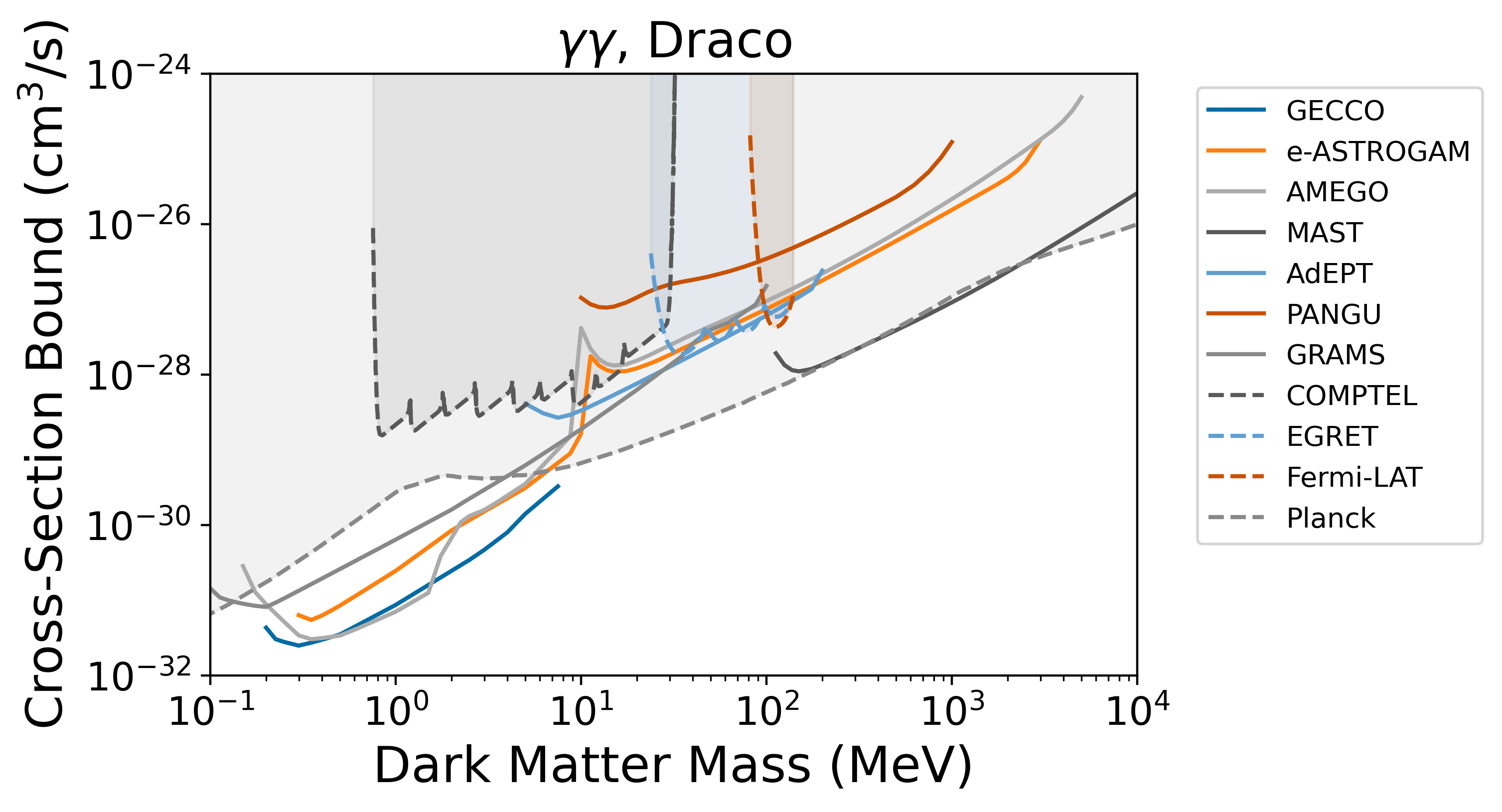}
    \includegraphics[width=.48\textwidth]{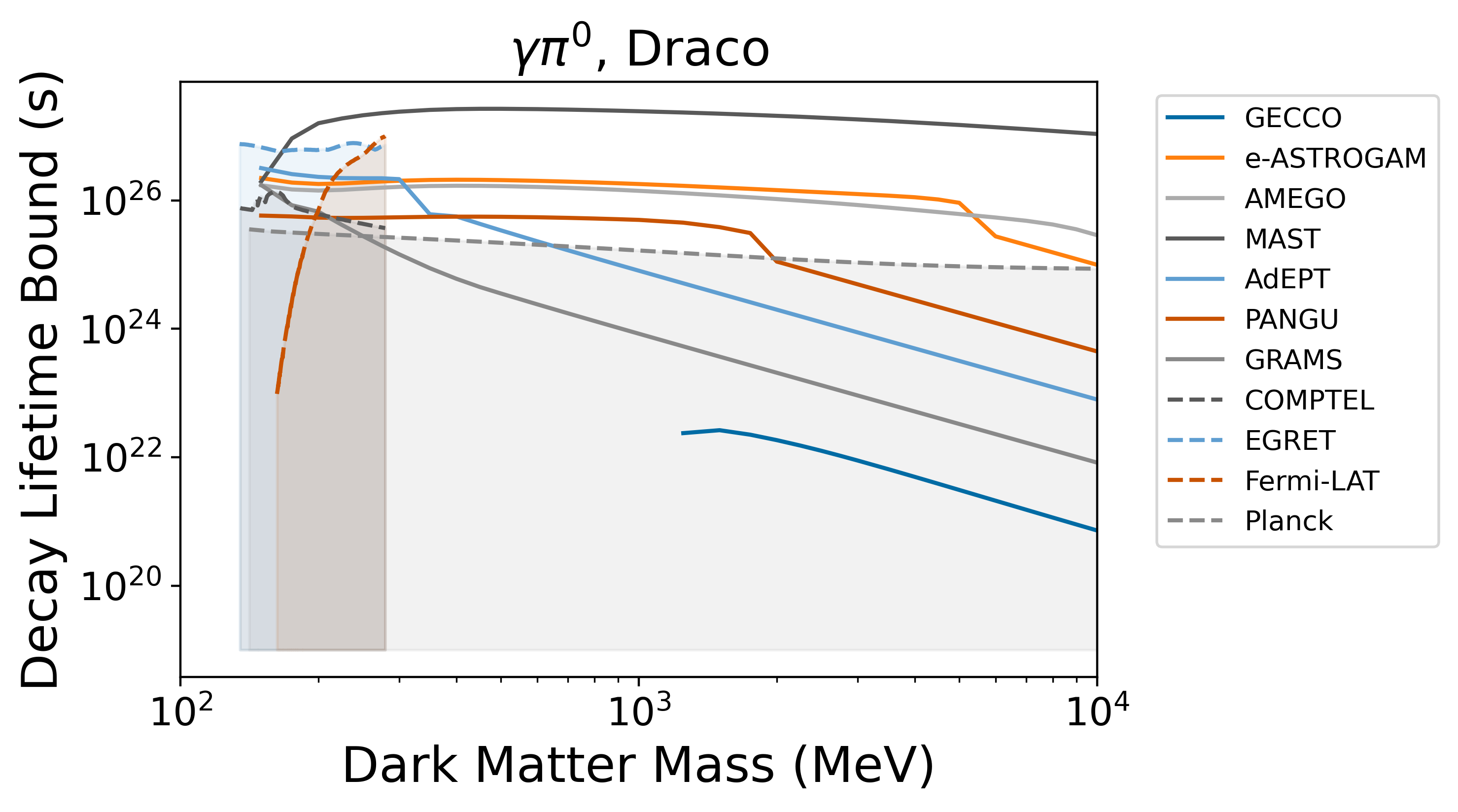} \hfill
    \includegraphics[width=.48\textwidth]{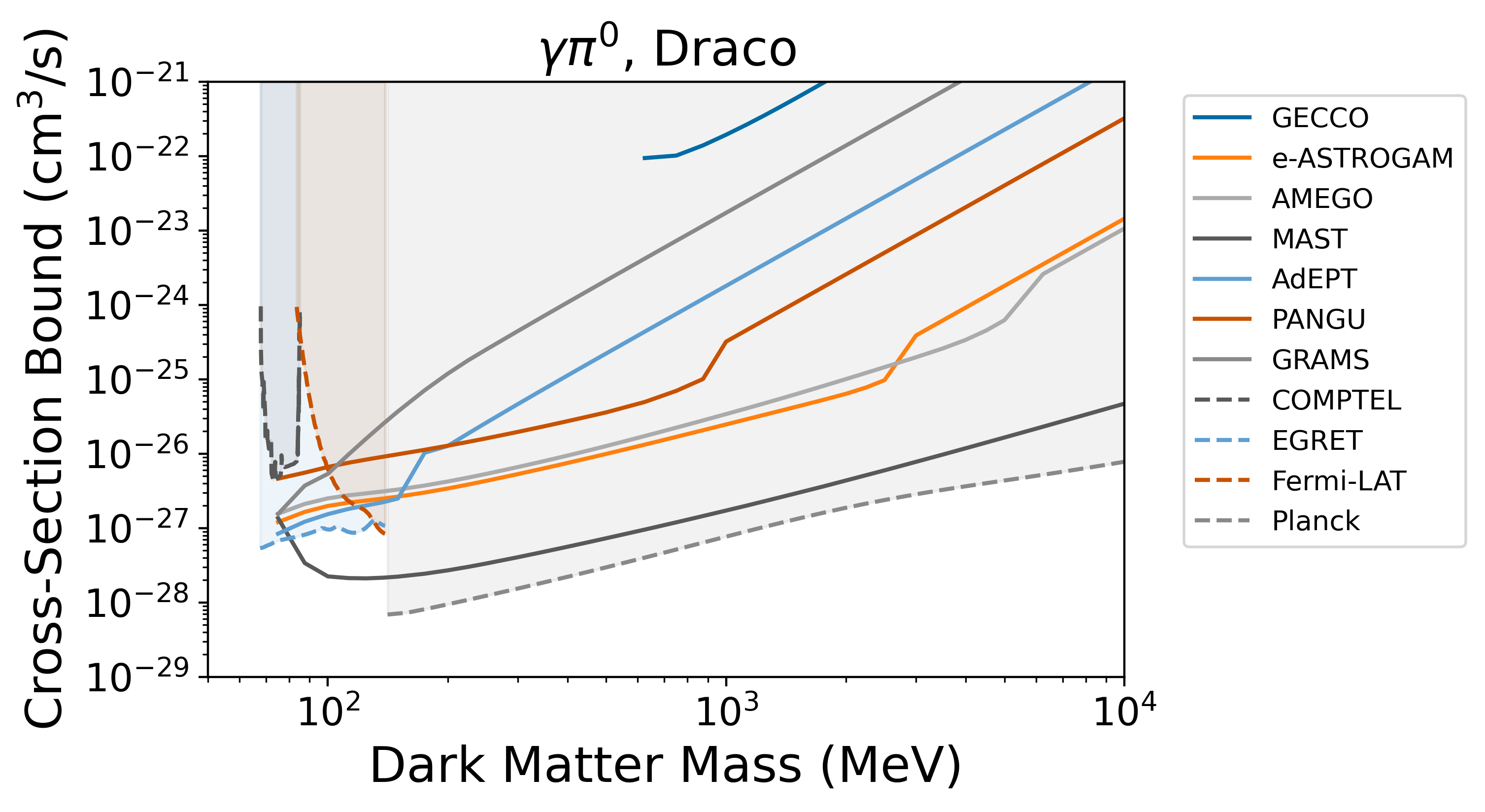}
    \includegraphics[width=.48\textwidth]{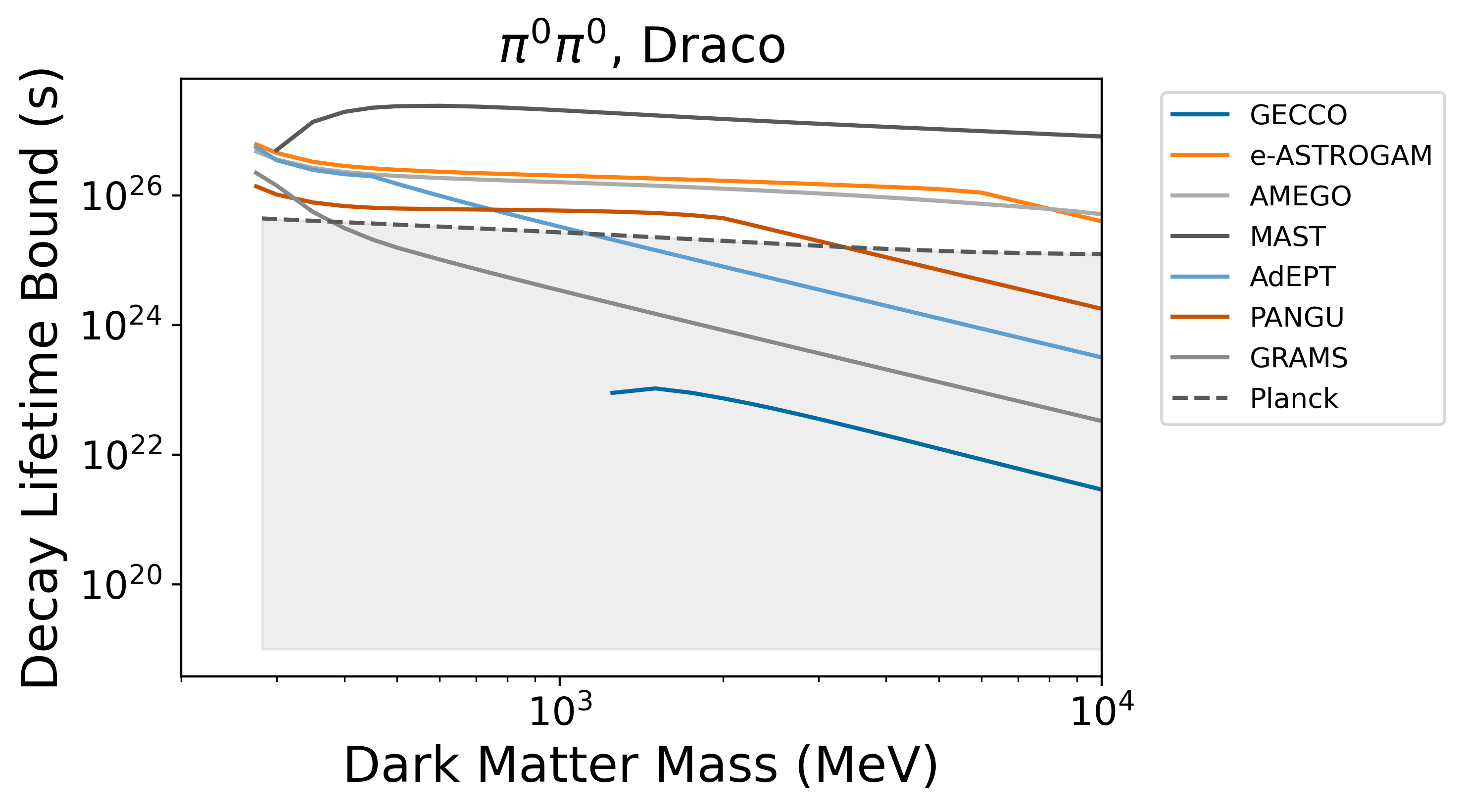} \hfill
    \includegraphics[width=.48\textwidth]{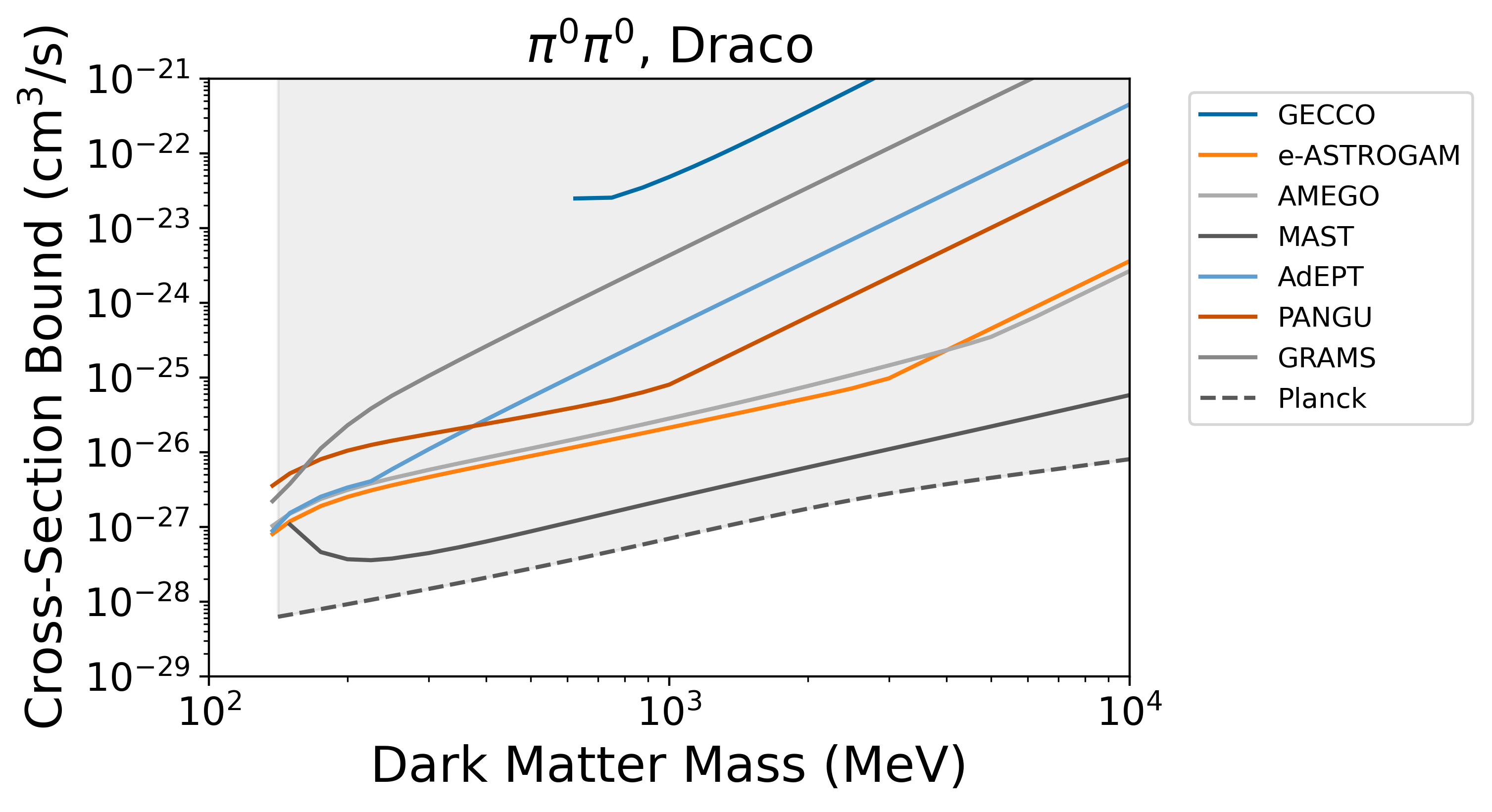}
    \caption{The constraints each instrument, pointed at Draco, can place on dark matter decay and annihilation into (from top to bottom) two photons, a photon and a neutral pion, and two neutral pions, assuming an observation time of $T_{\mathrm{obs}} = 10^6 \mathrm{\ s}$. These are compared against the current constraints described in Sec.~\ref{sec:results}.}
    \label{fig:draco_neutral}
\end{figure*}

\begin{figure*}
    \includegraphics[width=.48\textwidth]{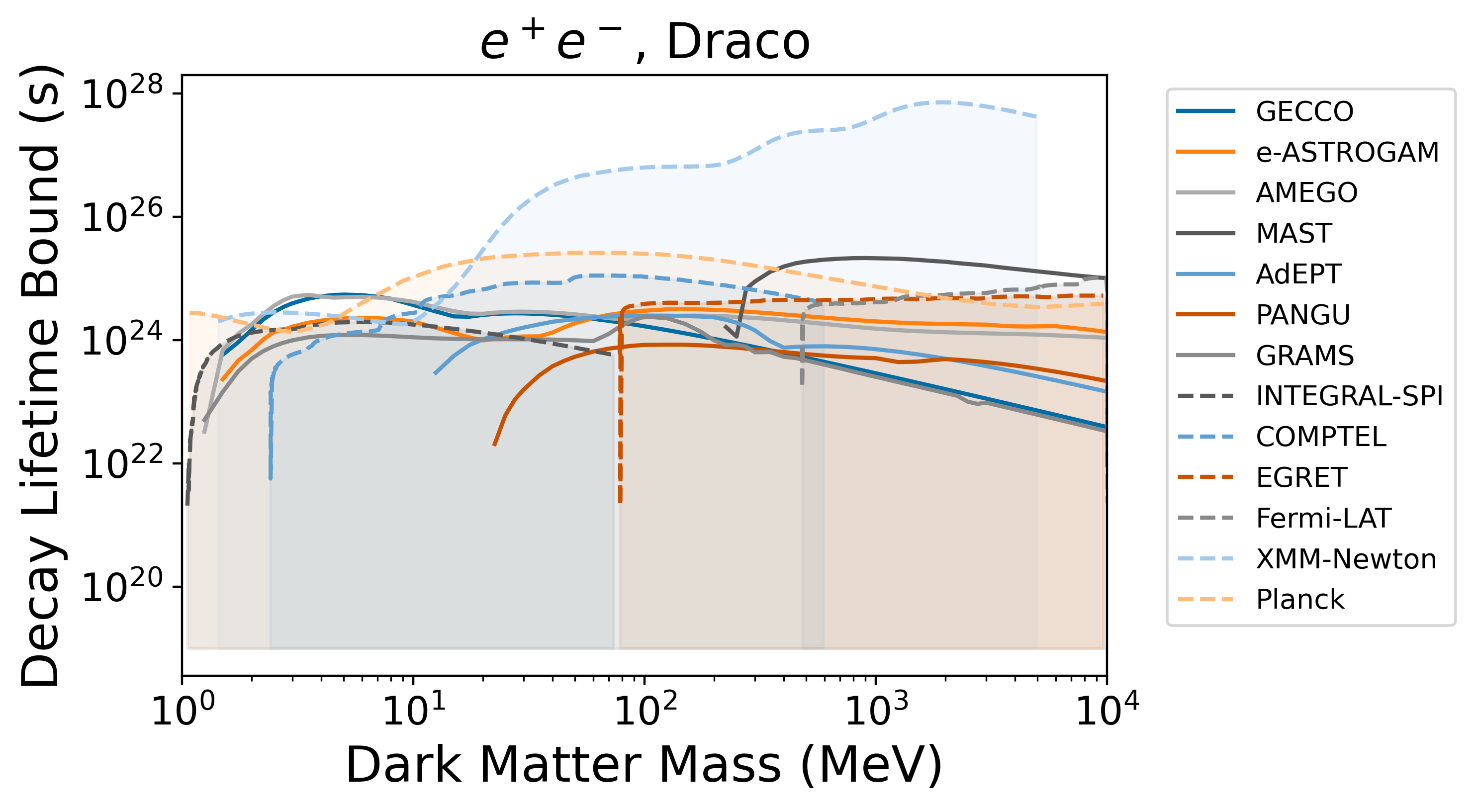} \hfill
    \includegraphics[width=.48\textwidth]{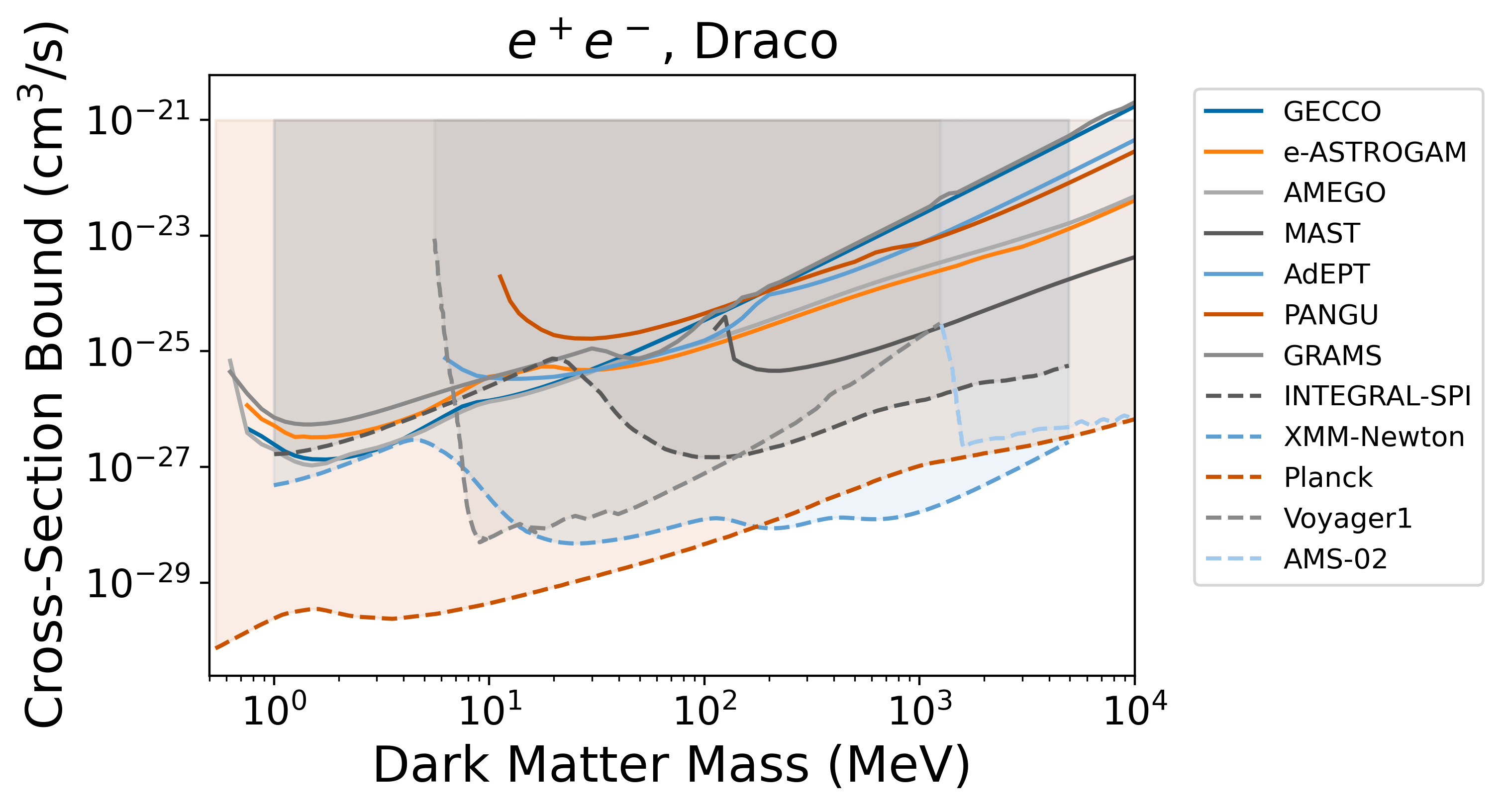}
    \includegraphics[width=.48\textwidth]{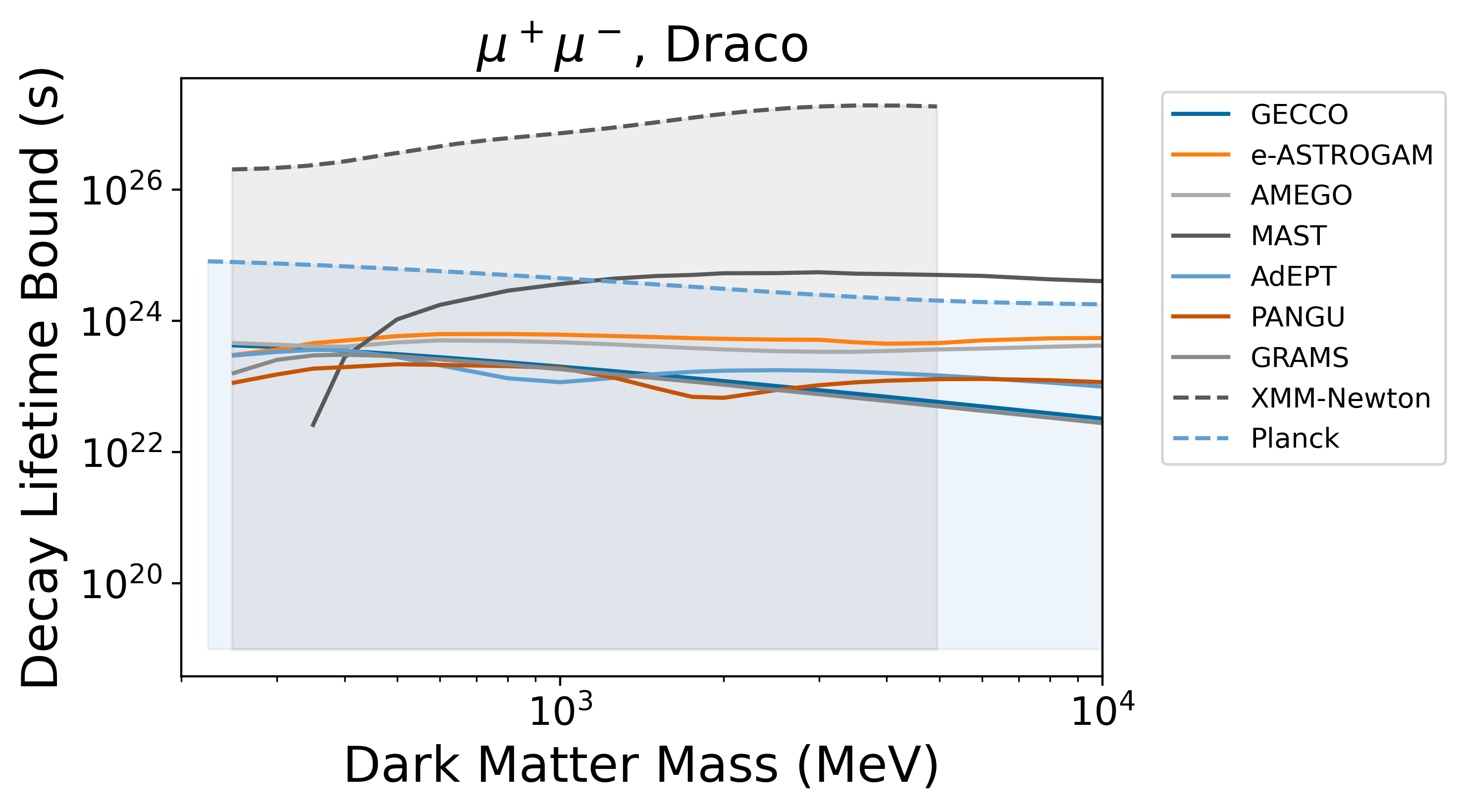} \hfill
    \includegraphics[width=.48\textwidth]{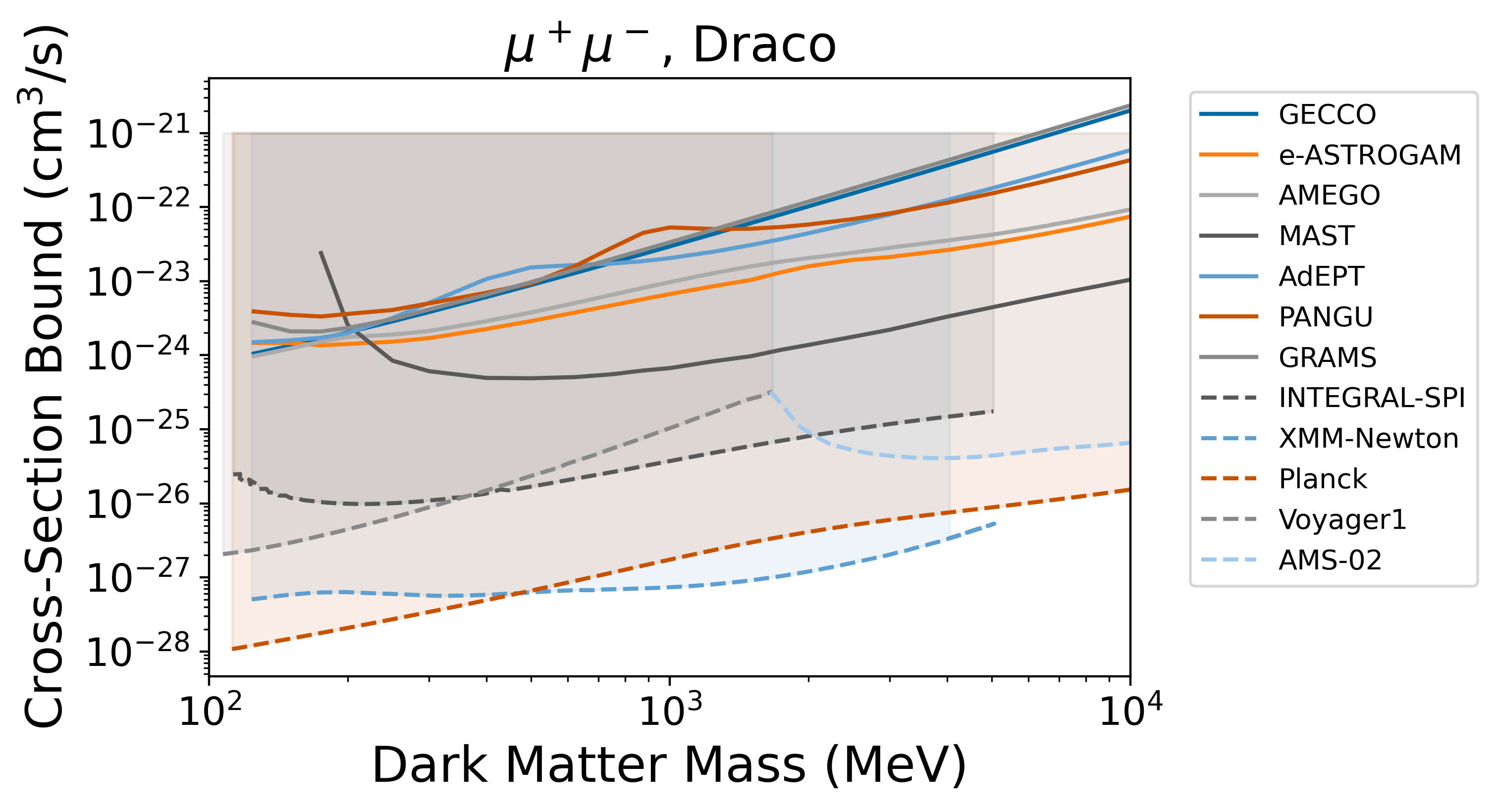}
    \includegraphics[width=.48\textwidth]{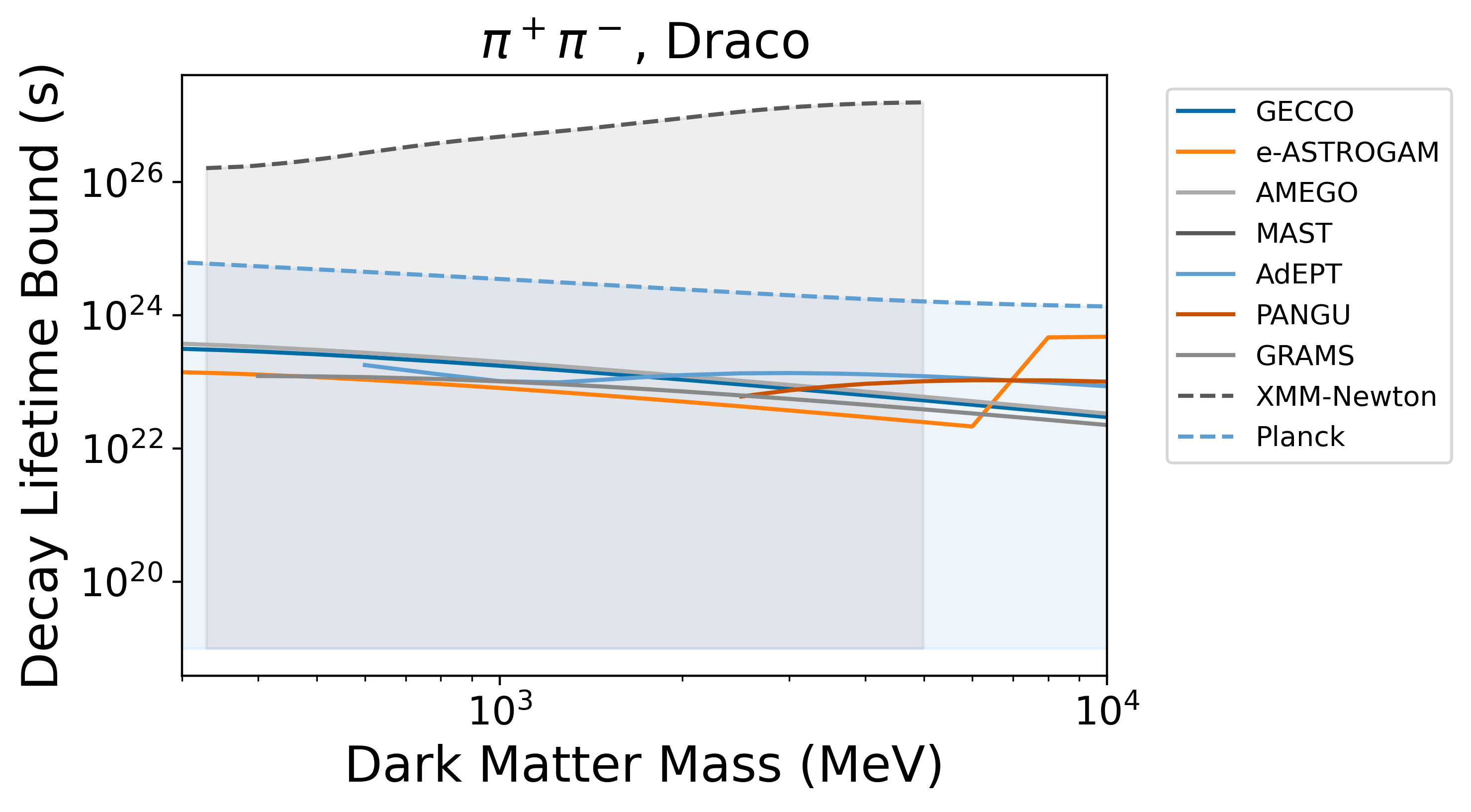} \hfill
    \includegraphics[width=.48\textwidth]{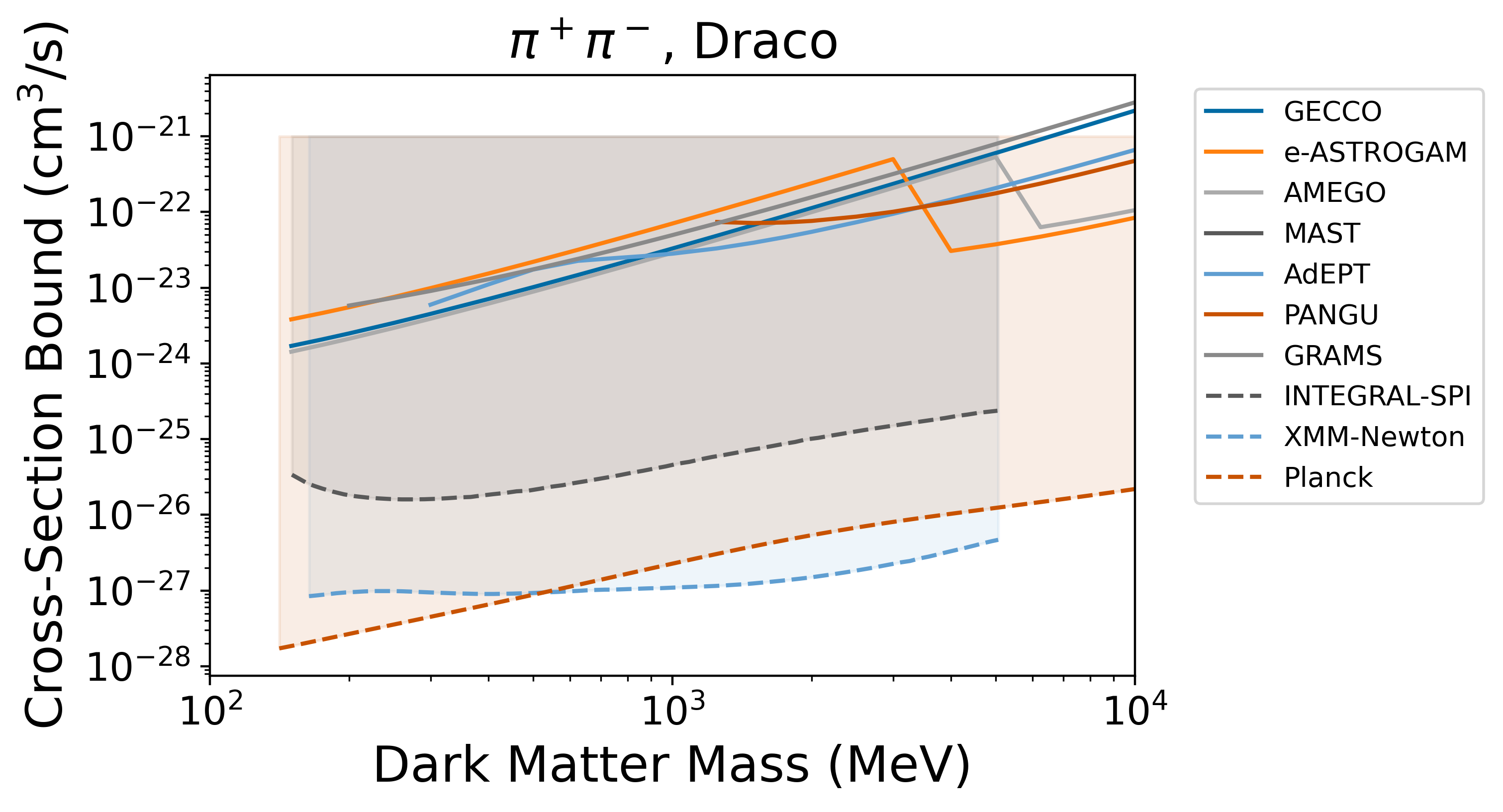}
    \caption{The constraints each instrument, pointed at Draco, can place on dark matter decay and annihilation into the final states (from top to bottom) $e^+e^-$, $\mu^+\mu^-$, and $\pi^+\pi^-$, assuming an observation time of $T_{\mathrm{obs}} = 10^6 \mathrm{\ s}$. These are compared against the current constraints described in Sec.~\ref{sec:results}.}
    \label{fig:draco_charged}
\end{figure*}
In Figs.~\ref{fig:draco_neutral} and \ref{fig:draco_charged}, we display our results for each instrument and decay/annihilation mode for observation of Draco. We note that these constraints are proportional to the decay factor, so the M31 constraints are simply a rescaling of the Draco constraints since we used the same background model for both targets.

\section{Atmospheric backgrounds}
\label{sec:atmospheric}
\begin{figure*}
    \includegraphics[width=.48\textwidth]{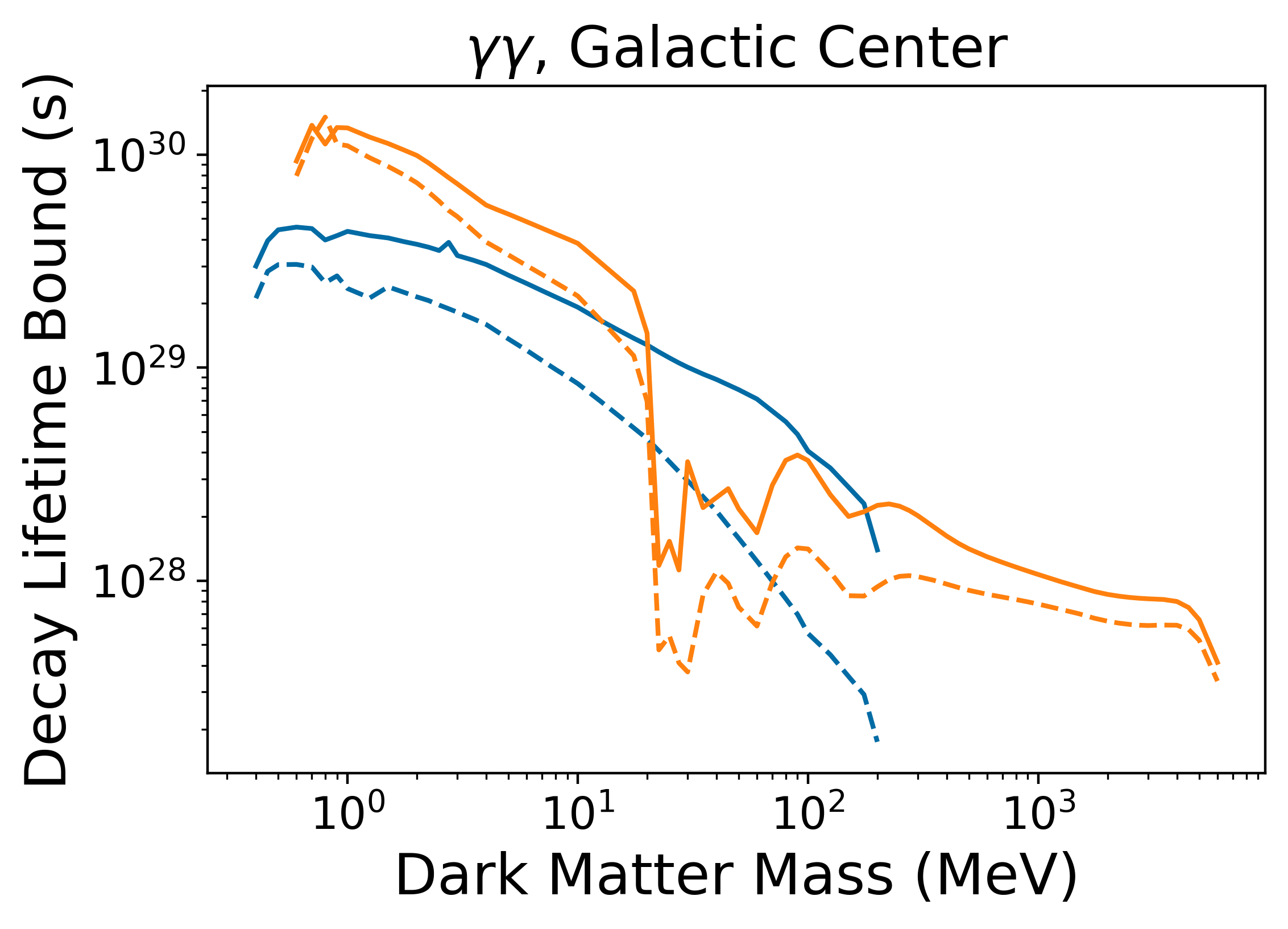} \hfill
    \includegraphics[width=.48\textwidth]{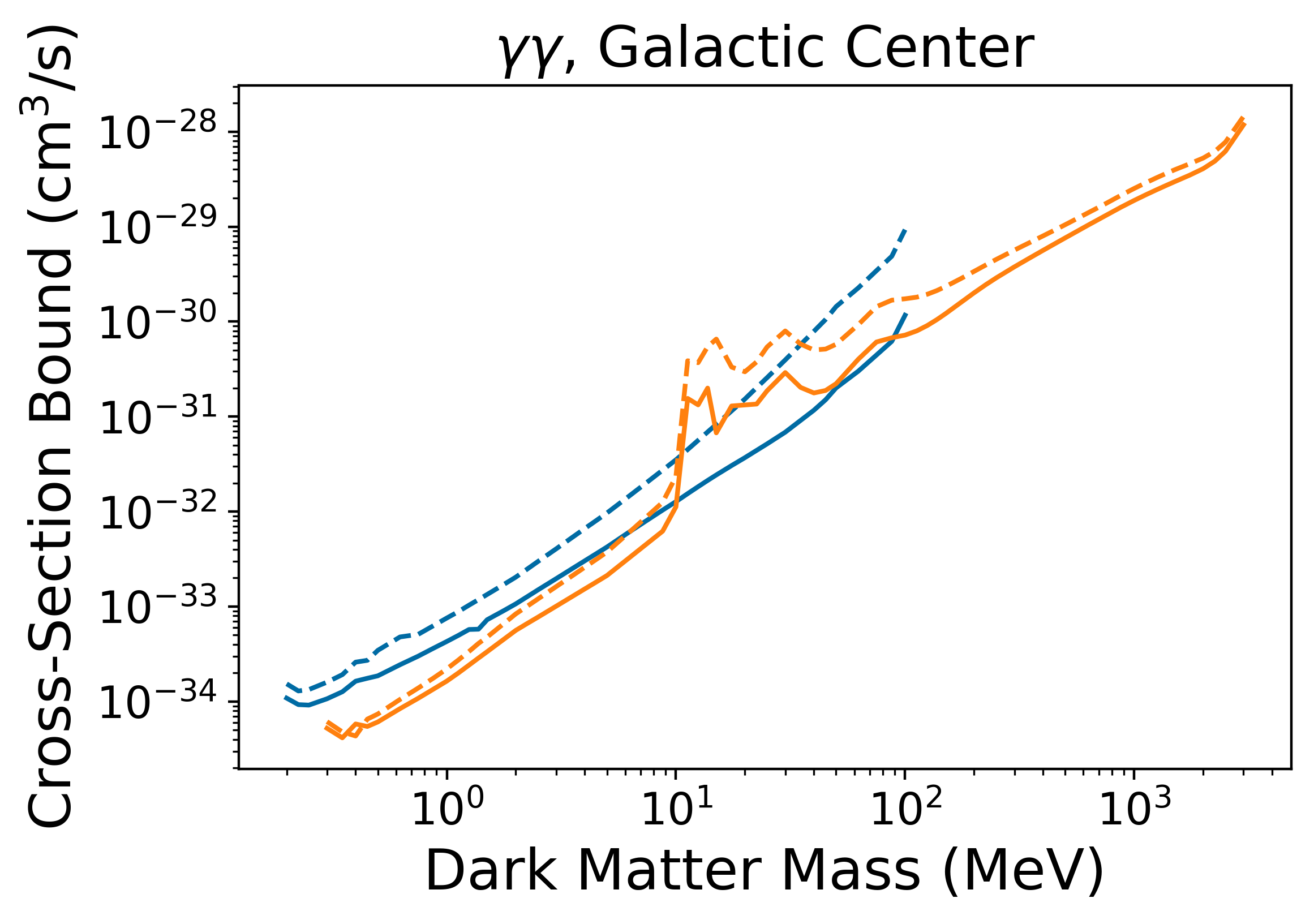}
    \includegraphics[width=.48\textwidth]{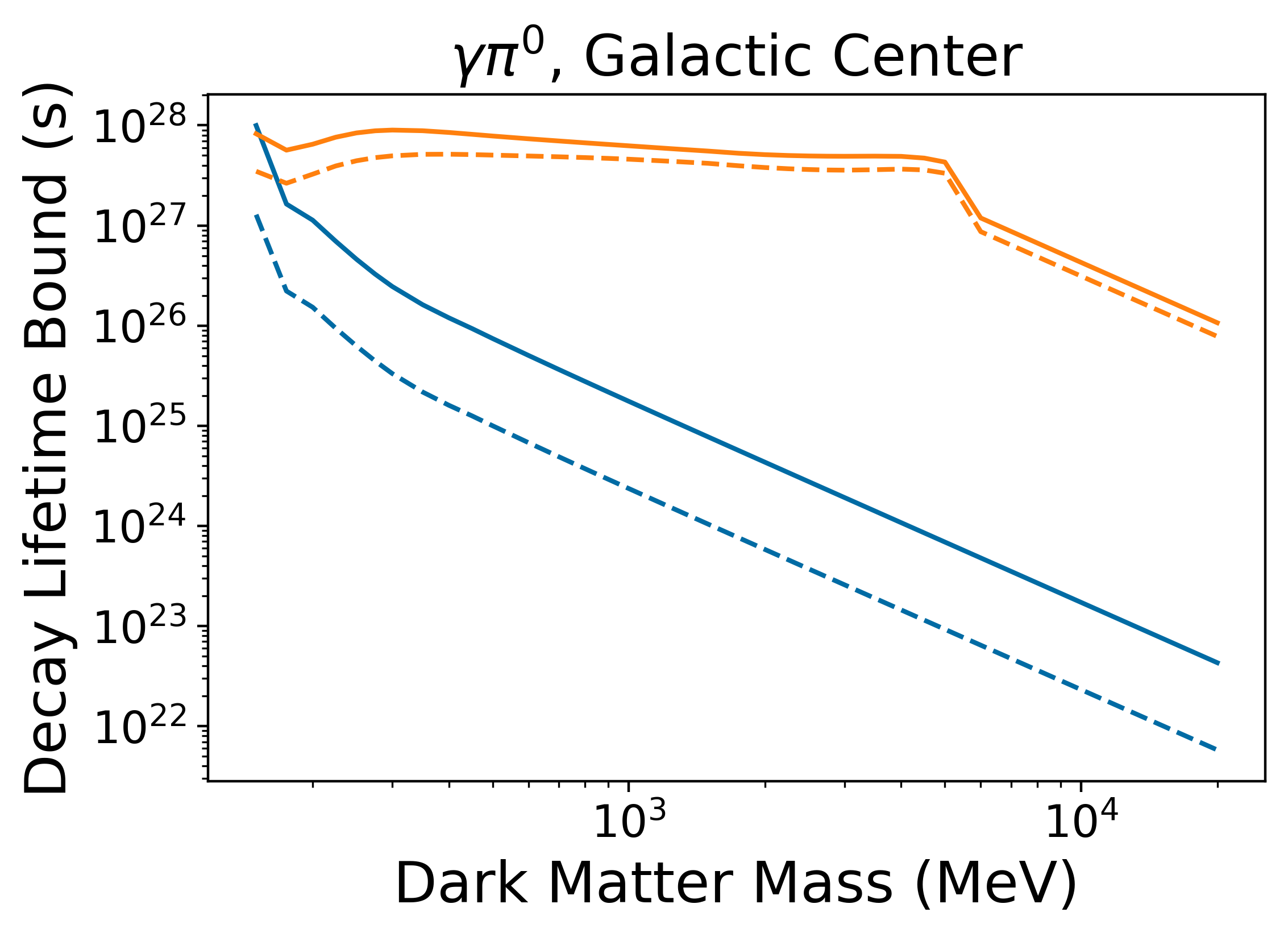} \hfill
    \includegraphics[width=.48\textwidth]{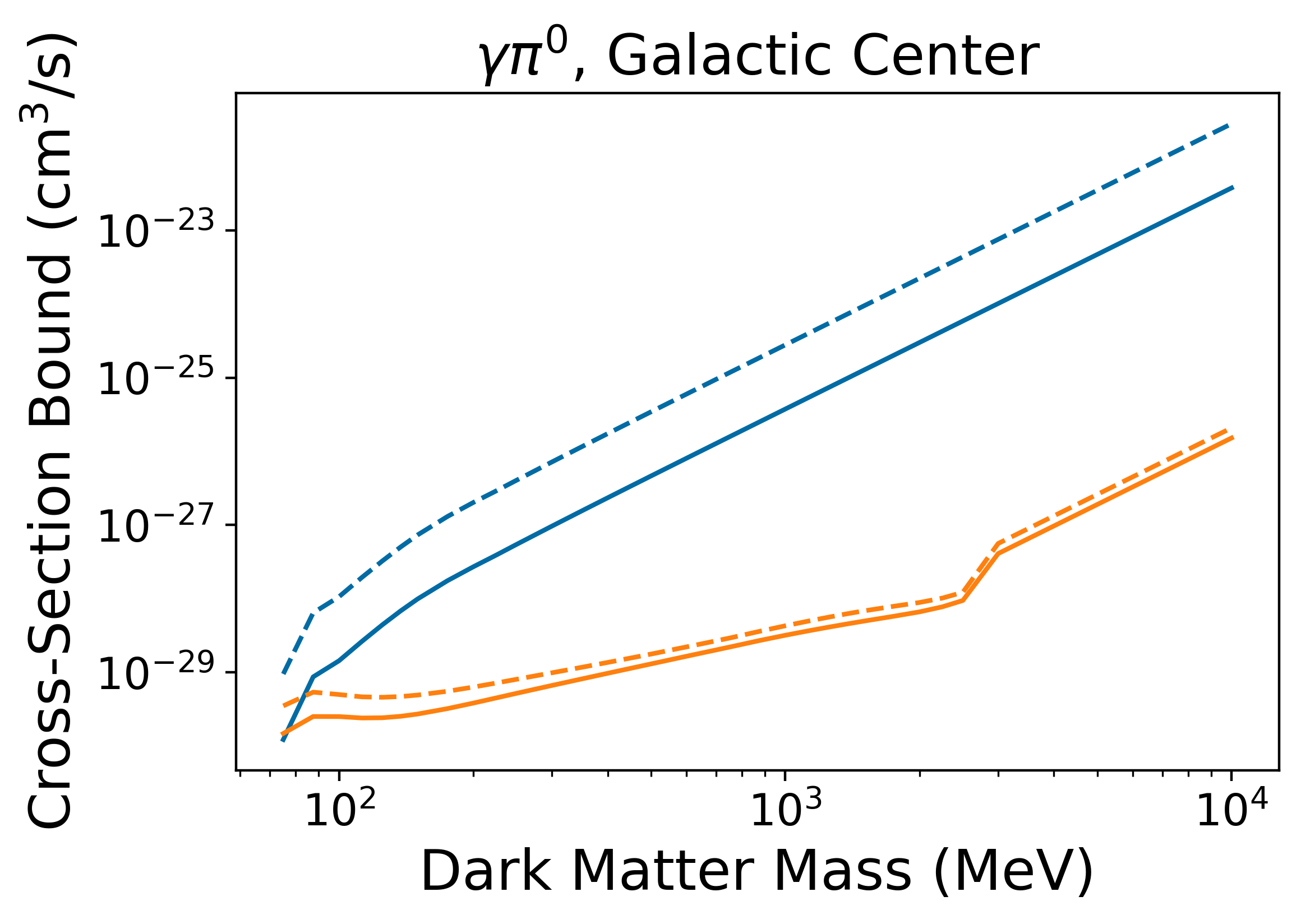}
    \includegraphics[width=.48\textwidth]{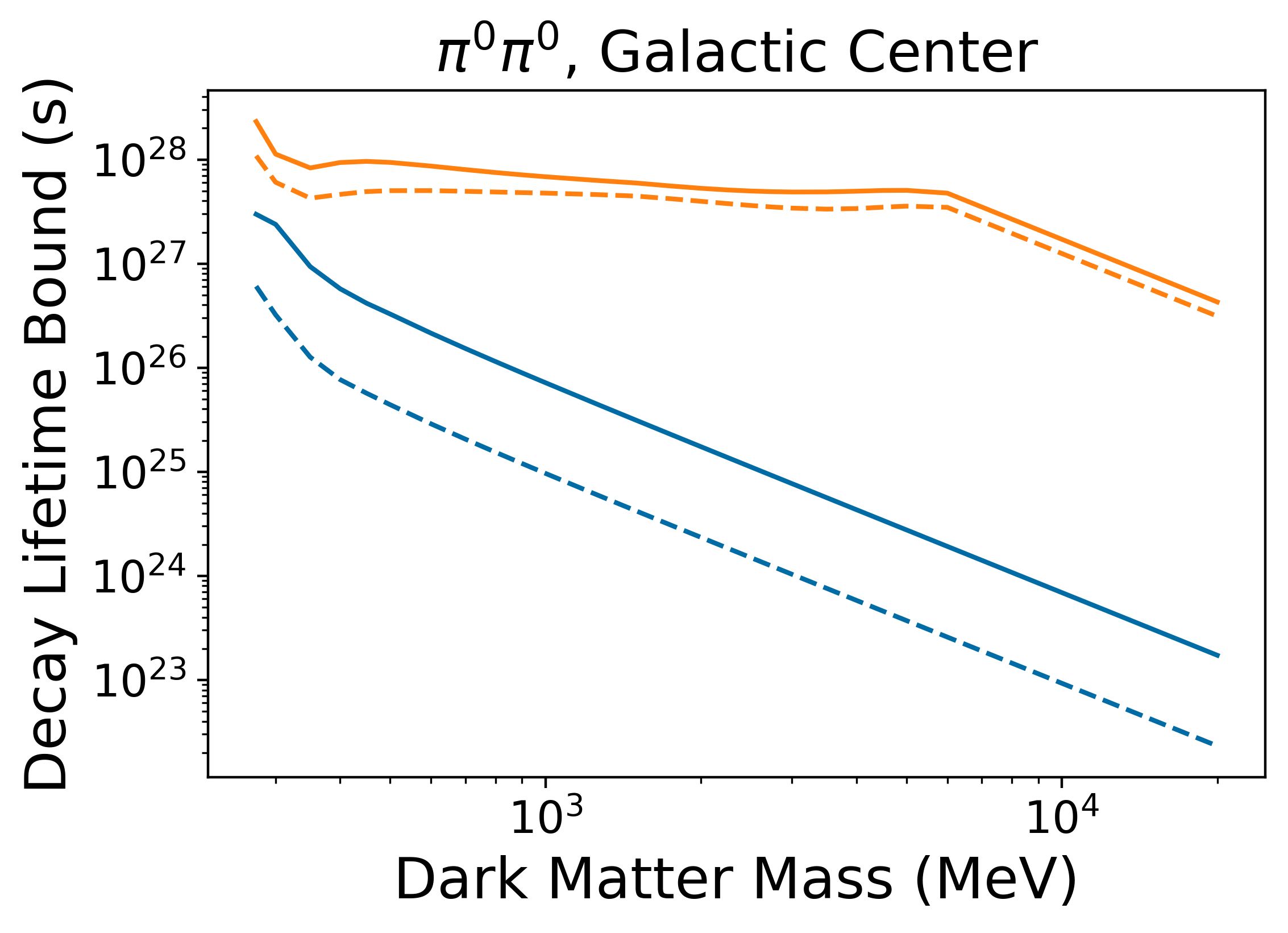} \hfill
    \includegraphics[width=.48\textwidth]{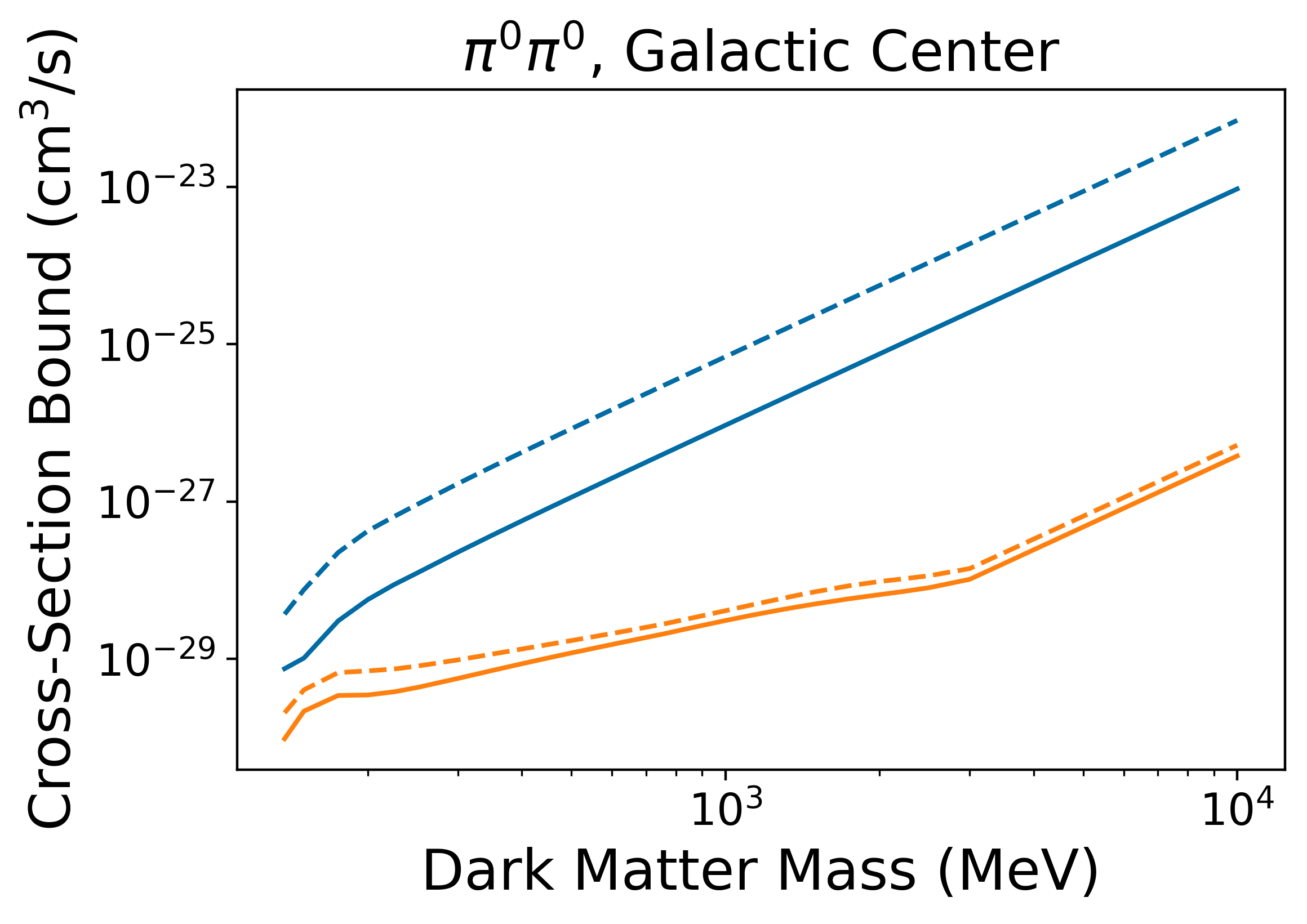}
    \caption{The constraints that the balloon version of GRAMS (blue) and e-ASTROGAM (orange), pointed at the Galactic Center, can place on dark matter decay and annihilation into (from top to bottom) two photons, a photon and a neutral pion, and two neutral pions, assuming an observation time of $T_{\mathrm{obs}} = 10^6 \mathrm{\ s}$, taking atmospheric backgrounds into consideration (dashed lines). These are compared against the results we obtained in Sec.~\ref{sec:results} without accounting for atmospheric backgrounds (solid lines).}
    \label{fig:atmospheric_neutral}
\end{figure*}

\begin{figure*}
    \includegraphics[width=.48\textwidth]{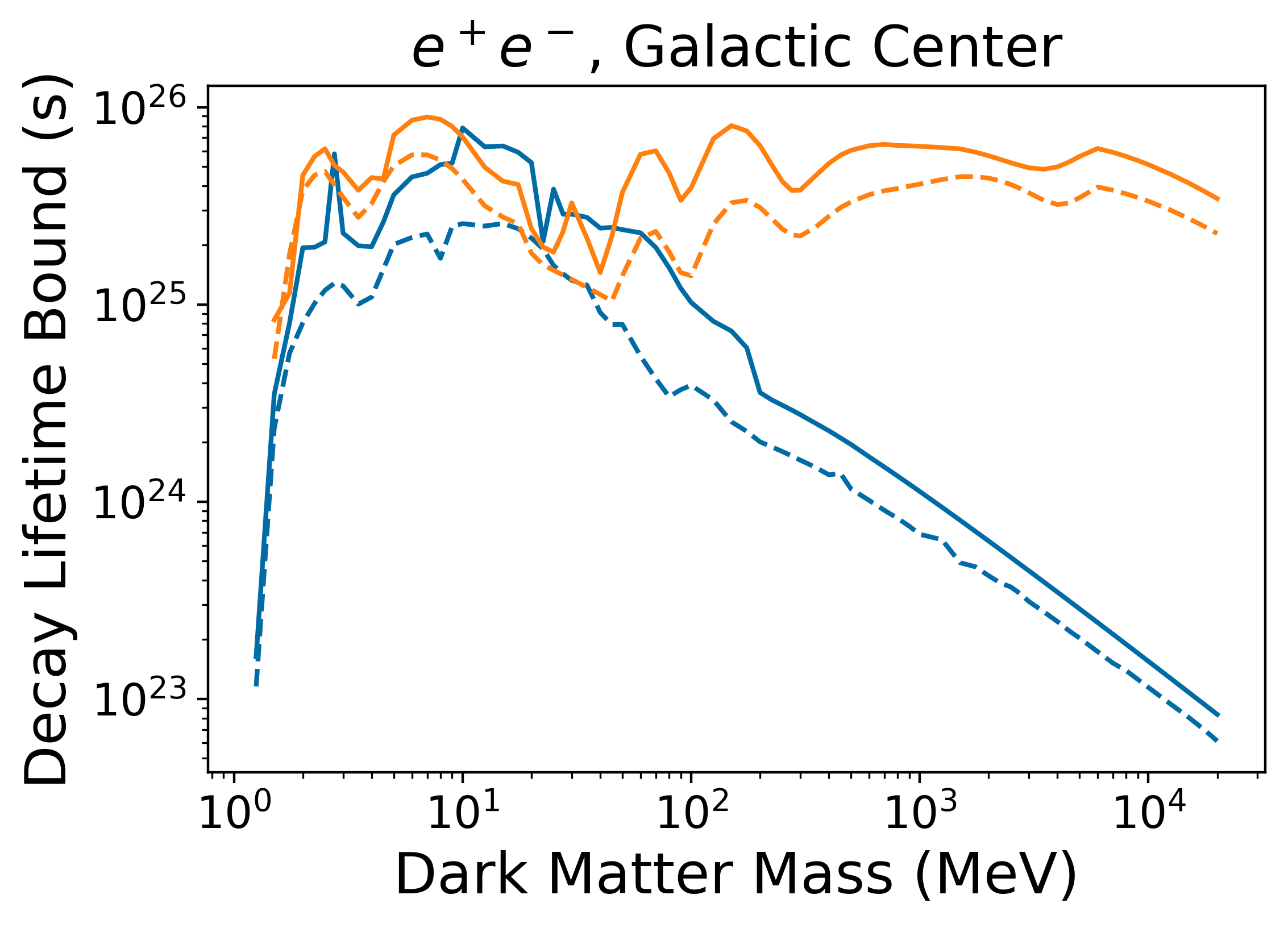} \hfill
    \includegraphics[width=.48\textwidth]{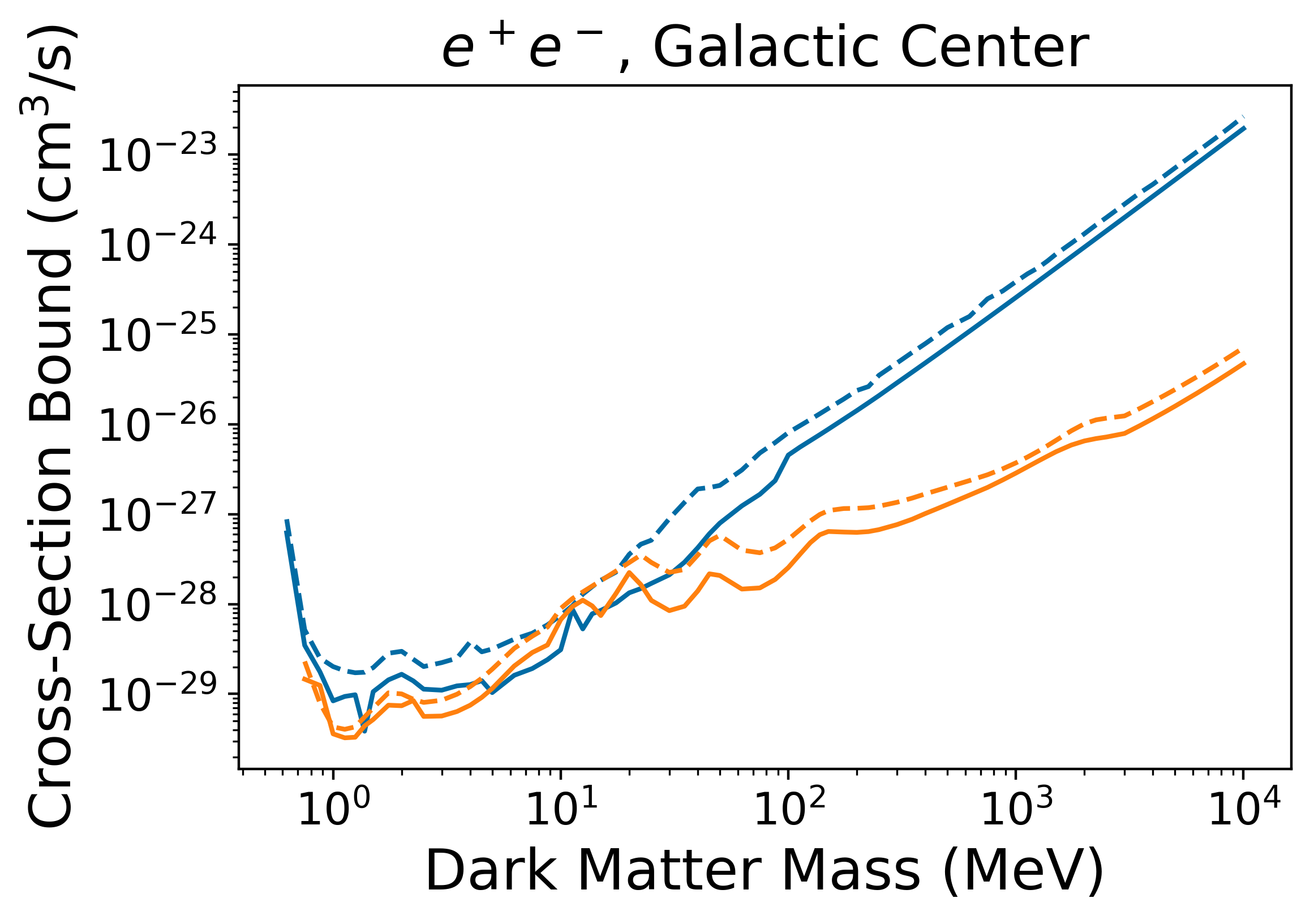}
    \includegraphics[width=.48\textwidth]{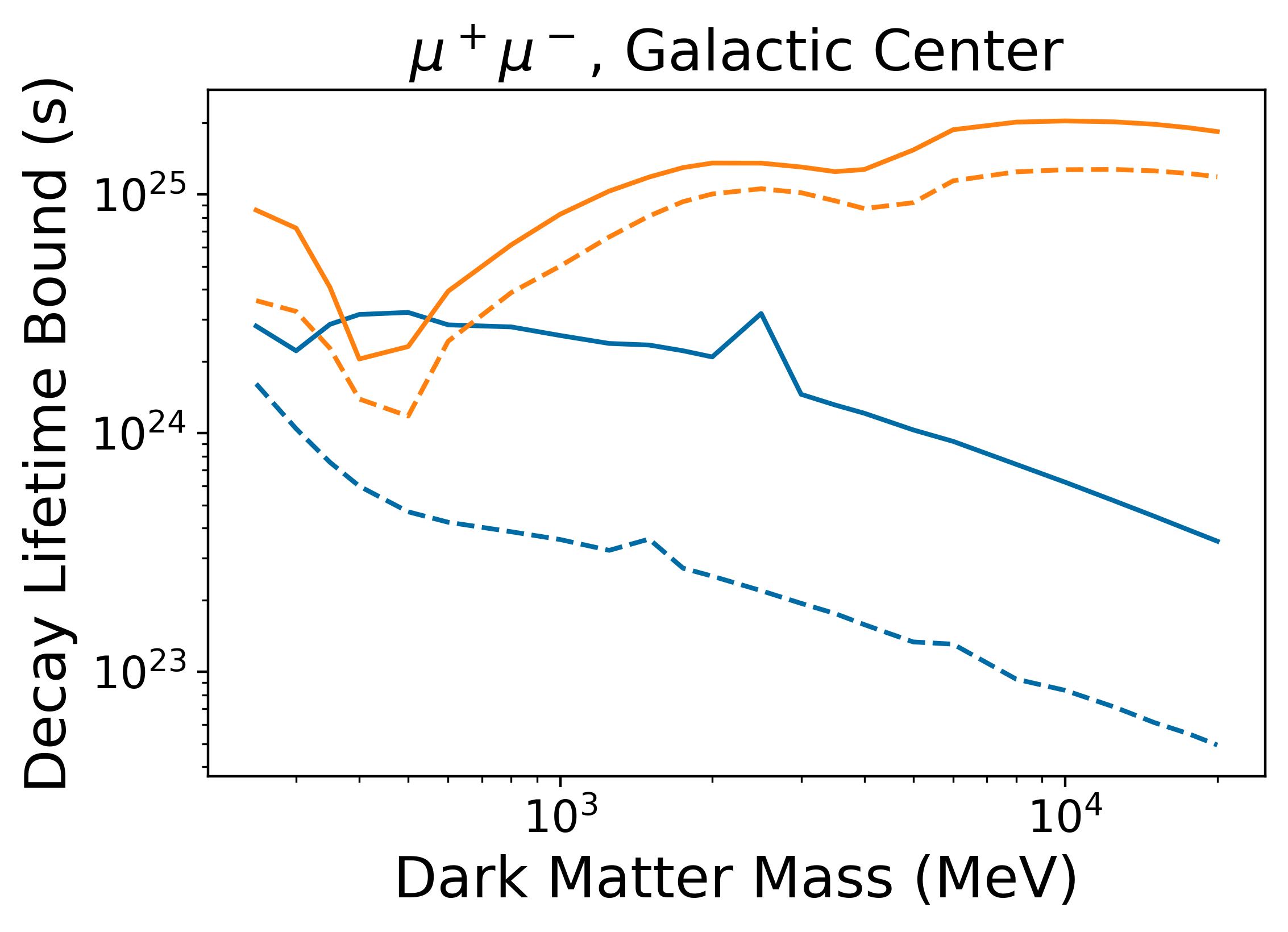} \hfill
    \includegraphics[width=.48\textwidth]{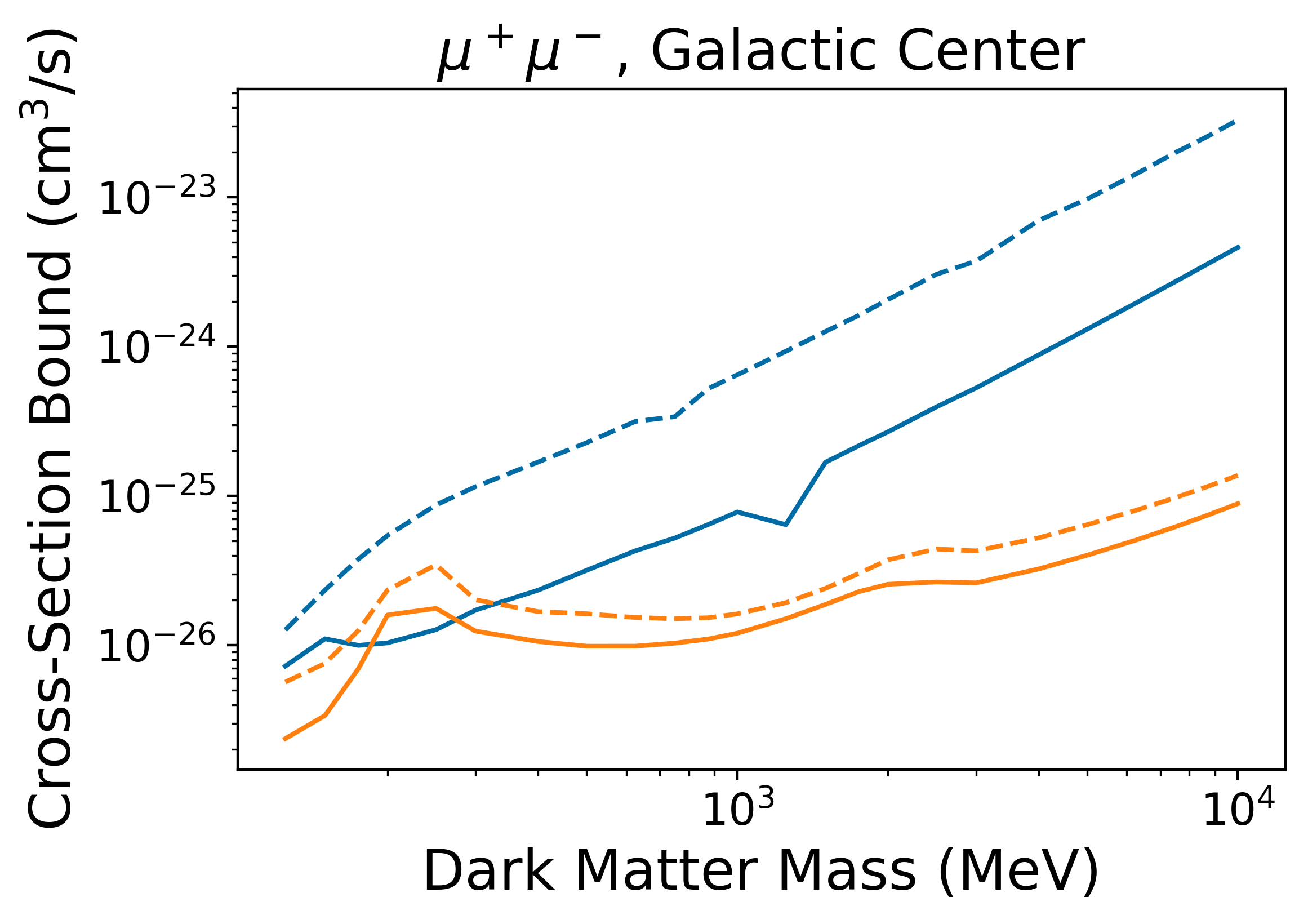}
    \includegraphics[width=.48\textwidth]{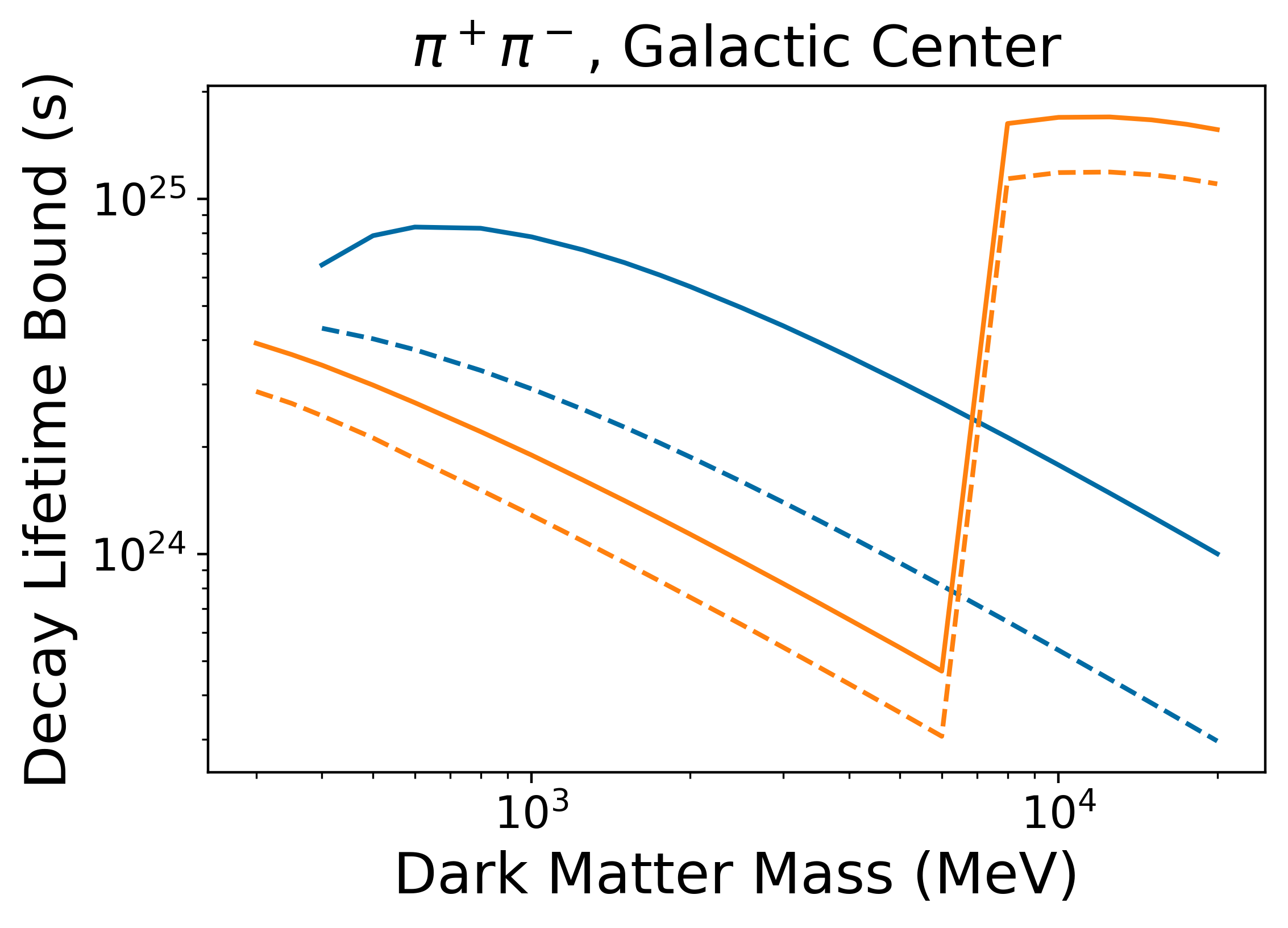} \hfill
    \includegraphics[width=.48\textwidth]{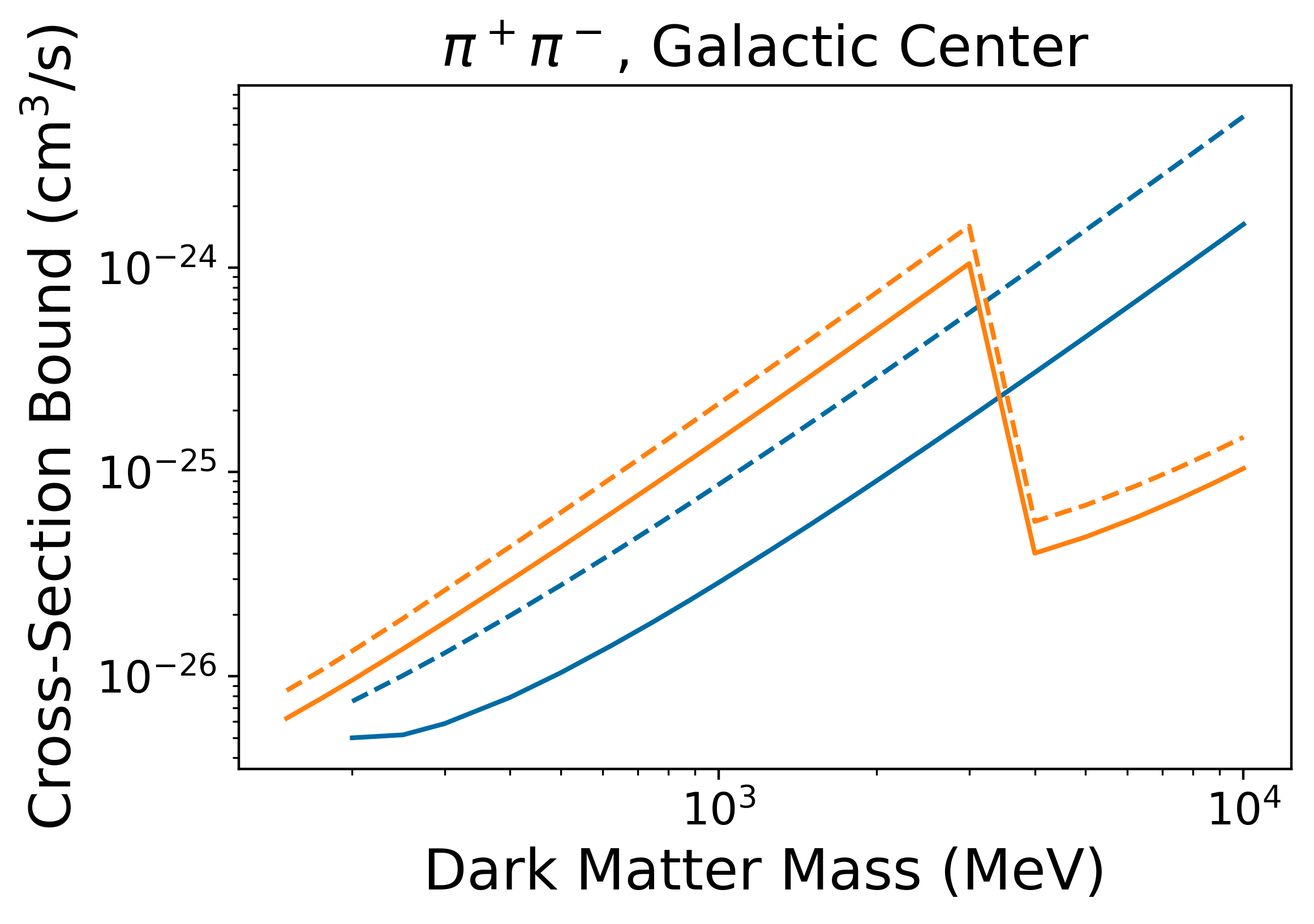}
    \caption{The constraints that the balloon version of GRAMS (blue) and e-ASTROGAM (orange), pointed at the Galactic Center, can place on dark matter decay and annihilation into the final states (from top to bottom) $e^+e^-$, $\mu^+\mu^-$, and $\pi^+\pi^-$, assuming an observation time of $T_{\mathrm{obs}} = 10^6\mathrm{\ s}$, taking atmospheric backgrounds into consideration (dashed lines). These are compared against the results we obtained in Sec.~\ref{sec:results} without accounting for atmospheric backgrounds (solid lines).}
    \label{fig:atmospheric_charged}
\end{figure*}

We now explore how the inclusion of atmospheric backgrounds affects our results. Since there do not yet exist atmospheric background estimates for most of the instruments we consider in this work, we focus on e-ASTROGAM and the balloon version of GRAMS in this section as examples of time-projection chamber and silicon-tracker technology respectively. An atmospheric background estimate for GRAMS is given in Fig.~6 of Ref.~\cite{ARAMAKI2020107}, and an atmospheric background estimate for e-ASTROGAM is given in Fig.~18 of Ref.~\cite{e-ASTROGAM:2016bph}. In both cases, we assume the atmospheric background is isotropic and well measured (i.e.~we do not include any free parameters associated with the atmospheric background in our Fisher analysis). The effect of including the atmospheric backgrounds is plotted in Figs.~\ref{fig:atmospheric_neutral} and \ref{fig:atmospheric_charged}. Overall, we find that accounting for atmospheric backgrounds changes our results by a factor of a few in the case of e-ASTROGAM and less than an order of magnitude in the case of GRAMS. Strategies to further reduce these backgrounds, or to effectively subtract them (e.g. by cuts or by direct measurement of the background in a region where it is bright) could mitigate this modest loss of sensitivity.

\end{document}